\documentclass[pra,letterpaper,twocolumn,aps,footinbib,superscriptaddress,showpacs,babel]{revtex4-2}
\usepackage{amsmath,mathtools,upgreek,amssymb,graphicx,paralist,dsfont,color,transparent, float}
\usepackage[colorlinks = true, linkcolor = blue, urlcolor  = blue, citecolor = blue, anchorcolor = blue]{hyperref}
\usepackage{comment}
\usepackage{color}

\newcommand{\ket}[1]{|#1\rangle}
\newcommand{\bra}[1]{\langle#1|}
\newcommand{\braket}[1]{\langle#1\rangle}
\newcommand{\tr}{\mathrm{Tr}}
\newcommand{\kett}[1]{|\negmedspace #1 \rangle}
\newcommand{\braa}[1]{\langle #1 \negmedspace|}
\newcommand{\dd}{\mathrm{d}}
\newcommand{\id}{\mathds{1}}

\begin{document}

\title{Bayesian Frequency Metrology with Optimal Ramsey Interferometry in Optical Atomic Clocks}

\author{T. Kielinski}
\affiliation{Institute for Theoretical Physics and Institute for Gravitational Physics (Albert-Einstein-Institute), Leibniz University Hannover, Appelstrasse 2, 30167 Hannover, Germany}

\author{K. Hammerer}
\email{klemens.hammerer@itp.uni-hannover.de}
\affiliation{Institute for Theoretical Physics and Institute for Gravitational Physics (Albert-Einstein-Institute), Leibniz University Hannover, Appelstrasse 2, 30167 Hannover, Germany}

\begin{abstract}
Frequency metrology is a cornerstone of modern precision measurements and optical atomic clocks have emerged as the most precise measurement devices. In this progress report, we explore various Ramsey interrogation schemes tailored to optical atomic clocks primarily limited by laser noise. To incorporate frequency fluctuations directly into the theoretical model, we consider a Bayesian framework. In this context, we review fundamental bounds arising in Bayesian estimation theory, which serve as a benchmark throughout this work. We investigate the trade-off between entanglement-enhanced sensitivity and robustness against laser noise in order to identify optimal initial states, measurement schemes and estimation strategies. Beside standard protocols based on coherent spin states, squeezed spin states and GHZ states, we consider variational Ramsey protocols implemented via low-depth quantum circuits based on one-axis twisting operations to approach optimal stability. In particular, we review known and identify new optimal interrogation schemes for a variety of scenarios, including different experimental platforms, ensemble sizes and regimes characterized by a wide range of interrogation durations and dead times. Hence, this work establishes a comprehensive theoretical framework for optimizing Ramsey interrogation schemes, providing guidance for the development of next-generation optical atomic clocks.
\end{abstract}

\maketitle

\tableofcontents

\begin{table}[h]
    \centering
    
    \begin{tabular}{l|l|l}
        \textbf{Acronym} & \textbf{Definition} & \textbf{Reference} \\
        \hline
        ADEV & Allan deviation & Sec.~\ref{sec:ADEV} \\
        BMSE & Bayesian mean squared error & Sec.~\ref{sec:BayesianPhaseEstimation} \\
        BCRB & Bayesian Cram\'{e}r-Rao bound & Sec.~\ref{sec:Bounds} \\
        BQCRB & Bayesian quantum Cram\'{e}r-Rao Bound & Sec.~\ref{sec:Bounds} \\
        CRB & Cram\'{e}r-Rao bound & Sec.~\ref{sec:Bounds} \\
        CSS & coherent spin states & Sec.~\ref{sec:standard_protocols} \\
        CTL & coherence time limit & Sec.~\ref{sec:standard_protocols} \\
        GHZ & Greenberger-Horne-Zeilinger & Sec.~\ref{sec:standard_protocols} \\
        HL & Heisenberg limit & Sec.~\ref{sec:standard_protocols} \\
        $\pi$HL & $\pi$-corrected Heisenberg limit & Sec.~\ref{sec:Bounds} \\
        LO & local oscillator & Sec.~\ref{sec:Motivation} \\
        MSE & mean squared error & Sec.~\ref{sec:BayesianPhaseEstimation} \\
        OAT & one-axis-twisting & Sec.~\ref{sec:standard_protocols} \\
        OQI & optimal quantum interferometer & Sec.~\ref{sec:Bounds} \\
        POI & phase operator based interferometer & Sec.~\ref{sec:Bounds} \\
        QFI & quantum Fisher information & Sec.~\ref{sec:Bounds} \\
        QPN & quantum projection noise & Sec.~\ref{sec:Estimators} \\
        SLD & symmetric logarithmic derivative & Sec.~\ref{sec:Estimators} \\
        SQL & standard quantum limit & Sec.~\ref{sec:Estimators} \\
        SSS & squeezed spin states & Sec.~\ref{sec:standard_protocols} \\
    \end{tabular}
    \caption{Acronyms.}
\end{table}

\section{Introduction}

    Frequency metrology constitutes a fundamental pillar in modern precision measurements, driving advancements across a wide range of scientific and technological fields~\cite{Pezze2018,Ye2024,Lombardi2003,Levine1999,Audoin2001,Riehle2003}. At the forefront of this discipline are optical atomic clocks, which exploit narrow-linewidth atomic transitions in the optical domain~\cite{Ludlow2015,Colombo2022}. This new generation of clocks was spurred by technological advances over the past decades, including breakthroughs in laser technology~\cite{Ludlow2015}, the invention of the optical frequency comb~\cite{Udem2002,BookCombs2005}, and the development of highly controllable platforms such as ion-traps~\cite{steinel2023,pelzer2023,Hausser2024}, tweezer-arrays~\cite{madjarov2019,norcia2019,young2020,shaw2024}, and optical lattices~\cite{Takamoto2015,Katori2003,Masoudi2015,Katori2011}. Today, state-of-the-art optical atomic clocks represent the most precise measurement devices ever built, achieving stabilities on the order of $10^{-18}$ and below~\cite{Guideline2019,Oelker2019,Schioppo2016,Nicholson2015,Bloom2014,PedrozoPeafiel2020,Aepelli2024,Keller2019,Hausser2024,Aharon2019,Sanner2019,Huang2017}. To illustrate this incredible precision, such clocks would gain or lose less than a second over the age of the universe. They have surpassed traditional microwave-based Caesium atomic clocks, which long served as the standard for timekeeping, paving the way for the redefinition of the SI second~\cite{Ludlow2015,Dimarcq2024}. This unprecedented stability renders optical clocks indispensable tools for a broad spectrum of applications. In research, they are instrumental in probing fundamental physics, from testing general relativity through gravitational redshift measurements~\cite{Takamoto2020,Chou2010,Bothwell2022,Dimarcq2024,Dreissen2022,Sanner2019} to exploring variations in fundamental constants~\cite{Safronova2019,Roberts2020} and searching for new physics beyond the Standard Model~\cite{Derevianko2014,Safronova2018,Kennedy2020}. In technology, optical atomic clocks foster potential applications ranging from enhancing global navigation satellite systems~\cite{Schuldt2021,Schmidt2023} and synchronizing large-scale networks~\cite{Droste2013} to supporting precision geodesy~\cite{Lisdat2016,Grotti2018,Mehlstubler2018,Grotti2024}.

    Quantum projection noise (QPN) is the most fundamental process limiting clock stability, arising from the stochastic nature of quantum measurements and the discrete outcomes inherent in finite-size ensembles~\cite{Itano1993,Gardiner2004}. For uncorrelated atoms, the standard quantum limit (SQL) imposes a fundamental bound on QPN~\cite{Ludlow2015,Pezze2018,Colombo2022}. However, stability beyond this classical limit can be achieved by introducing entanglement within the atomic ensemble~\cite{Pezze2018,Colombo2022}. Three decades ago, Wineland \textit{et al.} proposed in seminal works~\cite{Wineland1992,Wineland1994} to entangle cold ions via their common coupling to collective modes of motion to suppress projection noise in frequency spectroscopy, thereby overcoming the SQL and enhancing atomic clock stability. With the momentous advancements in optical atomic clocks and programmable quantum processors since then, this vision now encounters new opportunities and challenges. In recent years, entanglement on optical clock transitions has been demonstrated in various setups, including the generation of spin squeezing in trapped ions~\cite{franke2023} and in neutral atoms mediated by cavities~\cite{PedrozoPeafiel2020,Robinson2024} or Rydberg interactions~\cite{Eckner2023}. Recently, also maximally entangled GHZ states and cascades thereof  have been realized in optical clocks based on tweezer-arrays~\cite{Finkelstein2024,Cao2024}. In an ideal scenario, these GHZ states saturate the Heisenberg limit, representing the ultimate bound on projection noise and yielding a quadratic improvement over the SQL~\cite{Pezze2018}.

    However, in realistic scenarios, decoherence processes and external noise degrade the coherence of the quantum system, impairing the stability and preventing the achievement of the Heisenberg limit~\cite{Shaji2007,Demkowicz_Dobrza_ski_2012,Escher2011,Knysh2014}. While entanglement promises to overcome the SQL and thereby improving clock stability, the detrimental effects of decoherence are particularly pronounced in entangled states, since they are highly susceptible to the loss of coherence. In particular, Huelga \textit{et al.} have demonstrated that GHZ protocols suffer significantly from dephasing associated with random phase changes caused by stray fields or laser noise, ultimately showing no improvement over the SQL~\cite{Huelga1997}. In contrast, GHZ-like states offer optimal stability for clocks with small ensembles that are primarily limited by spontaneous decay~\cite{Kielinski2024}. As a consequence, incorporating decoherence effects and external noise is crucial for identifying optimal interrogation protocols in frequency metrology.

    Unlike magnetic field fluctuations or laser noise, the finite lifetime of qubits in the excited state represents a fundamental limit rather than an external noise source, a limitation examined in detail in Ref.~\cite{Kielinski2024}. State-of-the-art clock lasers achieve coherence times of several seconds~\cite{Matei2017}, entering the regime of the excited-state lifetime of various clock candidates, such as Sr$^+$-ions ($0.4\,\mathrm{s}$) or Ca$^+$-ions ($1.1\,\mathrm{s}$). With further technological improvements in the short-term laser stabilization, coherence times will potentially approach lifetimes of further clock species as Yb-atoms ($15.9\,\mathrm{s}$) or Al$^+$-ions ($20.7\,\mathrm{s}$). Nevertheless, the excited-state lifetimes of several clock candidates remain far beyond the regime of laser coherence times, ranging from minutes (Sr-atoms) to years (Yb$^+$-ions)~\cite{Guideline2019}. Moreover, the impressive level of laser coherence  is often degraded during propagation from the cavity to the location of the qubits. Consequently, a significant number of setups are currently, and will remain in the future, limited by laser noise.

    Naively, frequency fluctuations and the associated laser noise could be regarded as a purely technical problem. However, stabilizing the laser is precisely the central objective of an atomic clock, making frequency fluctuations the primary measurand~\cite{Riehle2003}. Disregarding laser noise as a mere technical issue would thus contradict the fundamental concept of atomic clocks. In principle, one might also ask how laser noise can impose a limiting factor although, by definition, it is the measurand, the quantity to be stabilized. To be precise, only the component of laser noise that cannot be corrected through interrogation of the atomic reference ultimately limits clock stability. Since Ramsey protocols have a finite range within which they can unambiguously interpret frequency fluctuations, errors arise when laser noise exceeds this range, fundamentally constraining stability. Additionally, dead time in clock operation leads to undetected aliased frequency deviations, further degrading performance. Investigating the impact of frequency fluctuations is therefore essential for advancing next-generation clocks. To address this challenge, various approaches have been developed to account for frequency noise and to determine optimal interrogation schemes for specific experimental setups~\cite{Leroux2017,Schulte2020,Fraas2016,Demkowicz_Dobrza_ski_2015}. A particularly promising framework in this endeavor is Bayesian frequency metrology~\cite{Han2024,Macieszczak2014,Kaubruegger2021,Thurtell2024}, which leverages Bayesian estimation theory to incorporate laser noise directly into the theoretical model.

    As a consequence, determining optimal interrogation protocols involves a trade-off between achieving entanglement-enhanced sensitivity, which enables surpassing the SQL, and ensuring robustness against noise. In recent years, operationally motivated echo protocols and variational quantum circuits have attracted significant interest, as they allow for a diverse range of interrogation schemes~\cite{SchulteEcho2020,Scharnagl2023,Kaubruegger2021,Thurtell2024,Marciniak2022,Li2023,Leibfried2004,Davis2016,Fröwis2016,Macri2016,Nolan2017,Colombo2022a,Burd2019,Anders2018}. In particular, these approaches have the potential to generate a high degree of entanglement while maintaining resilience to noise~\cite{SchulteEcho2020,Scharnagl2023,Kaubruegger2021,Thurtell2024,Marciniak2022,Li2023}.  One-axis-twisting (OAT)~\cite{Kitagawa1993} interactions serve as a versatile tool for implementing such protocols as they give rise to a variety of entangled states, ranging from squeezed spin states (SSS) to GHZ states, and facilitate variational classes of generalized Ramsey protocols~\cite{SchulteEcho2020,Scharnagl2023,Kaubruegger2021,Thurtell2024,Marciniak2022,Li2023}. Furthermore, OAT interactions are accessible in several setups as in ion-traps via Mølmer-Sørensen gates~\cite{Benhelm2008,Blatt2008,Leibfried2004}, in tweezer-arrays via Rydberg interactions~\cite{Eckner2023,Hines2023} or Bose-Einstein condensates via elastic collisions~\cite{Srensen2001,Estve2008,Gross2010,Riedel2010}.

    This work presents a progress report on frequency metrology tailored to optical atomic clocks primarily limited by laser noise. The objective is to outline potential advancements and challenges across various Ramsey interrogation schemes, effectively providing a theoretical guide for clock operation on different experimental platforms. In particular, we systematically examine a broad range of ensemble sizes and regimes defined by interrogation duration and dead time. To incorporate frequency fluctuations into the theoretical model, we employ a Bayesian framework for single-ensemble clocks, where the atomic reference is periodically interrogated using the same protocol in each clock cycle, while more general schemes are addressed in the outlook. To establish a theoretical foundation, we review Bayesian estimation theory and the corresponding fundamental bounds on clock stability. Additionally, we incorporate previous findings on clocks limited by laser noise, such as those in Refs.~\cite{Schulte2020,Leroux2017}, within the Bayesian framework and extend them in certain regimes. Building on pioneering work in Refs.~\cite{Kaubruegger2021,Marciniak2022,Thurtell2024} on variational quantum circuits, we identify optimal Ramsey schemes for various experimental platforms.\\

    In the following, we provide a brief overview of each section and outline the primary results:
    \begin{itemize}
        \item Sec.~\ref{sec:BayesianFrequencyMetrology}: We introduce the fundamental principles of atomic clocks and establish the connection between frequency metrology and phase estimation theory. By highlighting the impact of local oscillator noise on clock stability, we motivate the Bayesian approach as an effective framework for atomic clocks primarily limited by frequency fluctuations of the local oscillator. To start with, the framework of Bayesian phase estimation theory is introduced and a hierachy of lower bounds on the estimation uncertainty is collected, drawing an analogy to the local (frequentist) approach. In particular, the ultimate lower bound is derived, denoted as the optimal quantum interferometer (OQI), which represents the primary benchmark in this work. Additionally, the linear estimation strategy is discussed and the optimal Bayesian estimator is determined. By introducing the Allan deviation, we explicitly connect Bayesian phase estimation theory to frequency metrology and establish a relation between interrogation time and prior knowledge of the phase. Furthermore, we discuss general trade-offs in the context of frequency metrology. Hence, this section introduces the fundamental concepts and establishes the theoretical framework of this work.
        \item Sec.~\ref{sec:Theory}: This section aims to saturate the ultimate lower bound imposed by the optimal quantum interferometer (OQI). Initially, the standard protocols, utilizing coherent spin states (CSS), squeezed spin states (SSS) and GHZ states, are compared to the OQI. While GHZ states saturate the OQI at short interrogation times and SSS perform close to it at intermediate durations, substantial potential for enhancement remains across a broad range of interrogation times, particularly at long durations. To address this, especially considering small ensemble sizes characteristic of ion-traps and tweezer-arrays, we introduce generalized Ramsey protocols based on variational quantum circuits and identify optimal interrogation schemes. We demonstrate that in this regime, even low-depth quantum circuits suffice to approximate the OQI, which is crucial for maintaining reasonable operational complexity and thus enabling near-term experimental implementation. While the required circuit depth to achieve OQI stability increases with $N$, the performance gain diminishes with complexity, leading to a trade-off between reduced instability and increased experimental overhead, further motivating a focus on low circuit-complexity approaches.
        \item Sec.~\ref{sec:Simulations}: To validate theoretical predictions on clock stability, we perform Monte Carlo simulations of the full feedback loop in an atomic clock, cf. Fig.~\ref{fig:schematic}, from which we can infer its long-term stability as quantified by the Allan deviation. In this context, fringe hops emerge as a significant limitation. In particular, for small ensembles ($N\lesssim20$), characteristic of ion-traps, fringe hops impose a stricter constraint on clock stability than the coherence time limit (CTL) of the local oscillator. As a consequence, for long interrogation times, variational protocols provide marginal to no advantage over SSS, while GHZ states remain optimal at short interrogation times. In contrast, for ensembles sizes in the regime of tweezer-arrays ($N\gtrsim20$), fringe hops and the CTL impose comparable limitations on clock stability at long interrogation times. Consequently, variational Ramsey protocols provide a substantial improvement over SSS. Nevertheless, the variation in stability across different clock runs, due to the stochastic nature of atomic clocks, and the relative reduction in enhancement with increasing circuit depth, further supports the focus on low-depth quantum circuits.
        \item Sec.~\ref{sec:deadtime}: Incorporating dead time into atomic clock operation within the framework of Bayesian frequency metrology, we investigate the trade-off between quantum projection noise (QPN), the coherence time limit (CTL), and dead time effects. While clock stability for short dead times or small ensembles closely resembles the dead time-free scenario, dead time effects become increasingly significant with growing ensemble size or dead time, ultimately limiting clock performance. Following a general analysis, we examine specific examples with state-of-the-art parameters relevant to different experimental platforms, such as ion-traps, tweezer-arrays and lattice clocks. While GHZ states and SSS remain optimal for ion-traps utilizing only a few ions, the potential gain from variational quantum circuits in tweezer-arrays with several tens of atoms is substantially diminished. Specifically, SSS perform close to the optimal quantum interferometer (OQI) across a wide range of interrogation times, whereas variational quantum circuits offer an enhancement only at long interrogation times. However, this improvement is significantly reduced compared to the dead time-free case. Additionally, in the presence of dead time, fringe hops remain the dominant limitation in this regime, whereas in the dead time-free case, they constrain clock stability only at the same level as the CTL. As a consequence, SSS emerge as the preferred choice due to their robustness and practicality. For lattice clocks with hundreds or thousands of atoms, dead time effects strictly constrain clock stability and thus CSS suffice to approximate the OQI.
    \end{itemize}

\section{Bayesian frequency metrology}\label{sec:BayesianFrequencyMetrology}

    \subsection{Motivation}\label{sec:Motivation}

        A clock, at its core, consists of two essential components: a frequency standard, a device which generates a continuous and consistent frequency signal, and a mechanism that counts the oscillations over time. While the clockwork device essentially translates the frequency signal into measurable time intervals, the frequency standard represents the true heart of a clock.~\cite{Riehle2003} Frequency standards are commonly classified as either active or passive, depending on their operational principle. Active frequency standards generate their own oscillation at a given frequency, as the hydrogen maser or the Helium-Neon laser, where stimulated emission results in a highly coherent signal. Conversely, passive frequency standards require an external source to stimulate their oscillation. While active frequency standards typically excel in short-time stability, passive frequency standards often achieve superior long-term stability and accuracy, because the frequency can be precisely monitored and corrected against the reference response over time. Consequently, passive frequency standards are commonly preferred for clocks.~\cite{Levine1999,Audoin2001,Riehle2003}

        The concept of a passive frequency standard can be illustrated by imagining two pendulums. The first pendulum is our primary, noisy pendulum, whose fluctuating frequency we aim to stabilize. The second pendulum serves as an (almost) ideal reference, though it does not oscillate on its own. Hence, the task of a passive frequency standard is to periodically adjust the primary pendulum’s frequency to match that of the reference pendulum by repeatedly measuring the frequency difference between the two.~\cite{Riehle2003} However, each measurement introduces some noise into the system. Hence, it is desirable to extend the interrogation time as long as possible, thereby reducing the relative impact of this measurement noise and ultimately enhancing stability. If the interrogation time is extended too far, however, we risk missing a “tick” of the reference, leading to synchronization errors that may accumulate over repeated measurements. Therefore, while longer interrogation times improve stability, there is an optimal duration beyond which stability is compromised.~\cite{Ludlow2015,Riehle2003,Audoin2001}

\begin{figure*}[tbp]
            \centering
                \includegraphics[width=0.9\textwidth]{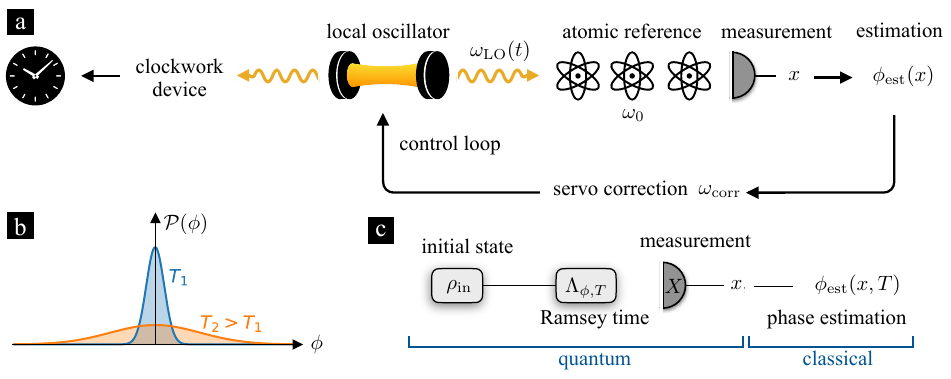}
         
            \caption{(a) Basic principle of an atomic clock: A local oscillator (LO) with fluctuating frequency $\omega_\mathrm{LO}(t)$ is stabilized in a control loop to an atomic transition $\omega_0$. During the free evolution time $T$, the probe state accumulates a phase $\phi$ arising from the frequency deviation. Based on the measurement outcome $x$, the phase is estimated by $\phi_\mathrm{est}$ and the LO frequency is corrected according to $\omega_\mathrm{corr}$ by the servo. (b) Broadening of the prior distribution with increasing interrogation time. (c) Generalized Ramsey interferometry: The phase $\phi$ is encoded during the interrogation time $T$ onto the initial state $\rho_\mathrm{in}$ via $\Lambda_{\phi,T}$. Based on the measurement outcome $x$ of the observable $X$, an estimation $\phi_\mathrm{est}$ of the phase is conducted.}
                \label{fig:schematic}
        \end{figure*}

        In (passive) atomic clocks (cf. Fig.~\ref{fig:schematic}(a)), the local oscillator (LO), representing the primary pendulum, generates an inherently noisy frequency signal $\omega_\mathrm{LO}(t)$ that varies over time $t$. The LO is stabilized to an atomic transition frequency $\omega_0$, acting as the reference pendulum, through repeated interrogations of the atomic ensemble according to a particular Ramsey interferometry scheme. During the Ramsey time $T$, the atoms accumulate a phase $\phi = \int_{t}^{t+T}\dd t'\omega(t')$, which effectively reflects the average of the frequency deviation $\omega(t) = \omega_0 - \omega_{LO}(t)$ over the interrogation period. At the end of each interrogation sequence, a measurement with outcome $x$ is performed, from which an estimate $\phi_\mathrm{est}(x)$ of the monitored phase $\phi$ is derived. The control cycle is completed by the servo that applies feedback to correct the LO frequency by $\omega_\mathrm{corr}$, based on the phase estimate $ \phi_\mathrm{est}(x)$, resulting in a stabilized LO signal. Consequently, frequency metrology is directly connected to phase estimation theory.~\cite{Pezze2018,Ludlow2015,Riehle2003,Audoin2001,Levine1999}

        Local oscillator noise plays a critical role in limiting the clock stability, even though it may seem counterintuitive given that its frequency is constantly measured and corrected based on feedback from the atomic reference. The phase fluctuations, arising from the LO's frequency noise, can be modeled by a distribution $\mathcal{P}(\phi)$, depending on the particular noise profile, by interpreting the phase $\phi$ as a random variable. As the interrogation time $T$ increases, the LO noise grows, causing the distribution to broaden as illustrated in Fig.~\ref{fig:schematic}(b). Typically, the phase can only be estimated unambiguously within a limited range. Hence, if the invertible domain of the main fringe of the signal is exceeded, estimation errors are accumulated and ultimately limit the stability. In the worst case, the feedback loop passes to an adjacent Ramsey fringe, resulting in the clock running systematically wrong and severely degrading the clock stability. Consequently, frequency metrology features a trade-off: longer interrogation times enhance stability, but LO frequency fluctuations introduce limitations. An appropriate framework to investigate different interrogation schemes respecting this trade-off is represented by Bayesian estimation. This approach essentially bridges between high sensitivity at the transition frequency and large dynamic range, based on the regime of frequency deviations the local oscillator is likely to generate.~\cite{Leroux2017,Schulte2020,Fraas2016}

        Building on the motivation outlined above, it becomes evident that phase estimation is a fundamental aspect of frequency metrology, particularly in the context of atomic clocks. In this section, we begin by introducing general phase estimation theory within the Bayesian framework. By reviewing the literature, we establish a hierarchy of lower bounds on the estimation error, drawing an analogy to the local (frequentist) approach, and analyze different estimation strategies. Finally, we link phase estimation to frequency metrology in the context of atomic clocks through the Allan deviation and discuss emerging trade-offs and their implications for the clock stability.

    \subsection{Bayesian phase estimation}\label{sec:BayesianPhaseEstimation}

        In interferometry, the objective is to estimate an unknown parameter $\phi$ as precise and accurate as possible.  In generalized Ramsey spectroscopy (cf. Fig.~\ref{fig:schematic}(c)), the phase $\phi$ is encoded onto the initial probe state $\rho_\mathrm{in}$ during the free evolution time $T$ (Ramsey dark time) via a completely-positive trace-preserving map $\Lambda_{\phi,T}$. Additionally, this quantum channel $\Lambda_{\phi,T}$ may also include decoherence processes such as dephasing or spontaneous decay, with their impact depending on $T$. After the free evolution, an appropriately chosen observable $X$ is measured. The measurement is described by a positive operator-valued measure (POVM) $\{\Pi_x\}$, where $\Pi_x\geq 0$ and $\sum_x\Pi_x=\id$, and $x$ denotes the measurement outcome. Finally, an estimate $\phi_\mathrm{est}(x)$ of the parameter $\phi$ is performed, based on the measurement outcome $x$ (of $X$). In the context of an atomic clock, the phase $\phi = \omega T$ originates from the average frequency difference $\omega = \frac{1}{T}\int_0^T\dd t' [\omega_0-\omega_\mathrm{LO}(t')]$ between the atomic transition $\omega_0$ and the local oscillator $\omega_\mathrm{LO}$, while $X$ typically corresponds to a spin measurement. Due to the inherent indeterministic nature of quantum measurements, the outcomes $x$ are random and occur with conditional probability
        \begin{align}
            P(x|\phi) = \tr\left(\Pi_x\Lambda_{\phi,T}[\rho_\mathrm{in}]\right).
        \end{align}
        Consequently, the estimator $\phi_\mathrm{est}$ likewise is a random variable, as it depends on the measurement outcome $x$. Unlike state preparation, free evolution, and measurement, which are governed by quantum mechanics, the estimation process involves classical post-processing of measurement data and is thus addressed within the framework of classical phase estimation theory.

        In local (or frequentist) phase estimation, it is typically assumed that the phase $\phi$ is tightly centered around a fixed working point $\phi_0$, such that $(\phi -\phi_0)^2\ll 1$, and that the estimator is locally unbiased. Furthermore, probabilities are defined as the infinite-sample limit of an event. However, these assumptions are often not valid in the context of optical atomic clocks. When the finite coherence time of the laser becomes the dominant limitation on clock stability, fluctuations in the accumulated phase during the free interrogation time become relevant and in principle can take arbitrary values $-\infty < \phi < \infty$. Additionally, these fluctuations require phase estimation based on single measurements to ensure unambiguous determination of $\phi$, as the phase may change significantly between measurements of consecutive cycles, potentially preventing a unique estimation or assignment. This constraint makes asymptotic estimation, i.e. the collection and averaging of large amounts of data, impossible. Consequently, Bayesian phase estimation represents the more appropriate framework. 
        
        In Bayesian phase estimation, the phase $\phi$ is treated as a continuous random variable and the posterior knowledge of $\phi$, represented by the posterior distribution $P(\phi|x)$, is updated according to Bayes theorem
        \begin{align}
            P(\phi|x) = \frac{\mathcal{P}(\phi)P(x|\phi)}{P(x)}\label{eq:BayesTheorem}
        \end{align}
        based on the statistical model $P(x|\phi)$ and a prior distribution $\mathcal{P}(\phi)$, reflecting the knowledge on the phase before any measurement. The marginal likelihood $P(x) = \int \dd\phi\, \mathcal{P}(\phi)P(x|\phi)$ represents the probability of observing outcome $x$, averaged over all possible values of $\phi$, and thus, basically provides a normalization of the posterior distribution. The interplay between prior information and measurement data already becomes evident at this stage. If $\mathcal{P}(\phi)$ varies slowly compared to $P(x|\phi)$, for example in the case of a flat prior or in the asymptotic limit of large ensembles, it has minimal influence on the posterior knowledge, and the statistical model primarily governs the inference strategy. Conversely, if the prior is sharply peaked, prior information dominates the estimation process and significantly shapes the posterior distribution.

        A common cost function to quantify the phase estimation uncertainty is the Bayesian mean squared error (BMSE) defined as
        \begin{align}
            (\Delta\phi)^2 &=\int_{-\infty}^{+\infty}  \dd\phi\,  \mathcal{P}(\phi) \sum_x P(x|\phi) \left[\phi - \phi_\mathrm{est}(x)\right]^2.\label{eq:BMSE}
        \end{align}
        The BMSE corresponds to the mean squared error (MSE) of the estimated phase $\phi_\mathrm{est}(x)$ with respect to the true phase value $\phi$, the typical cost function of local phase estimation, averaged over the prior distribution $\mathcal{P}(\phi)$. This reflects a global approach by incorporating all possible values of $\phi$, which additionally makes unbiasedness redundant. Moreover, this approach is well-suited for arbitrary signals and estimation strategies, as it assesses the overall performance by averaging over the entire prior distribution, eliminating the need for specific assumptions about the signal structure or the estimation method. In general, for a proper estimation strategy, information about the phase is gained through the measurement. Consequently, the BMSE is bounded by $0\leq (\Delta\phi)^2 \leq (\delta\phi)^2$. In the limit of narrow prior distributions, where $\mathcal{P}(\phi)$ approximates a delta distribution centered at the optimal working point $\phi_0$, the BMSE reduces to the MSE. Due to its global averaging, the BMSE is always lower bounded by the MSE evaluated at the optimal working point $\phi_0$, where the MSE attains its maximum. Using Bayes theorem, we can express the BMSE in terms of the posterior distribution according to
        \begin{align}
            (\Delta\phi)^2 &=\sum_x P(x) \int_{-\infty}^{+\infty}  \dd\phi\,  P(\phi|x) \left[\phi - \phi_\mathrm{est}(x)\right]^2.\label{eq:BMSE2}
        \end{align}
        
        For the primary investigations in this work, we assume a unitary phase evolution through the quantum channel
        \begin{align}
            \Lambda_{\phi,T}[\rho_\mathrm{in}] = \mathcal{R}_z(\phi) \rho_\mathrm{in} \mathcal{R}_z^\dagger(\phi)\label{eq:unitaryEvolution}
        \end{align}
        with rotation $\mathcal{R}_z(\phi) = e^{-i\phi S_z}$, where $S_{x,y,z}$ denote the collective spin operators of $N$ two level systems. Consequently, the quantum channel, and thus the statistical model $P(x|\phi)$, is $2\pi$-periodic with respect to the phase, i.e. $\Lambda_{\phi,T} = \Lambda_{\phi+2\pi,T}$. In this case, it is common to use a periodic cost function. However, in the context of atomic clocks, we explicitly adopt a global definition of the phase spanning  $-\infty < \phi < \infty$, since $\phi + 2\pi k$ (with $k \in \mathbb{Z}$) originates from a different frequency deviation $\omega$ than $\phi$, and thus has a distinct physical interpretation. This distinction proves particularly useful to quantify the coherence time limit of the local oscillator (cf. Sec.~\ref{sec:Bounds}) and to discuss fringe hops within this framework (cf. Sec.~\ref{sec:Simulations}). Furthermore, we assume a Gaussian prior distribution
        \begin{align}
            \mathcal{P}(\phi) &= \frac{1}{\sqrt{2\pi(\delta\phi)^2}}\exp\left(-\frac{\phi^2}{2(\delta\phi)^2}\right)\label{eq:prior}
        \end{align}
        with zero mean and width $\delta\phi$, which is a reasonable approximation for the full feedback loop of an atomic clock~\cite{Leroux2017}.

    \subsection{Bounds}\label{sec:Bounds}
        The goal of Bayesian estimation is to minimize the cost function, the Bayesian mean squared error (BMSE). For a given prior distribution $\mathcal{P}(\phi)$, there are three degrees of freedom to optimize: the initial state $\rho_\mathrm{in}$, the measurement $\{\Pi_x\}$, and the estimation strategy $\phi_\mathrm{est}(x)$. Based on these degrees of freedom and building on Refs.~\cite{VanTrees1968,Macieszczak2013,Jarzyna2015,Demkowicz_Dobrza_ski_2015}, we collect a hierarchy of lower bounds for the BMSE (see App.~\ref{app:bounds} for detailed proofs), analogous to the local estimation approach.  The discussion in this section remains general, allowing for arbitrary prior distributions $\mathcal{P}(\phi)$ and quantum channels $\Lambda_{\phi,T}$. Specific assumptions and asymptotic results will be explicitly noted.

        \paragraph*{Bayesian Cram\'{e}r-Rao Bound (BCRB)---} For a given initial state $\rho_\mathrm{in}$ and measurement $\{\Pi_x\}$, the Bayesian Cram\'{e}r-Rao Bound (BCRB) $(\Delta\phi_\mathrm{BCRB})^2$ represents a lower bound on the BMSE $(\Delta\phi)^2$ and thus, implicates an optimization over all possible estimators $\phi_\mathrm{est}$. Assuming standard regularity conditions $\sum_x \partial_\phi P(x|\phi)=0$ and vanishing of the prior at the boundaries $\lim_{\phi\to \pm\infty} \mathcal{P}(\phi)=0$, the BCRB results from the van Trees inequality~\cite{VanTrees1968} and reads~\cite{Gill1995}
        \begin{align}
            (\Delta\phi)^2\geq (\Delta\phi_\mathrm{BCRB})^2 = \min_{\phi_\mathrm{est}} (\Delta\phi)^2 = \frac{1}{\overline{\mathcal{F}}+ \mathcal{I}}\label{eq:BCRB}
        \end{align}
        (proof in App.~\ref{app:BCRB}). Here, the measurement contribution is represented by the Fisher information averaged over the prior distribution
        \begin{align}
            \overline{\mathcal{F}} =\int \dd\phi\, \mathcal{P}(\phi) \mathcal{F}(\phi) = \int \dd\phi\, \mathcal{P}(\phi) \sum_x \frac{1}{P(x|\phi)}\left(\frac{\dd P(x|\phi)}{ \dd\phi\,}\right)^2\label{eq:FI_average}
        \end{align}
        and
        \begin{align}
            \mathcal{I} = \int \dd\phi\, \frac{1}{ \mathcal{P}(\phi)}\left(\frac{\dd  \mathcal{P}(\phi)}{ \dd\phi\,}\right)^2\label{eq:prior_knowledge}
        \end{align}
        denotes the information contained in the prior knowledge, given by the Fisher information of the prior distribution. While $\mathcal{I} \geq 0$, the average Fisher information $\overline{\mathcal{F}}$ is upper bounded by its maximal value $\mathcal{F}_\mathrm{max} = \mathcal{F}(\phi_0)$ achieved at the optimal working point $\phi_0$, i.e. $\overline{\mathcal{F}}\leq \mathcal{F}_\mathrm{max}$. Hence, from Eq.~\eqref{eq:BCRB} it is evident that the BCRB in turn is lower bounded by the Cram\'{e}r-Rao bound (CRB), the corresponding bound in local phase estimation, $(\Delta\phi_\mathrm{BCRB})^2\geq (\Delta\phi_\mathrm{CRB})^2 = 1/\mathcal{F}_\mathrm{max}$. In contrast to the local approach, the optimal estimation strategy in the Bayesian framework can be derived explicitly as we will show in Sec.~\ref{sec:Estimators}.

        For a Gaussian prior distribution, the prior information simplifies to $\mathcal{I} = (\delta\phi)^{-2}$. Moreover, while $\overline{\mathcal{F}}$ typically increases with the ensemble size, the prior information $\mathcal{I}$ is independent of $N$. Consequently, in the asymptotic limit of large $N$, the prior knowledge primarily contributes in the averaging of the Fisher information and we obtain $(\Delta\phi_\mathrm{BCRB})^2 \simeq  \overline{\mathcal{F}}^{-1}$.

        \paragraph*{Bayesian Quantum Cram\'{e}r-Rao Bound (BQCRB)---} The Bayesian quantum Cram\'{e}r-Rao bound (BQCRB) extends the classical Bayesian Cram\'{e}r-Rao bound (BCRB) by including the optimization over all measurements $\{\Pi_x\}$. For a given initial state $\rho_\mathrm{in}$, the BQCRB
        \begin{align}
            (\Delta\phi_\mathrm{BQCRB})^2= \min_{\{\Pi_x\}} (\Delta\phi_\mathrm{BCRB})^2= \min_{\{\Pi_x\}, \phi_\mathrm{est}} (\Delta\phi)^2
        \end{align}
        provides a lower bound on the BCRB and thus establishes the hierachy
        \begin{align}
            (\Delta\phi)^2 \geq (\Delta\phi_\mathrm{BCRB})^2\geq (\Delta\phi_\mathrm{BQCRB})^2.
        \end{align}
        Naively, one might suggest to simply replace the average Fisher information $\overline{\mathcal{F}}$ in Eq.~\eqref{eq:BCRB} by the average quantum Fisher information $\overline{\mathcal{F}_Q} = \int \dd\phi\,  \mathcal{F}_Q(\Lambda_{\phi,T}[\rho_\mathrm{in}])$. However, in general, the optimal measurement depends on $\phi$ and thus, this approach would effectively correspond to averaging over a set of measurements, each optimized for a particular phase value $\phi$. By restricting the measurements, without loss of optimality, to the class of projection-valued measures (PVM) $\Pi_x = \ket{x}\bra{x}$, with orthonormal eigenstates $\ket{x}$, $\langle x|x'\rangle=\delta_{x,x'}$, of the observable $X$ with eigenvalue $x$, the BQCRB can be expressed as~\cite{Macieszczak2014}
        \begin{align}
                (\Delta\phi_\mathrm{BQCRB})^2 &= (\delta\phi)^2 - \tr(\overline{\rho}L^2) \label{eq:BQCRB}
        \end{align}
        (proof in App.~\ref{app:BQCRB}). Here, the double minimization over the measurement $\{\Pi_x\}$ and estimator $\phi_\mathrm{est}$ is combined in a single quantity $L = \sum_x \Pi_x \phi_\mathrm{est}(x)$. The optimal $L$ is determined by the implicit equation
        \begin{align}
                \overline{\rho}' = \frac{1}{2}\left(\overline{\rho}L + L \overline{\rho}\right)\label{eq:L_equation}
        \end{align}
        where $\overline{\rho} = \int \dd\phi\, \mathcal{P}(\phi)\Lambda_{\phi,T}[\rho_\mathrm{in}]$ denotes the average state and $\overline{\rho}' = \int \dd\phi\, \mathcal{P}(\phi)\Lambda_{\phi,T}[\rho_\mathrm{in}] \phi$. Unfortunately, the introduction of the operator $L$, while essential for deriving the BQCRB, comes with the limitation that the optimal measurement and estimator cannot be determined explicitly. Instead, only the evaluation of the bound itself is possible. Interestingly, Eq.~\eqref{eq:BQCRB} and Eq.~\eqref{eq:L_equation} have a similar structure as the quantum Fisher information (QFI) in local phase estimation. Indeed, assuming a unitary phase evolution according to Eq.~\eqref{eq:unitaryEvolution} and a Gaussian prior distribution as defined in Eq.~\eqref{eq:prior}, the BQCRB can be related to the QFI $\mathcal{F}_Q[\overline{\rho}]$ of the average state $\overline{\rho}$ by \cite{Jarzyna2015}
        \begin{align}
                (\Delta\phi_\mathrm{BQCRB})^2 = (\delta\phi)^2\left[1 - (\delta\phi)^2\mathcal{F}_Q[\overline{\rho}]\right]\label{eq:BQCRB_QFI}
        \end{align}
        (derivation in App.~\ref{app:BQCRB}). In this case, the optimal measurement corresponds to the symmetric logarithmic derivative (SLD) of the QFI approach associated with $\mathcal{F}_Q[\overline{\rho}]$, and the optimal Bayesian estimator can be determined explicitly (cf. Sec.~\ref{sec:Estimators}.). Evaluating the BQCRB thus becomes computationally equivalent to calculating the QFI of the average state $\overline{\rho}$.

        \paragraph*{Optimal Quantum Interferometer (OQI)---} The optimal quantum interferometer (OQI) represents the ultimate lower bound of the BMSE, completing the hierachy
        \begin{align}
            (\Delta\phi)^2 \geq (\Delta\phi_\mathrm{BCRB})^2\geq (\Delta\phi_\mathrm{BQCRB})^2\geq (\Delta\phi_\mathrm{OQI})^2.
        \end{align}
        The OQI simultaneously optimizes over all three degrees of freedom: the initial state $\rho_\mathrm{in}$, measurement $\{\Pi_x\}$ and estimator $\phi_\mathrm{est}$:
        \begin{align}
            (\Delta\phi_\mathrm{OQI})^2 &= \min_{\rho_\mathrm{in}} (\Delta\phi_\mathrm{BQCRB})^2 = \min_{\rho_\mathrm{in}, \{\Pi_x\}} (\Delta\phi_\mathrm{BCRB})^2 \nonumber\\
            &= \min_{\rho_\mathrm{in}, \{\Pi_x\}, \phi_\mathrm{est}} (\Delta\phi)^2.
        \end{align}
        Unfortunately, no general expressions for the OQI sensitivity for arbitrary ensemble sizes are available, but they rather require complex optimization procedures. An algorithm presented in Refs.~\cite{Macieszczak2013,Macieszczak2014} iteratively optimizes the initial probe state and the measurement (cf. App.~\ref{app:OQI}). While this algorithm enables an efficient computation for small ensembles, numerical optimization becomes challenging with increasing $N$.

        Considering a $2\pi$-periodic quantum channel with respect to the phase $\phi$ (cf. Eq.~\eqref{eq:unitaryEvolution}), the OQI allows for unambiguous phase estimation within the range $[-\pi,+\pi]$. However, for sufficiently broad prior distributions, the phase $\phi$ may exceed this invertible regime and an estimation error of $(2\pi)^2$ is accumulated, associated with transitions between adjacent Ramsey fringes. Although the Bayesian framework naturally accounts for this crossover, it is nevertheless instructive to examine their contribution separately. For a Gaussian prior distribution, the estimation error associated with these events can be modeled by (cf. App.~\ref{app:OQI})
        \begin{align}
            (\Delta\phi^{\mathrm{CTL}}_{\mathrm{OQI}})^2 = 4\pi^2 \left[1-\mathrm{erf}\left(\frac{\pi}{\sqrt{2}\delta\phi}\right)\right]\label{eq:CTL_OQI}
        \end{align}
        where $\mathrm{erf}(z)$ denotes the error function. In the context of an atomic clock, in this regime of long interrogation times, the coherence time of the local oscillator will become relevant and ultimately limits the clock stability. Consequently, we will denote Eq.~\eqref{eq:CTL_OQI} as the coherence time limit (CTL) of the OQI.

        In the asymptotic limit ($N \gg 1$), assuming unitary phase evolution as described by Eq.~\eqref{eq:unitaryEvolution} and restricting to the invertible range $[-\pi,+\pi]$, it has been shown for arbitrary prior distributions~\cite{Gorecki2020,Jarzyna2015,Buzek1999,Berry2000} that the ultimate lower bound is given by 
        \begin{align}
            (\Delta\phi_{\pi\mathrm{HL}})^2 =  \frac{\pi^2}{N^2}.\label{eq:piHL}
        \end{align}
        In the absence of decoherence, this asymptotic limit reflects Heisenberg scaling with an additional factor of $\pi$, and is therefore referred to as the $\pi$-corrected Heisenberg limit ($\pi$HL). Intuitively, the $\pi$HL can be interpreted as the maximal estimation error associated with estimating a phase within $[-\pi,+\pi]$ using $N+1$ evenly spaced measurement outcomes. Additionally taking into account the estimation error outside of the invertible range, as modeled by Eq.~\eqref{eq:CTL_OQI}, the overall asymptotic estimation error for the OQI reads
        \begin{align}
            (\Delta\phi_{\mathrm{OQI}}^\mathrm{asym})^2 &=  (\Delta\phi_{\pi\mathrm{HL}})^2 + (\Delta\phi^{\mathrm{CTL}}_{\mathrm{OQI}})^2\nonumber\\
            &= \frac{\pi^2}{N^2} + 4\pi^2 \left[1-\mathrm{erf}\left(\frac{\pi}{\sqrt{2}\delta\phi}\right)\right].\label{eq:OQi_asym}
        \end{align}
        This result combines the fundamental limit set by the $\pi$HL with the contributions from phase estimation errors associated with transitions between Ramsey fringes, offering a comprehensive characterization of the OQI's performance in the asymptotic regime. Notably, this bound can be saturated asymptotically by the phase operator based interferometer (POI)~\cite{Summy1990,Berry2000,Buzek1999,Pegg1988,Sanders1995,Derka1998,Luis1996,Kaubruegger2021,Demkowicz_Dobrza_ski_2015} (cf. App.~\ref{app:OQI}). 

    \subsection{Estimators}\label{sec:Estimators}
        Based on a statistical model $P(x|\phi)$, defined by an initial state $\rho_\mathrm{in}$, free evolution $\Lambda_{\phi,T}$ and measurement $\{\Pi_x\}$, various estimation strategies can be applied. In this work, we focus on two such strategies: the linear estimator and the optimal Bayesian estimator. The linear estimator is renowned for its simplicity and is both theoretically and experimentally commonly used and well understood. It often arises naturally in a local approach (\textit{i.e.} for a narrow prior distribution and linear error propagation), where it is the standard choice. In contrast, the optimal Bayesian estimator, as the name suggests, achieves the best possible performance in Bayesian phase estimation.

        \paragraph*{Linear estimator---} The linear estimator 
        \begin{align}
            \phi_\mathrm{est}^\mathrm{linear} (x) = a\cdot x,\label{eq:linear_est}
        \end{align}
        with scaling factor $a\in\mathbb{R}$, originates from local phase estimation. In this context, assuming an unbiased estimator and small deviations from the optimal working point $\phi_0$, the signal can be approximated linearly. In this case, the  estimation error is typically characterized by quantum projection noise (QPN)~\cite{Pezze2018}
        \begin{align}
            \Delta\phi_\mathrm{QPN}=\frac{\xi}{\sqrt{N}}=\frac{\Delta X(\phi_0)}{ |\partial_\phi\braket{X(\phi)}|_{\phi=\phi_0}},
        \end{align}
        where $\xi$ denotes the Wineland squeezing parameter~\cite{Wineland1992, Wineland1994}. This local result is obtained in the limit of narrow prior distributions ($\delta\phi\to 0 $) around the optimal working point $\phi_0$ and by choosing the particular scaling factor $a=\left(\partial_\phi\braket{X(\phi)}|_{\phi=\phi_0}\right)^{-1}$, corresponding to the inverse slope of the signal at $\phi_0$. 
        
        In the Bayesian framework, however, this approach is poorly suited. First, the assumption of narrow prior distributions fails for realistic fluctuations of the phase, as discussed above. Second, the prior information explicitly influences the cost function and thus, the scaling factor $a$ has to depend on the prior distribution. For an arbitrary prior distribution with zero mean $\int \dd\phi\, \mathcal{P}(\phi)\phi=0$ and variance $(\delta\phi)^2 = \int \dd\phi\, \mathcal{P}(\phi)\phi^2 $, the optimal scaling factor and corresponding BMSE are given by (cf. App.~\ref{app:linear_estimator})
        \begin{align}
            a &= \frac{\int  \dd\phi\, \mathcal{P}(\phi) \phi\braket{X(\phi)}}{\int  \dd\phi\, \mathcal{P}(\phi)\braket{X^2(\phi)}} \\
            (\Delta\phi)^2 &= (\delta\phi)^2 - \frac{\left[\int  \dd\phi\, \mathcal{P}(\phi) \phi\braket{X(\phi)}\right]^2}{\int  \dd\phi\, \mathcal{P}(\phi)\braket{X^2(\phi)}}.\label{eq:BMSE_linear}
        \end{align}
        As in the local approach, the linear estimator and its estimation error depend only on the first and second moments of the observable $X$, which typically are easier to evaluate than the full statistical model $P(x|\phi)$. This simplicity makes the linear estimator a practical choice for phase estimation. Nevertheless, despite its advantages and reliable performance in several situations, the linear estimation strategy in general does not saturate the BCRB and thus is not optimal.

        \paragraph*{Optimal Bayesian estimator---} In contrast to local phase estimation, where the Cram\'{e}r-Rao bound can in general only be approximated in the infinite-sample limit using the maximum-likelihood estimator~\cite{Pezze2018}, the optimal estimator in Bayesian phase estimation can be derived explicitly~\cite{Demkowicz_Dobrza_ski_2015}
        \begin{align}
            \phi_\mathrm{est}^\mathrm{opt}(x) &= \int  \dd\phi\,  P(\phi|x)\phi\label{eq:MMSE_est}
        \end{align}
        (derivation in App.~\ref{app:mmse_estimator}), saturating the BCRB with single shot measurements. This estimator corresponds to the average phase with respect to the posterior distribution $P(\phi|x)$, which can be expressed in terms of the statistical model $P(x|\phi)$ and prior distribution $\mathcal{P}(\phi)$ according to Bayes theorem Eq.~\eqref{eq:BayesTheorem}. As a consequence, the optimal Bayesian estimator can be highly non-linear. The associated BMSE is given by
        \begin{align}
            (\Delta\phi)^2 &=  (\delta\phi)^2 - \sum_x  \frac{\left[ \int  \dd\phi\, \mathcal{P}(\phi)  P(x|\phi) \phi\right]^2}{P(x)}.\label{eq:BMSE_MMSE}
        \end{align}
        Although this resembles the structure of Eq.~\eqref{eq:BMSE_linear}, the BMSE for the optimal Bayesian estimator explicitly depends on the statistical model, rather than merely on the first and second moments of the observable. Additionally, for the optimal Bayesian estimator, Eq.~\eqref{eq:BMSE2} reduces to the average posterior variance. Since the optimal Bayesian estimator saturates the BCRB, and thus minimizes the BMSE with respect to all estimation strategies, it is commonly referred to as the minimal mean squared error (MMSE) estimator. However, we continue to use the term `optimal Bayesian estimator' throughout this work for consistency and clarity.

    \subsection{Allan deviation}\label{sec:ADEV}
        The long-term stability of an atomic clock is quantified by the Allan deviation $\sigma_y(\tau)$~\cite{Allan1966,Handbook} (cf. App.~\ref{app:ADEV} for a brief introductory overview), characterizing the fluctuations of fractional frequency deviations $y(t) = \omega(t)/\omega_0$ averaged over $\tau\gg T_C=T+T_D$. Here, the total cycle duration $T_C$ accounts for the interrogation time $T$ and any potential dead time $T_D$, arising from preparation steps and application of the feedback. In local frequency metrology, assuming short interrogation times leading to narrow prior distributions, the Allan deviation is well approximated by~\cite{Pezze2018,Ludlow2015}
        \begin{align}
            \sigma_y(\tau) =\frac{1}{\omega_0 } \frac{\Delta\phi_\mathrm{QPN}}{T}\sqrt{\frac{T_C}{\tau}}= \frac{1}{\omega_0 } \frac{\xi}{\sqrt{N}}\frac{1}{T}\sqrt{\frac{T_C}{\tau}}.\label{eq:ADEV_local}
        \end{align}
        In this context, clock stability is determined by quantum projection noise $\Delta\phi_\mathrm{QPN}=\xi/\sqrt{N}$, characterizing the uncertainty associated with the measurement process.
        
        In Bayesian frequency metrology, however, the BMSE incorporates both measurement uncertainty and prior knowledge, preventing a straightforward substitution of $\Delta\phi_\mathrm{QPN}$ by $\Delta\phi$. To isolate the measurement contribution from the prior knowledge $\mathcal{I}$, we introduce the effective measurement uncertainty motivated by the Bayesian Cram\'{e}r-Rao Bound (BCRB) in Eq.~\eqref{eq:BCRB} and following Refs.~\cite{Leroux2017,Kaubruegger2021}
        \begin{align}
            \Delta\phi_M = \left(\frac{1}{(\Delta\phi)^2}-\mathcal{I}\right)^{-1/2} = \left(\frac{1}{(\Delta\phi)^2}-\frac{1}{(\delta\phi)^2}\right)^{-1/2}\label{eq:effMeasVar}
        \end{align}
        where $\mathcal{I} = (\delta\phi)^{-2}$ for a Gaussian distribution. Hence, $\Delta\phi_M$ quantifies the quality of the measurement process in a single interrogation cycle. According to the discussion of the BCRB in Sec.~\ref{sec:Bounds}, the effective measurement uncertainty  is lower bounded by the average Fisher information $(\Delta\phi_M)^2 \geq 1/\overline{\mathcal{F}}$ and thus, a connection to the local approach can be established, yielding
        \begin{align}
            (\Delta\phi_M)^2 \geq \frac{1}{\overline{\mathcal{F}}} \geq (\Delta\phi_\mathrm{CRB})^2= \frac{1}{\mathcal{F}_\mathrm{max}} = \frac{1}{\mathcal{F}(\phi_0)}.
        \end{align}
        As a consequence, the clock stability in local frequency metrology, quantified by Eq.~\eqref{eq:ADEV_local}, emerges in the limit of narrow prior distributions ($\delta\phi \ll 1$) or equivalently short interrogation times ($T \ll 1$).

        With the effective measurement uncertainty, the Allan deviation in Bayesian frequency metrology is given by
        \begin{align}
            \sigma_y(\tau) = \frac{1}{\omega_0}\frac{\Delta\phi_M}{T}\sqrt{\frac{T_C}{\tau}}.\label{eq:ADEV_BMSE}
        \end{align}
        Consequently, the three key quantities in Bayesian frequency metrology are the prior width $\delta\phi$, the BMSE $\Delta\phi$, and the effective measurement uncertainty $\Delta\phi_M$. These quantities directly reflect the sensitivity in Bayesian frequency metrology and ultimately determine the clock stability. In the following, we will examine their relation in a qualitative discussion.

        Due to noise in the local oscillator, the phase diffusion grows with Ramsey dark time. Thus, the prior width $\delta\phi$ of the relative phase will be monotonically increasing with the interrogation time $T$ (cf. Fig.~\ref{fig:schematic}(b)). At first glance, one might assume that $\delta\phi$ is solely determined by the characteristics of the free running LO frequency $\omega_{\mathrm{LO}}(t)$. However, it depends even more strongly on the noise of the stabilized frequency $\omega(t)$ and therefore on the details of interrogation, estimation and feedback. Although, in general, the prior phase distribution can vary over different clock cycles, it becomes stationary if the feedback loop stabilizes the LO reliably to the atomic reference. In this case, the residual noise can be considered to be white, to a good approximation, and thus can be modeled by a normal distribution characterized by the spread $\delta\phi$ (cf. Eq.~\eqref{eq:prior}).

        For a given finite $\delta\phi$, the interrogation protocol and estimation strategy can be optimized to minimize the estimation error $\Delta\phi$. At the same time, the effective measurement uncertainty, Eq.~\eqref{eq:effMeasVar}, and thus also the Allan deviation, Eq.~\eqref{eq:ADEV_BMSE}, are minimized. Consequently, $\Delta\phi$ will ultimately determine the stabilized frequency noise, which in turn affects $\delta\phi$. Therefore, in order to reflect the closed feedback loop of the atomic clock, $\Delta\phi$ has to be optimized iteratively for suitably chosen $\delta\phi$, as detailed in Sec.~\ref{sec:Prior}.

        The average error in phase estimation $\Delta\phi$ depends on the prior width $\delta\phi$ as well as the particular interrogation sequence and estimation strategy. As discussed before, $\Delta\phi\leq\delta\phi$ and thus the estimation error $\Delta\phi$ is reduced compared to the prior width $\delta\phi$, since a proper Ramsey protocol increases the information about the phase. Here, equality $\Delta\phi=\delta\phi$ corresponds a worst case scenario in which the effective measurement variance diverges $\Delta\phi_M\rightarrow\infty$. This case represents an ineffective interrogation scheme, where the information gained through measurement and estimation fails to improve the characterization of residual noise. Conversely, a hypothetical perfect phase estimation (precluded by quantum mechanics due to its intrinsic indeterminism) would result in a vanishing estimation error $\Delta\phi\rightarrow 0$. Likewise, the effective measurement uncertainty would also vanish $\Delta\phi_M\rightarrow 0$, since this scenario implies a perfect measurement.

        The form of the Allan deviation in Eq.~\eqref{eq:ADEV_BMSE} suggests that the stability can be improved by increasing the interrogation time $T$. However, this is only true as long as the coherence time limit (CTL) of the LO remains negligible and quantum projection noise $\Delta\phi_\mathrm{QPN} = \frac{\xi}{\sqrt{N}}$ of the measurement dominates the effective measurement uncertainty. In general, three regimes can be distinguished based on the relation between the prior width $\delta\phi$ and the QPN $\Delta\phi_\mathrm{QPN}$:
        
        (\textit{i}) Considering small prior widths $\delta\phi\ll 1$, the measurement and estimation protocol cannot significantly improve the knowledge of the phase distribution, since $\Delta\phi_\mathrm{QPN}\gg \delta\phi$ and thus $\Delta\phi \simeq \delta\phi$. In this case, $\Delta\phi_M \simeq \Delta\phi_\mathrm{QPN} $ and the local form of the Allan deviation, Eq.~\eqref{eq:ADEV_local}, is reproduced. 
        
        (\textit{ii}) With increasing interrogation time $T$, the prior width surpasses QPN $\delta\phi > \Delta\phi_\mathrm{QPN}$. Nevertheless, in this regime, the information gain on the phase distribution resulting from the measurement and estimation strategy leads to $\Delta\phi\ll\delta\phi$ and thus $\Delta\phi_M < \delta\phi$. Hence, the optimal working point of the atomic clock is located in this region. 
        
        (\textit{iii}) At long interrogation times, the coherence time of the local oscillator will become relevant and ultimately limits the clock stability. Here, the phase noise exceeds the domain of the measurement scheme where an unambiguous estimation is possible, giving $\Delta\phi_M\gg\delta\phi \simeq \Delta\phi \gg \Delta\phi_\mathrm{QPN}$. 
        
        Consequently, the Allan deviation features a trade-off between increased stability achieved through long interrogation times and limitations imposed by the coherence time of the local oscillator. Fortunately, as previously discussed, this trade-off is inherently addressed within the framework of Bayesian frequency metrology. 

    \subsection{Interrogation time and prior width}\label{sec:Prior}
        In the previous section, we linked the clock stability at interrogation time $T$, characterized by the Allan deviation, with Bayesian phase estimation with prior width $\delta\phi$, described by the BMSE. Furthermore, we qualitatively discussed that the prior width increases with the interrogation time. To complete the connection between Bayesian phase estimation and frequency metrology, this section aims to establish an explicit relation between $\delta\phi$ and $T$. This relation serves as a bridge between the frequency fluctuations of the laser in an experiment and a theoretical representation in terms of a Gaussian prior distribution with a specific width. Establishing this connection is essential for accurately modeling experiments and ensuring the applicability of theoretical predictions to realistic scenarios.
        
        State-of-the-art clock lasers are characterized by the spectral noise density $S_y(f) = \sum_\alpha h_\alpha f^{-(1+\alpha)}$, which can be modeled by a power law \cite{Ludlow2015,Rutman1991,Handbook,Barnes1971}, where $\alpha=-1, 0, 1$ corresponds to white, flicker and random walk frequency noise, respectively. Accordingly, the Allan variance of the free-running LO can be expressed as $\sigma^2_{y,\text{LO}}(\tau)=\tilde h_\alpha \tau^\alpha$. To compare different local oscillators, a single timescale is defined characterizing the stability. Following Ref.~\cite{Leroux2017}, we define the laser coherence time $Z$ implicitly by
        \begin{align}
            \sigma_{y,\text{LO}}(Z_C) 2\pi\nu_0 Z = 1\text{ rad}.\label{eq:CoherenceTime}
        \end{align}
        Here, $\sigma_{y,\text{LO}}(Z_C)$ denotes the Allan deviation of the local oscillator averaging over a single cycle duration $Z_C=Z+T_D$ with dead time $T_D$.

        In Ref.~\cite{Leroux2017}, it was demonstrated that the prior width of the full feedback loop can be approximated by the power law
        \begin{align}
            (\delta\phi)^2 \simeq\chi(\alpha) \left(\frac{T}{Z}\right)^{2+\alpha}\label{eq:PowerLaw}
        \end{align}
        depending solely on the ratio of interrogation time $T$ and coherence time of the local oscillator $Z$, and the numerically determined factor $\chi(\alpha)=1,1.7,2$ for $\alpha=-1,0,1$. This approximation was derived in the limit of large ensembles and long interrogation times using the conventional Ramsey protocol in the framework of local phase estimation, and successfully applied in Refs.~\cite{Kaubruegger2021,Schulte2020}. However, in the full feedback loop of an atomic clock, the prior width $\delta\phi$ and estimation error $\Delta\phi$ mutually influence each other. Therefore, $\delta\phi$ has to be adjusted iteratively to account for the closed feedback loop dynamics, as motivated in the previous section and detailed in App.~\ref{app:iterated_prior}. This iterative procedure is employed in Sec.~\ref{sec:Simulations}, where realistic Monte Carlo simulations of the full feedback loop of an atomic clock are performed. Nevertheless, Eq.~\eqref{eq:PowerLaw} remains a convenient approximation for general investigations and is thus adopted in Sec.~\ref{sec:Theory}.
        
        In the following, motivated by state-of-the-art clock lasers~\cite{Matei2017}, we assume a local oscillator predominantly limited by flicker frequency noise. Additionally, we neglect systematic shifts in the atomic transition frequency $\omega_0$. Moreover, we will assume the atomic excited-state lifetime $t_\mathrm{exc}$ to be substantially longer than the clock cycle duration $T_C$, such that $t_\mathrm{exc}\gg T_C$.

\section{Quantum frequency metrology}\label{sec:Theory}
    In this section, we aim to saturate the OQI in the context of atomic clocks. We begin by analyzing standard protocols and compare their performance to the OQI. Afterwards, we introduce variational classes of quantum circuits and investigate the associated optimal Ramsey protocols.

    \begin{figure*}[tbp]
        \centering
            \includegraphics[width=1\textwidth]{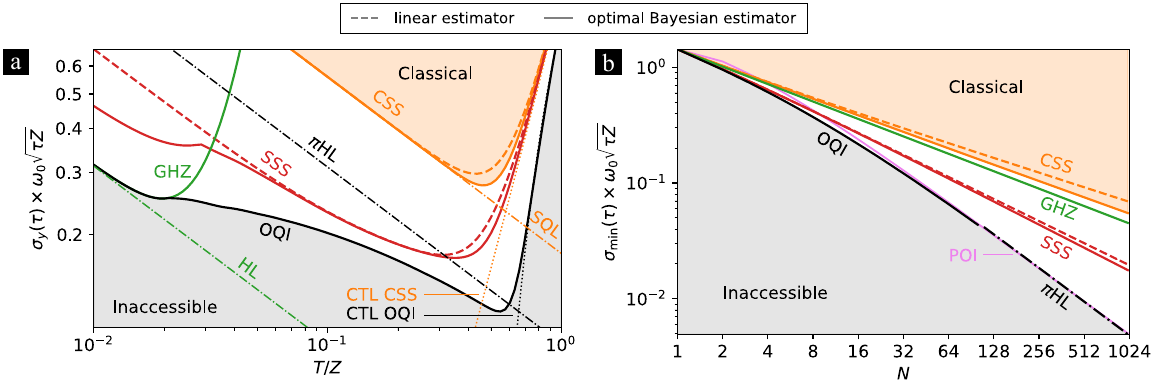}
     
        \caption{(a) Generic scaling of the dimensionless Allan deviation $\sigma_y(\tau)\times \omega_0\sqrt{\tau Z}$ with the interrogation time $T$ for the example of $N=32$, rescaled by the averaging time $\tau$, laser coherence time $Z$ and transition frequency $\omega_0$. Stabilities for the CSS (orange), SSS (red) and GHZ (green) protocols are compared to the performance of the OQI (black). For CSS and SSS, both the linear (dashed) and optimal Bayesian estimator (solid) are depicted.  The gray shaded area represents the inaccessible stability region set by the OQI limit, while the orange shaded area indicates achievable stabilities using uncorrelated atoms. Dotted lines correspond to the CTL for OQI (black) and for CSS and SSS with the linear estimator (orange). Additionally, benchmarks such as the SQL (orange), HL (green), and $\pi$HL (black) are included as dashed-dotted lines. (b) Scaling of the dimensionless minimal Allan deviation $\sigma_\mathrm{min}\times \omega_0\sqrt{\tau Z}$ with the ensemble size $N$. In addition to the standard protocols, the POI performance (violet) is presented. For the OQI and POI, numerical optimization is performed for $N\leq 100$, while the asymptotic behavior, represented by the $\pi$HL (black dashed-dotted), is shown for $N>100$.}
            \label{fig:standard_protocols}
    \end{figure*}
    
    \subsection{Standard protocols}\label{sec:standard_protocols}
        To start with, we examine the effective measurement uncertainty and corresponding clock stability of standard Ramsey protocols (see App.~\ref{app:standard_protocols} for detailed derivations). Specifically, we focus on coherent spin states (CSS), squeezed spin states (SSS), and GHZ states, as well as the ultimate lower bound defined by the optimal quantum interferometer (OQI). For all Ramsey schemes, the dependence of clock stability on the interrogation time $T$ reflects the three regimes discussed in Sec.~\ref{sec:ADEV}, representing a trade-off between enhanced stability achieved through longer interrogation times and the coherence time limit. These distinct regimes are depicted in Fig.~\ref{fig:standard_protocols}(a), which illustrates the generic dependence of stability on interrogation time. Furthermore, the scaling of the minimal Allan deviation $\sigma_\mathrm{min}$ with ensemble size $N$ at the optimal interrogation time $T_\mathrm{min}$ is presented in Fig.~\ref{fig:standard_protocols}(b). These figures are based on the work of Kaubrügger et al.~\cite{Kaubruegger2021} and are adapted here within the framework defined above. In particular, emphasis is placed on comparing the linear and optimal Bayesian estimation strategies across different protocols, with performance benchmarked against the OQI. To enable comparability between various setups, the achievable Allan deviations $\sigma_y(\tau)$ are rescaled with respect to the atomic transition frequency $\omega_0$, the total averaging time $\tau$ and the laser coherence time $Z$. This rescaling ensures that the results are transferable to specific experimental parameters.

        \paragraph*{Coherent Spin States (CSS)---} The conventional clock protocol employs Ramsey interferometry with coherent spin states (CSS)~\cite{Arecchi1972,Radcliffe1971,Hioe1974} as initial states, a collective projective spin measurement and a linear estimation strategy. In this scenario, the effective measurement uncertainty can be evaluated analytically~\cite{Leroux2017} as
        \begin{align}
            (\Delta\phi_M^\mathrm{CSS})^2 = \frac{\cosh((\delta\phi)^2)}{N} + \sinh((\delta\phi)^2)-(\delta\phi)^2.\label{eq:CSS_efm}
        \end{align}
        For short interrogation times $T/Z\ll 1$, leading to narrow prior widths $\delta\phi\ll 1$, the conventional standard quantum limit (SQL) $\Delta\phi_\mathrm{SQL} = 1/\sqrt{N}$ is recovered. Conversely, for long interrogation times $T/Z\sim 1$, frequency fluctuations of the local oscillator dominate and the first term in Eq.~\eqref{eq:CSS_efm} becomes negligible. This regime defines the coherence time limit for CSS with a linear estimator,
        \begin{align}
            (\Delta\phi_\mathrm{CTL}^\mathrm{CSS})^2 =  \sinh((\delta\phi)^2)-(\delta\phi)^2.
        \end{align}
        Hence, the stability reflects a trade-off between these two regimes, as illustrated in Fig.~\ref{fig:standard_protocols}(a), determining the minimal Allan deviation $\sigma_\mathrm{min}$. As the ensemble size $N$ increases, the first term in Eq.~\eqref{eq:CSS_efm} decreases, leading to shorter optimal interrogation times $T_\mathrm{min}$ to achieve $\sigma_\mathrm{min}$.
        
        For the optimal Bayesian estimator, an explicit evaluation of the conditional probabilities $P(x|\phi)$ is required, as discussed in Sec.~\ref{sec:Estimators}. Although $P(x|\phi)$ can be determined analytically for the CSS, the integrals in Eq.~\eqref{eq:BMSE_MMSE} generally have to be evaluated numerically. For short interrogation times, the narrow prior phase distribution allows for a good approximation by linearizing the signal. Thus, the optimal Bayesian estimator reproduces the linear estimator in this regime. In contrast, for interrogation times in the region of the minimal Allan deviation, higher-order contributions of the sinusoidal signal become relevant and the curvature of the signal has to be considered. In this case, the optimal Bayesian estimator approximates the \textit{arcsin} estimator, which directly inverts the signal and thus allows to estimate the phase unambiguously in the range $[-\pi/2,+\pi/2]$. This results in an extended dynamic range compared to the linear estimator, which cannot account for any non-linearity of the signal and thus exhibits a higher minimal instability. As a consequence, the optimal Bayesian estimator improves the scaling of $\Delta\phi_M$ with the ensemble size $N$ to $\mathcal{O}(N^{-0.47})$, compared to $\mathcal{O}(N^{-0.42})$ for the linear estimator, as shown in Fig.~\ref{fig:standard_protocols}(b). While the choice of estimator has limited impact for small ensembles, the stability gain from the optimal Bayesian estimator becomes significant for large ensembles $N\gg 1$. Importantly, this improvement arises solely from classical post-processing of the measurement outcomes, while the quantum circuit remains unchanged. Nevertheless, the CTL prevents both estimation strategies from achieving the SQL of $1/\sqrt{N}$.

        \paragraph*{Squeezed Spin States (SSS)---} Extending the conventional Ramsey protocol with a single one-axis-twisting (OAT) interaction~\cite{Kitagawa1993} for state preparation, various entangled states can be generated. Here, OAT interactions are denoted by $\mathcal{T}_\mathbf{k}(\mu) = \exp\left(-i\tfrac{\mu}{2}S_\mathbf{k}^2\right)$ with twisting stength $\mu$ around axis $\mathbf{k}$, where $S_\mathbf{k} = k_1 S_x + k_2 S_y + k_3 S_z$ is the spin projection along direction $\mathbf{k}$. In particular, for small twisting strengths $\mu$, one-axis-twisting generates squeezed spin states (SSS) by shearing the initial CSS around the twisting axis, characterized by a squeezing parameter $\xi < 1$. Using the linear estimator, the effective measurement uncertainty is given by
        \begin{align}
            (\Delta\phi_M^\mathrm{SSS})^2 =  \frac{\braket{S_y^2}}{\braket{S_x}^2}\cosh((\delta\phi)^2) + \frac{\braket{S_x^2}}{\braket{S_x}^2}\sinh((\delta\phi)^2)-(\delta\phi)^2,
        \end{align}
        with expectation values provided in App.~\ref{app:SSS}. SSS show enhanced stability compared to CSS due to reduced fluctuations in the measured spin observable. However, the gain comes at the cost of smaller dynamic range, as the minimal Allan deviation is achieved at shorter interrogation times compared to CSS (cf. Fig.~\ref{fig:standard_protocols}(a)). This is a direct consequence of SSS sharing the same coherence time limit as CSS, since $\braket{S_y^2}/\braket{S_x}^2\ll 1$ and $\braket{S_x^2}/\braket{S_x}^2\simeq 1$ for large prior widths $\delta\phi$ and optimal twisting strength. Similar to the conventional Ramsey protocol, SSS with the optimal Bayesian estimator achieve a slightly extended dynamic range at long interrogation times. For large ensembles $N\gg 1$, the asymptotic scaling of the effective measurement uncertainty with the optimal Bayesian estimator approximates $\mathcal{O}(N^{-2/3})$, depicted in Fig.~\ref{fig:standard_protocols}(b), reflecting the scaling observed in local phase estimation~\cite{Kitagawa1993,Pezze2018}. In contrast, the linear estimator exhibits a scaling $\mathcal{O}(N^{-0.63})$. Furthermore, the optimal Bayesian estimation strategy offers a remarkable advantage at short interrogation times, where $\delta\phi \lesssim 1/N$, as shown in Fig.~\ref{fig:standard_protocols}(a). In this regime, the estimator becomes highly non-linear, allowing for substantially stronger twisting strengths $\mu$, resulting in stronger squeezing and enhanced stability. However, as $\delta\phi$ approaches $1/N$, the scaling of the Allan deviation with the interrogation time $T$ stagnates and converges towards the stability achieved with the linear estimator.
        
        \paragraph*{GHZ States---} The maximally entangled Greenberger-Horne-Zeilinger (GHZ) state $\ket{\mathrm{GHZ}} = \left[\kett{\downarrow}^{\otimes N} + \kett{\uparrow}^{\otimes N}\right] /\sqrt{2}$~\cite{GHZ} represents an equal superposition of the collective ground and excited states. The corresponding Ramsey sequence, initially proposed by Wineland \textit{et al.}~\cite{Wineland1996}, is referred to as the GHZ protocol. During the free evolution time, the accumulated phase is amplified by a factor of $N$ due to the maximal entanglement of the GHZ state. Subsequently, the parity $\Pi$ is measured, resulting in a binary outcome $\pm 1$ that indicates whether the number of atoms in the ground state is even or odd. Since a binary outcome inevitably results in a linear estimator, both estimation strategies coincide and result in the effective measurement uncertainty
        \begin{align}
            (\Delta\phi_M^\mathrm{GHZ})^2 = \frac{e^{N^2(\delta\phi)^2}}{N^2} - (\delta\phi)^2.\label{eq:GHZ_efm}
        \end{align}
        However, the optimal Bayesian estimator allows to avoid a parity measurement and perform a conventional projective spin measurement instead. In this case, the optimal Bayesian estimator effectively maps even and odd numbers of atoms in the ground state to the parity $\pm 1$, thereby mimicking a parity measurement and achieving the same sensitivity (cf. App.~\ref{app:GHZ}). This strategy was essentially employed in a different framework in Ref.~\cite{Monz2011}. Both measurement and estimation strategies are optimal, since Eq.~\eqref{eq:GHZ_efm} aligns with the BQCRB. For short interrogation times, where $\delta\phi \lesssim 1/N$, the GHZ protocol achieves the conventional Heisenberg limit (HL) $\Delta\phi_\mathrm{HL} = 1 /N$, as illustrated in Fig.~\ref{fig:standard_protocols}(a), which corresponds to the OQI in a decoherence-free local phase estimation scenario. However, the sensitivity of the GHZ protocol decreases $N$-times faster than that of CSS as the prior width increases. For a parity measurement, this is attributed to the $N$-times increased oscillation frequency of the sinusoidal signal, yielding a reduced dynamic range. Ultimately, the resulting ambiguities in phase estimation cause the GHZ protocol to be effectively insensitive to phases $\phi \gtrsim 1/N$. Consequently, the optimal interrogation time scales approximately as $1/(NZ)$, leading to a scaling of the effective measurement uncertainty of $\mathcal{O}(N^{-1/2})$, equivalent to the SQL (cf. Fig.~\ref{fig:standard_protocols}(b)). While the minimal Allan deviation of the GHZ protocol provides only a minor improvement over CSS and is outperformed by SSS, its shorter optimal interrogation time offers practical advantages.  For instance, reduced probe times mitigate contrast losses and time dilation shifts caused by motional heating in ion crystals, thus improving the signal-to-noise ratio and the accuracy of such a clock~\cite{Ludlow2015}.

        \paragraph*{OQI---} As discussed in Sec.~\ref{sec:Bounds}, the OQI requires numerical optimization, as no analytical expressions are available for arbitrary ensemble sizes. Instead, we investigate the general scaling based on Fig.~\ref{fig:standard_protocols}. For short interrogation times $T/Z\ll 1$, where $\delta\phi \lesssim 1/N$, the OQI is saturated by the GHZ protocol, achieving the Heisenberg limit $\sigma_\mathrm{HL}(\tau) = 1 /(N\sqrt{T\tau})$, as shown in Fig.~\ref{fig:standard_protocols}(a). As the interrogation time increases and $\delta\phi \gtrsim N$, a characteristic \textit{plateau} emerges in which the Allan deviation decreases only marginally with $T$. This plateau shifts to shorter interrogation times as the ensemble size $N$ increases, reflecting the coherence time limit of the GHZ protocol. Beyond this plateau, as the interrogation time increases further, the scaling of the Allan deviation with $T$ converges back to $1/\sqrt{T}$, ultimately reaching its minimum $\sigma_\mathrm{min}$ at $T_\mathrm{min}$. In the limit of large ensembles ($N \gg 1$), this minimum is determined by the $\pi$-corrected Heisenberg limit. While the OQI significantly outperforms SSS at $T_\mathrm{min}$, SSS perform close to the OQI in the transition regime, especially for small ensembles. The relative gain of the OQI over SSS at $T_\mathrm{min}$ increases with the ensemble size, as illustrated in Fig.~\ref{fig:standard_protocols}(b). In the asymptotic limit $N \gg 1$, the POI, introduced in Sec.~\ref{sec:Bounds} and further detailed in App.~\ref{app:OQI}, is optimal, saturating the $\pi$HL. In this regime, the OQI scales as $\mathcal{O}(N^{-0.97})$, closely approximating Heisenberg scaling. Ultimately, the coherence time limit is approached at long interrogation times.\\

        In the following, we essentially distinguish between two different regimes concerning the ensemble size $N$: The first regime covers systems ranging from $N=1$ to some tens of atoms, as relevant for ion-traps~\cite{steinel2023,pelzer2023,Hausser2024} or tweezer-arrays~\cite{madjarov2019,norcia2019,young2020,shaw2024}. The remainder of this section primarily focuses on bridging the gap between SSS and the OQI by identifying Ramsey protocols of increasing complexity that approximate the OQI within this regime. In contrast, for large ensembles ($N\gtrsim 100$), the regime of lattice clocks is reached and the asymptotic scaling is approximated~\cite{Ludlow2015,Takamoto2015,Katori2003,Masoudi2015,Katori2011}. In this regime, dead time typically emerges as the dominant limitation~\cite{Schulte2020}, as discussed in detail in Sec.~\ref{sec:deadtime}.

    \subsection{Variational classes}\label{sec:VariationalClass}
        Recent advances in quantum information have inspired the development of variational quantum circuits as versatile tools for implementing interferometers with setup-specific quantum gates. Typically, each layer in these circuits comprises an entanglement-generating interaction and (single qubit) rotations that provide geometric flexibility. One-axis twisting (OAT)~\cite{Kitagawa1993} interactions have gathered significant attention, as they can be implemented in several experimental platforms~\cite{Benhelm2008,Blatt2008,Leibfried2004,Eckner2023,Hines2023,Srensen2001,Estve2008,Gross2010,Riedel2010} and corresponding circuits represent a natural extension of squeezed spin states (SSS). Combined with collective rotations, OAT interactions form the building blocks of several variational quantum circuits~\cite{SchulteEcho2020,Scharnagl2023,Kaubruegger2021,Thurtell2024,Marciniak2022,Li2023}. While Ref.~\cite{Scharnagl2023} offers a unified framework for generalized echo protocols in local phase estimation, encompassing numerous previously documented approaches~\cite{Li2023,Leibfried2004,Davis2016,Fröwis2016,Macri2016,Nolan2017,Colombo2022a,Burd2019,Anders2018}, this work investigates variational classes specifically tailored to Bayesian frequency metrology.
        
        In general, any variational Ramsey protocol can be expressed as
        \begin{align}
            P(x|\phi) &= \tr\left(\ket{x}\bra{x}\Lambda_{\phi,T}[\rho_\mathrm{in}]\right)\\
            \rho_\mathrm{in} &= \mathcal{U}_\mathrm{prep}\ket{\psi_0}\bra{\psi_0}\,\mathcal{U}_\mathrm{prep}^\dag\\
            \ket{x}\bra{x} &= \mathcal{U}_\mathrm{meas}^\dag\ket{M}\bra{M}\,\mathcal{U}_\mathrm{meas},
        \end{align}
        with arbitrary unitary preparation and measurement operations $\mathcal{U}_\mathrm{prep}$ and $\mathcal{U}_\mathrm{meas}$, respectively. While $\mathcal{U}_\mathrm{prep}$ generates the initial state by acting on the ground state $\ket{\psi_0} = \kett{\downarrow}^{\otimes N}$, $\mathcal{U}_\mathrm{meas}$ effectively determines the measurement $X$ by transforming the projectors $\ket{M}\bra{M}$ associated with Dicke states of spin $S=N/2$ and eigenvalue $M$ of $S_z$. Since any alternative choice of $\ket{\psi_0}$ and Dicke basis $\{\ket{M}\bra{M}\}$ can be incorporated into $\mathcal{U}_\mathrm{prep}$ and $\mathcal{U}_\mathrm{meas}$ by additional transformations, fixing $\ket{\psi_0}$ and $\{\ket{M}\bra{M}\}$ does not limit the generality of the protocol. The unitaries $\mathcal{U}_\mathrm{prep}$ and $\mathcal{U}_\mathrm{meas}$ are constructed from $n$ and $m$ layers of the variational circuit, respectively. Consequently, $n$ effectively determines the level of entanglement in the initial state, while $m$ governs the measurement strategy and dynamic range, ultimately determining the minimal Allan deviation $\sigma_\mathrm{min}$.
        
        \paragraph*{Previous results---} Pioneering work on Bayesian variational Ramsey protocols was conducted in Refs.~\cite{Kaubruegger2021,Marciniak2022}.  These studies introduced a variational class constrained to be invariant under the $x$-parity transformation, resulting in an anti-symmetric signal. Each layer of the quantum circuit consisted of two OAT interactions applied along orthogonal directions, combined with a collective rotation about one of these axes. While this choice provided a diverse class of entanglement-generating unitaries in each layer, it imposed significant geometric constraints. Nevertheless, the quality of phase estimation was not compromised, as the main objective was to saturate the OQI in the asymptotic limit of deep circuits. The analysis primarily focused on ensembles with several tens of qubits and employed linear estimation strategies. Kaubruegger \textit{et al.} demonstrated that the minimal Allan deviation $\sigma_\mathrm{min}$ could be achieved with sufficiently deep circuits. However, this approach had two key drawbacks: the reliance on deep circuits due to restricted geometric flexibility, and the inclusion of two OAT interactions per layer, which are experimentally more challenging to implement than collective rotations.

        Thurtell \textit{et al.}  in Ref.~\cite{Thurtell2024} addressed these limitations by proposing a variational class where each layer comprises a single OAT interaction around the $z$-axis combined with global rotations.  These rotations are designed to effectively transform the OAT interaction with respect to an arbitrary axis, thereby eliminating geometric constraints. This approach reduced both the circuit depth and the number of OAT interactions, while achieving results comparable to those in Ref.~\cite{Kaubruegger2021}. Nevertheless, a considerable number of OAT interactions remained necessary. Moreover, the analysis was conducted within the framework of general Bayesian phase estimation and thus, did not consider the trade-off with respect to the interrogation time in frequency metrology.

        \paragraph*{Variational Ramsey protocols---} In the following, we aim to approximate the OQI within the framework of Bayesian frequency metrology. Instead of exploring the convergence towards the OQI with many layers for state preparation and measurement, we focus on variational quantum circuits with minimal depth. We primarily consider small ensembles relevant to ion-traps and additionally investigate the transition toward tweezer-arrays, which have been predominantly studied in Refs.~\cite{Kaubruegger2021,Thurtell2024}. Given the high degree of control achievable in these systems, this represents the regime where near-term experimental implementation is most likely. Moreover, variational protocols are less favorable in setups with many atoms, such as in lattice clocks, as we will discuss below. In contrast to earlier studies, relying on linear estimation strategies, we employ the optimal Bayesian estimator to fully exploit the potential of variational Ramsey protocols. This choice is motivated by the substantial improvements observed for the standard protocols, including enhanced squeezing for SSS at short interrogation times, an extended dynamic range for CSS and SSS, and an effective reduction in the circuit depth required to implement the GHZ protocol. Additionally, by using the optimal Bayesian estimator, we ensure saturation of the BCRB for any given initial state and measurement and thus minimizing the required circuit depth. A comprehensive comparison with the linear estimation strategy is provided in Sec.~\ref{sec:linearEstimator}.

        Building on the advancements in Refs.~\cite{Kaubruegger2021,Thurtell2024}, we define the variational class of generalized Ramsey protocols considered in this work, as illustrated in Fig.~\ref{fig:variational}(a), by
        \begin{align}
            \mathcal{U}_\mathrm{prep} &= \mathcal{R}_\mathbf{n}\Big[\bigotimes_{j=1}^{n} \mathcal{T}_j\Big]\mathcal{R}_{\frac{\pi}{2}}\nonumber\\
            \mathcal{U}_\mathrm{meas} &= \mathcal{R}_\mathbf{m}\Big[\bigotimes_{j=n+ 1}^{n+m} \mathcal{T}_j\Big]\mathcal{R}_\mathbf{n}^\dag,\label{eq:variational_class}
        \end{align}
        where we introduced the abbreviations $\mathcal{T}_j = \mathcal{T}_{\mathbf{k}_j}(\mu_j)$ and $\mathcal{R}_{\frac{\pi}{2}}=\mathcal{R}_y\left(-\tfrac{\pi}{2}\right)$. The $\pi/2$-pulse $\mathcal{R}_{\frac{\pi}{2}}$ in $\mathcal{U}_\mathrm{en}$ generates the CSS polarized in $x$-direction $\ket{+}^{\otimes N} = (\kett{\uparrow} + \kett{\downarrow})^{\otimes N} / \sqrt{2}^N = \mathcal{R}_{\frac{\pi}{2}}\kett{\downarrow}^{\otimes N}$. The rotations $\mathcal{R}_\mathbf{n}$ and $\mathcal{R}_\mathbf{m}$ result in an effective phase evolution around an arbitrary axis $\mathbf{n}$, $S_\mathbf{n}=\mathcal{R}_\mathbf{n}^\dag S_z \mathcal{R}_\mathbf{n} $, and an effective measurement of $S_\mathbf{m}=\mathcal{R}_\mathbf{m}^\dag S_z \mathcal{R}_\mathbf{m}$, respectively. Similarly, each one-axis-twisting $\mathcal{T}_\mathbf{k}(\mu) = \mathcal{R}_\mathbf{k}^\dag\mathcal{T}_z(\mu)\mathcal{R}_\mathbf{k}$ can be expressed as an OAT with respect to the $z$-axis and a rotation $\mathcal{R}_\mathbf{k}$. The resulting variational classes are not restricted by any geometric constraints. By employing the optimal Bayesian estimator, which can take arbitrary non-linear forms, there are likewise no restrictions imposed on the signals. Consequently, the signals often exhibit strongly non-sinusoidal shapes without any symmetry and no apparent relation to the phase (cf. App.~\ref{app:signals}), which may initially seem counterintuitive. However, when combined with the corresponding estimator, this approach achieves a low phase estimation uncertainty, as we will see in this section. A similar behavior has already been observed for the GHZ protocol, where the estimator effectively mimics a parity measurement while the signal itself vanishes. In contrast, the use of a linear estimator, as defined in Eq.~\eqref{eq:linear_est}, typically results in signals that are anti-symmetric, at least within the range of the prior distribution. 
        
        For a given protocol class $[n,m]$, the quantum circuit contains $n + m$ OAT interactions with associated twisting strengths $\mu_j$. Together with the rotations $\mathcal{R}_\mathbf{n},\mathcal{R}_\mathbf{m},\mathcal{R}_{\mathbf{k}_j}$, which ensure geometric generality and are each characterized by two variational parameters, the total number of variational parameters is $4 + 3(n+m)$. Notably, the particular choice of the CSS $\ket{+}^{\otimes N}$ as the initial state allows us to fix the first OAT of $\mathcal{U}_\mathrm{prep}$ along the $z$-axis without losing any generality. This simplification reduces the total number of variational parameters by two.

        The variational class defined in Eq.~\eqref{eq:variational_class} contains the standard Ramsey protocols as limiting cases. Coherent spin states (CSS) are recovered in the $[0,0]$-protocol, while squeezed spin states (SSS) are implemented within the $[1,0]$-class. The GHZ protocol emerges as a special case, either within the $[1,0]$-class using the optimal Bayesian estimator, as discussed in the previous section, or as part of the $[1,1]$-class, as implemented in Ref.~\cite{Leibfried2004}.

    \subsection{Optimal protocols}
\begin{figure*}[tbp]
            \centering
                \includegraphics[width=1\textwidth]{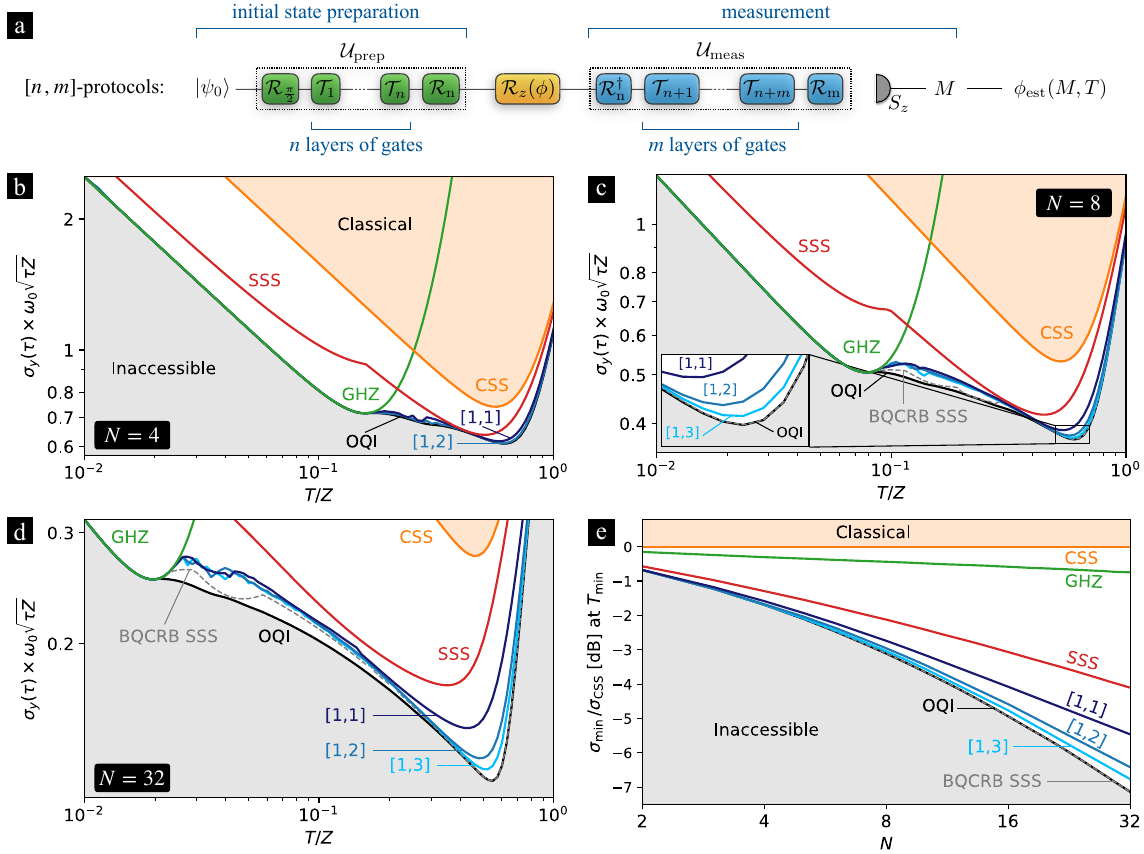}
         
            \caption{(a) Visualization of the variational Ramsey protocols defined in Eq.~\eqref{eq:variational_class}. The $\pi/2$-pulse $\mathcal{R}_\frac{\pi}{2} = \mathcal{R}_y(-\pi/2)$ generates the coherent spin state (CSS) polarized in $x$-direction from the ground state $\ket{\psi_0}=\kett{\downarrow}^{\otimes N}$. Entanglement in the initial state and measurement is introduced via one-axis-twisting (OAT) interactions, denoted by $\mathcal{T}_j = \mathcal{T}_{\mathbf{k}_j}(\mu_j)$ with twisting strength $\mu_j$ around axis $\mathbf{k}_j$. During the free evolution time $T$, the phase $\phi$ is imprinted onto the initial state via a rotation around the $z$-axis. The rotations $\mathcal{R}_\mathbf{n}$ and $\mathcal{R}_\mathbf{m}$ result in an effective phase evolution around an arbitrary axis $\mathbf{n}$, $S_\mathbf{n}=\mathcal{R}_\mathbf{n}^\dag S_z \mathcal{R}_\mathbf{n} $, and an effective measurement of $S_\mathbf{m}=\mathcal{R}_\mathbf{m}^\dag S_z \mathcal{R}_\mathbf{m}$, respectively. Finally, the phase is estimated based on measurement outcome $M$ of observable $S_z$. (b-d) Approximating the OQI using variational $[1,m]$-classes (blue) for (a) $N=4$, (b) $N=8$ and (c) $N=32$.  For comparison, the standard protocols are shown as they naturally emerge as specific quantum circuits within the variational classes. Additionally, the BQCRB of SSS is shown (dashed gray). With increasing $N$, the complexity of the variational circuits required to approach OQI performance increases. (e)  Scaling of the gain in clock stability compared to CSS at the optimal interrogation time $T_\mathrm{min}$ with $N$.}
                \label{fig:variational}
        \end{figure*}

        For fixed ensemble size $N$, circuit depth $[n,m]$ and prior phase width $\delta\phi$, the optimization of the quantum circuits introduced above is performed over all variational parameters. To enable a general discussion, we adopt the power-law scaling of the prior width with interrogation time $T$, as defined in Eq.~\eqref{eq:PowerLaw}. Results for exemplary ensemble sizes $N$ as well as the scaling of the stability with $N$ are presented in Fig.~\ref{fig:variational}. The variational protocols are primarily compared to the OQI, as saturating it is the central goal of this section. Additionally, comparisons to standard Ramsey protocols are provided where relevant to highlight specific advantages and limitations.

        \paragraph*{General results---} We begin by examining the general behavior and scaling of the variational classes, with a particular focus on the number of layers $n$ and $m$. While $[n,0]$-protocols yield collective spin measurements with sinusoidal signals, increasing $m$ allows for arbitrary signal shapes (cf. App.~\ref{app:signals}), since no geometric constraints are imposed. As for the standard Ramsey protocols, variational protocols exhibit a clear trade-off between enhanced stability for increasing interrogation times and the coherence time limit of the local oscillator.

        At long interrogation times close to the minimal Allan deviation $\sigma_\mathrm{min}$, the OQI is saturated by the BQCRB of SSS for any ensemble size $N$. Within the variational framework, this can be implemented using the $[1,m]$-classes, as the optimal measurement of the BQCRB is approximated in the limit $m \gg 1$. Increasing the number of entangling layers $n$ yields $\sigma_\mathrm{min}$ comparable to that of the corresponding $[1,m]$-protocols, consistent with findings in Ref.~\cite{Thurtell2024}. Consequently, to extend the dynamic range and approximate the OQI at long interrogation times requires to increase $m$. 
        
        In contrast, at short interrogation times, the dynamic range is negligible, and increasing the entanglement depth of the initial state, effectively determined by $n$, becomes beneficial. However, GHZ states are already optimal in this regime and saturate the Heisenberg limit. Thus, $n=1$ remains sufficient.

        At the plateau of the OQI, where GHZ states become ineffective, the $[1,m]$-classes in general do not saturate the OQI. This regime becomes broader for larger ensembles, since the dynamic range of GHZ states reduces with $N$. Here, achieving the OQI requires asymptotically deep quantum circuits, which, however, is unfavorable. Even in the limit of $n+m\gg 1$, as considered in Refs.~\cite{Kaubruegger2021,Marciniak2022}, the variational class only gradually approximates the OQI with increasing complexity. Furthermore, in the plateau regime, the optimal variational parameters strongly depend on the prior width, causing  substantial variations in the interferometer sequence. As a consequence, even minor modifications in the interrogation time can lead to profound changes in the form of both the signals and the associated estimation strategies. In particular, as the regime of GHZ states is exceeded, the twisting strengths decrease significantly, effectively reducing the degree of entanglement to adapt to increased LO noise. Interestingly, this susceptibility diminishes with increasing circuit complexity $m$. This can be interpreted as follows: For low depth quantum circuits, the variational degree of freedom is limited and thus, the optimal states and measurements have to be extremely well tailored to a specific prior width to ensure a sufficiently high degree of entanglement and dynamic range at the same time. As the variational complexity increases, the variational space grows and reduces the susceptibility to small variations in the prior width. Additionally, this dependence gives rise to a large number of local minima, making global optimization tedious and facilitating numerical errors, indicated by the non-smooth curves.

        As a consequence, we focus on approximating the OQI in all regimes except  the plateau using variational $[1,m]$-classes and strive for a minimal circuit depth $m$.

        \paragraph*{Protocol complexity and ensemble size---}For the simplest case, $N=2$, the GHZ protocol is optimal across most interrogation times because the critical prior width, $\delta\phi \sim 1/N$, is relatively large. Consequently, the region between the plateau and the minimal Allan deviation is narrow. In this transitional regime, SSS achieve the OQI, while the minimal ADEV as well as the plateau of the OQI are saturated by the simplest non-trivial variational class, the $[1,1]$-protocols. Hence, standard protocols are sufficient to saturate the OQI across a wide range of interrogation times.

        For $N=4$, illustrated in Fig.~\ref{fig:variational}(b), the plateau of the OQI already broadens significantly. SSS perform close to the OQI and the variational protocols over a narrow range of interrogation times in the transition regime. The $[1,1]$-protocols remain optimal for a broad range of interrogation times, closely approaching the OQI at the plateau and at the minimal ADEV. Increasing the circuit complexity, the $[1,2]$-class approximates the OQI.

        For $N=8$, depicted in Fig.~\ref{fig:variational}(c), the discrepancy between low-depth quantum circuits and the OQI at the plateau becomes more pronounced. Even the BQCRB of SSS, effectively represented by $[1,m]$-classes for $m\gg 1$, approximates the OQI only closely. In the transition regime, where the OQI scaling reverts to $\sim 1/\sqrt{T}$, SSS approximate both the variational classes and the OQI, but the deviation grows with $N$, as discussed in Sec.~\ref{sec:standard_protocols}. At the minimal ADEV, $[1,1]$-protocols substantially extend the dynamic range, but nevertheless leave a noticeable gap to the OQI, which is largely closed by $[1,2]$-classes. Increasing the variational complexity further, the $[1,3]$-class approximates the OQI in the vicinity of $\sigma_\mathrm{min}$, but does not fully saturate it. Consequently, already for $N=8$ relatively deep quantum circuits are required to saturate the OQI entirely. Since the gain diminishes with increasing $m$, and in order to keep the quantum circuit comparably simple, we do not increase the circuit depth further.

        As $N$ increases, this trend continues, as shown for $N=32$ in Fig.~\ref{fig:variational}(d). In this case, even the BQCRB of SSS exhibits significant deviations from the OQI at the plateau. In contrast, in the scaling regime of $\sim 1/\sqrt{T}$, and particularly at the minimal Allan deviation, the BQCRB saturates the OQI. The overall minimum is approximated by increasing $m$, but diminishing gains make deeper circuits less advantageous. Hence, we restrict our analysis to variational classes $[1,m]$ with $m\leq 3$ as before. In general, the variational complexity required to saturate the OQI grows with $N$ (cf. Fig.~\ref{fig:variational}(e)). These results align with the asymptotic analysis of OQI saturation in Refs.~\cite{Kaubruegger2021,Marciniak2022,Thurtell2024}.
        
        For large ensemble sizes ($N \gtrsim 100$), reaching the regime of optical lattice clocks, atom number fluctuations during interrogation become relevant~\cite{Ludlow2015}. Variational Ramsey protocols, optimized for fixed $N$, are highly sensitive to such fluctuations, making them less favorable in this regime. Instead, the POI emerges as a robust alternative, saturating the OQI in the limit of large $N$.\\

        In summary, for systems with small ensembles $N$, such as ion-traps and tweezer-arrays, low-depth variational classes are sufficient to approximate the OQI. These protocols generally achieve optimal performance across all interrogation times, except at the OQI plateau. Already the simplest variational protocols from the $[1,1]$-class significantly enhance the stability at long interrogation times, particularly in the regime of the minimal Allan deviation. The circuit depth of $[1,m]$ protocols required to actually saturate the OQI at long interrogation times increases with $N$. However, the performance gain diminishes with $m$, presenting a trade-off between reduced instability and increasing complexity. To maintain a reasonable balance between dynamic range and circuit depth, we restrict our analysis to $m\leq 3$, acknowledging that the OQI can be fully saturated in the limit of deep circuits $m\gg 1$, as quantified by the BQCRB of SSS.

\section{Application in the full feedback loop of an atomic clock}\label{sec:Simulations}
    The Bayesian approach captures key aspects of atomic clock operation, including finite prior information, single-shot measurements, and the trade-off between enhanced stability achieved through longer interrogation times and the coherence time limit of the local oscillator. However, it models only a single clock cycle and neglects cumulative effects that arise in a full feedback loop. In particular, in the regime where the invertible domain of the main fringe is exceeded by the prior distribution and thus an unambiguous phase estimation is no longer possible, so called \textit{fringe hops} might occur. In this scenario, the feedback loop passes to an adjacent fringe resulting in the clock running systematically wrong and consequently degrading the clock stability. Whether fringe hops or the coherence time limit impose the dominant constraint depends on the specific Ramsey protocol and interrogation time. Since fringe hops are a feature only emerging in the context of a full feedback loop, they are not captured by the theoretical model presented above. While existing approaches, such as those in Refs.~\cite{Schulte2020,Li2022}, provide rough estimates for the effects of fringe hops based on single cycle properties, they are typically limited to sinusoidal signals and lack general applicability. A rigorous treatment of fringe hops requires modeling the complete feedback loop, as pursued in Ref.~\cite{Fraas2016}, but adapting this framework to variational Ramsey protocols lies beyond the scope of this work. Instead, we perform realistic Monte Carlo simulations of the full feedback loop to validate our theoretical predictions on clock stability. These numerical simulations reflect the basic principles of atomic clock operation (cf. Sec.~\ref{sec:Motivation}), building on the methods presented in Refs.~\cite{Leroux2017,Schulte2020}. Further implementation details are provided in the App.~\ref{app:clock_simulation}. The prior width in the full feedback loop is determined iteratively, as discussed in Sec.~\ref{sec:Prior} and App.~\ref{app:iterated_prior}.

    To start with, in Sec.~\ref{sec:limiation_FH} we examine the limitations imposed by fringe hops and discuss the associated deviations between theoretical predictions and numerical simulations. In Sec.~\ref{sec:simulations_clock}, we investigate the clock stability within the full feedback loop of an atomic clock for various Ramsey protocols and ensemble sizes, identifying the protocols that perform best in the respective regimes. Furthermore, in Sec.~\ref{sec:linearEstimator}, we compare the linear estimation strategy with the optimal Bayesian estimator,  focusing particularly on variational quantum circuits and the limitations imposed by fringe hops.

    \subsection{Limitation due to fringe hops}\label{sec:limiation_FH}
        Results of numerical simulations, presented in Fig.~\ref{fig:simulations}, show good agreement with theoretical predictions across a wide range of interrogation times. However, significant deviations arise in two regimes.
        
        First, for small ensembles at long interrogation times, fringe hops limit the clock stability rather than the coherence time limit. As a result, the minimal Allan deviation $\sigma_\mathrm{min}$ is not achieved for the standard protocols and variational classes. Instead, the best stability is observed at $T_\mathrm{sim} < T_\mathrm{min}$, lying within the transition regime between the plateau of the OQI and $\sigma_\mathrm{min}$. However, as $N$ increases, $T_\mathrm{sim}$ approaches the coherence time limit at $T_\mathrm{min}$, resulting in improved stability. In particular, for $N\gtrsim 20$, fringe hops and the coherence time limit spoil the stability at the same level and thus, the minimal Allan deviation is achieved for the standard protocols and variational classes. Notably, GHZ protocols remain limited by fringe hops regardless of $N$ due to their inherently narrow dynamic range which decreases with the ensemble size.

        Second, deviations arise in the regime of the plateau of the OQI, which primarily can be explained by three arguments: (i) Similar to long interrogation times, fringe hops can occur in this regime. For instance, for $\delta\phi \lesssim 1/N$, the optimal variational protocols resemble the GHZ protocol. However, as argued before, fringe hops prevent GHZ protocols to achieve its minimal Allan deviation. Likewise, the optimal variational protocols may not attain the theoretical prediction as $\delta\phi \sim 1/N$. In this regime, $m \neq 0$ typically generates highly non-sinusoidal signals with reduced dynamic range compared to CSS and SSS (cf. App.~\ref{app:signals}), resulting in severe limitations due to fringe hops. (ii) As described in the previous section, in the regime of the OQI plateau, the optimal variational parameters are highly sensitive to small changes in the interrogation time, where this susceptibility diminishes with the circuit complexity $m$. With increasing ensemble size, this limitation increases, since the plateau gets broader with $N$. Although the prior width is determined iteratively, modeling the actual prior distribution solely based on the width remains a simplified parametrization of the prior knowledge. Furthermore, this iterative evaluation of the prior width relies on a fixed interrogation sequence (cf. App.~\ref{app:iterated_prior}), which may not capture the true prior width of variational protocols sufficiently accurate. Consequently, the optimization can lead to variational protocols that are more susceptible to the true residual noise than predicted by the model, resulting in deviations between theoretical predictions and numerical simulations. (iii) Additionally, the assumption of a Gaussian prior distribution for the residual noise in each cycle may not reproduce the true dynamics appropriately. In particular, for small ensembles, the number of possible measurement outcomes is small and thus, the central limit theorem justifies this assumption to a reasonable level only in the asymptotic limit of many repetitions. Consequently, corresponding deviations reduce with increasing $N$.    
    
        As a consequence, stability can be compromised in both regimes. To address these limitations, we simulate the clock performance of several protocols for a fixed interrogation time $T$ and variational class $[n,m]$, corresponding to distinct local minima in the parameter landscape, and select the protocol that achieves the best stability (see App.~\ref{app:clock_simulation}). Consequently, the best-performing protocol identified in simulations may differ from the theoretical optimum, leading to deviations between simulation and theory. Furthermore, the sensitivity landscape typically features numerous local minima and thus,  it is not feasible to simulate all emerging protocols. In extreme cases, fringe hops may affect all simulated protocols, leading to complete stability loss. Hence, we show the least complex variational class $[1,m]$ that achieves theoretical predictions at the OQI plateau. At long interrogation times, we include simulation results of deeper quantum circuits where a substantial gain is observed.

    \subsection{Clock stability}\label{sec:simulations_clock}
\begin{figure}[tbp]
            \centering
                \includegraphics[width=\linewidth]{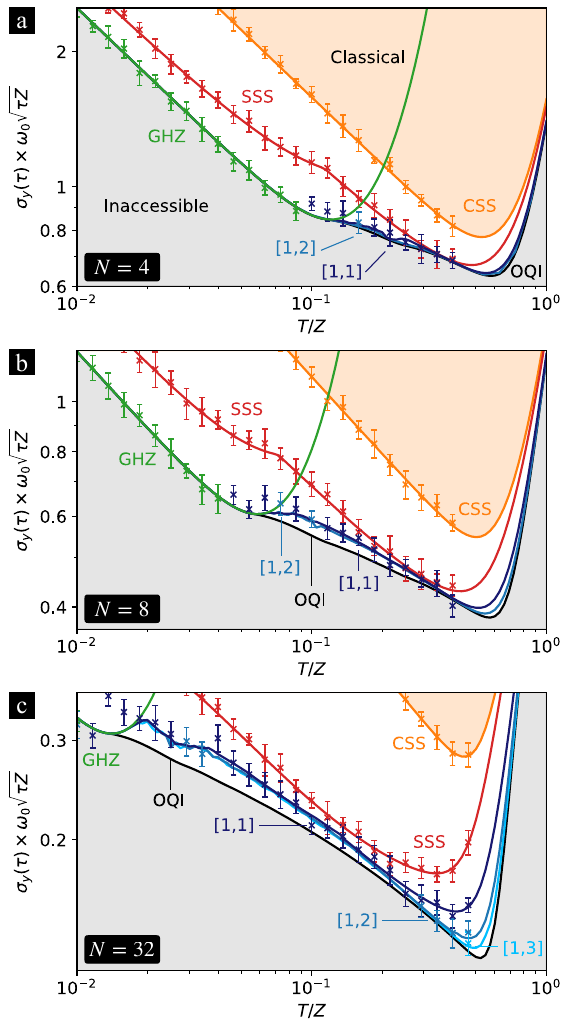}
         
            \caption{Numerical simulations of the full feedback loop in an atomic clock compared to the theoretical predictions of the Allan deviation for ensemble sizes (a) $N=4$, (b) $N=8$ and (c) $N=32$. Symbols represent the mean clock stability, while error bars indicate fluctuations over independent clock runs, arising from the stochastic nature of the Monte Carlo simulations. Fringe hops limit stability in the plateau regime and at long interrogation times, as discussed in the main text. The associated prior width is obtained iteratively. Further details on the numerical simulations are provided in App.~\ref{app:clock_simulation}.}
                \label{fig:simulations}
        \end{figure}

        Overall, numerical simulations align closely with theoretical predictions across a wide range of interrogation times. However, as discussed in the previous section, fringe hops impose the primary limitation at the OQI plateau. Additionally, for ensembles with $N\lesssim 20$ and long interrogation times, fringe hops limit the clock stability rather than the coherence time limit of the local oscillator. As a consequence, the minimal Allan deviation $\sigma_\mathrm{min}$ is not achieved for small ensembles, and variational protocols provide marginal to no advantage over SSS in this regime. In particular, we distinguish between three regimes based on the ensemble size: 
        
        (i) For very small ensembles with $N\lesssim 4$ (cf. Fig.~\ref{fig:simulations}(a)), the GHZ protocol saturates the ultimate lower limit, represented by the OQI, for short interrogation times, while at long interrogation times, approaching the fringe hop limit $T_\mathrm{sim}$, SSS become optimal. Hence, variational protocols provide an advantage over standard protocols only in the regime of the OQI plateau. In this regime, typically the simplest $[1,1]$-class already is sufficient to saturate the OQI. However, the optimal protocols vary significantly with interrogation time, increasing their susceptibility to fringe hops. Furthermore, the OQI plateau is relatively narrow for $N\lesssim 4$. Given the trade-off between potential stability gains and the experimental challenges involved in implementing more complex Ramsey protocols, the GHZ protocol at short interrogation times and SSS at longer interrogation times remain the preferable choices for ensembles with $N\lesssim 4$.
        
        (ii) For intermediate ensembles with $N\lesssim 20$ (cf. Fig.~\ref{fig:simulations}(b)), fringe hops continue to impose the fundamental limitation at long interrogation times. Furthermore, the regime of the OQI plateau expands, which in turn increases the region where variational protocols provide an advantage over SSS. Nevertheless, $[1,m]$ protocols remain fragile to fringe hops in this regime and additionally do not suffice to actually saturate the OQI. At long interrogation times approaching $T_\mathrm{sim}$, which itself approximates $T_\mathrm{min}$ with increasing $N$, variational protocols offer improved stability compared to SSS. In this regime, again the simplest $[1,1]$-class is sufficient to achieve a relevant improvement, while the additional benefit of $[1,m]$ protocols with $m>1$ is negligible when considering the fluctuations over independent clock runs. However, to achieve a gain compared to SSS at long interrogation times requires $T\sim T_\mathrm{sim}$. For practical implementation in an experiment, Ref.~\cite{Leroux2017} suggests to choose an interrogation time slightly shorter than $T_\mathrm{sim}$, effectively providing a safety margin against fringe hops. As a result, similar to $N\leq 4$, variational protocols for $N\lesssim 20$ effectively enhance clock stability primarily within the OQI plateau regime, which remains less favorable in experimental settings, while GHZ states and SSS are beneficial at short and long interrogation times, respectively.
        
        (iii) As the ensemble size increases to $N\gtrsim 20$ (cf. Fig.~\ref{fig:simulations}(c)), the limitations imposed by fringe hops and the coherence time limit become comparable. In this regime, variational protocols succeed to achieve $\sigma_\mathrm{min}$, resulting in a substantial gain in stability over SSS. Furthermore, as $N$ grows, increasing the circuit complexity $m$ of the $[1,m]$ protocols provides relevant gains in stability.

        To conclude, variational protocols for clocks with only a few atoms, characteristic of ion-traps, primarily enhance stability within the OQI plateau. However, this regime is less favorable due to the strong dependence of variational parameters on interrogation time and increased susceptibility to fringe hops. In contrast, for clocks with several tens of atoms, typical of tweezer-arrays, variational Ramsey protocols offer a significant improvement in clock stability, particularly at long interrogation times. Here, low-depth quantum circuits are sufficient, as the benefits diminish with increasing $m$, resulting in a trade-off between increased complexity and extended dynamic range.

    \subsection{Comparison of linear and optimal Bayesian estimation}\label{sec:linearEstimator}
        In Sec.~\ref{sec:standard_protocols}, we observed that the optimal Bayesian estimator achieves significant improvements over the linear estimator in several regimes for standard protocols. For instance, it provides a larger dynamic range at long interrogation times and enables stronger squeezed spin states at short interrogation times due to its non-linearity. Nevertheless, the linear estimator delivers equivalent results at interrogation times where the signal can be linearized within the extent of the prior distribution, which typically corresponds to $T/Z\ll 1$.  Moreover, the linear estimator simplifies numerical studies (see Sec.~\ref{sec:Estimators}) and has delivered remarkable results in previous works~\cite{Leroux2017,Schulte2020}, including applications in variational Ramsey interferometry~\cite{Kaubruegger2021,Marciniak2022,Thurtell2024}. Therefore, we compare the performance of the linear and optimal Bayesian estimators in the context of variational interrogation protocols to determine whether the potential advantages of the optimal Bayesian estimator, while significant in some regimes for standard protocols, translate into meaningful improvements in the case of variational quantum circuits.

\begin{figure}[!tbp]
            \centering
                \includegraphics[width=\linewidth]{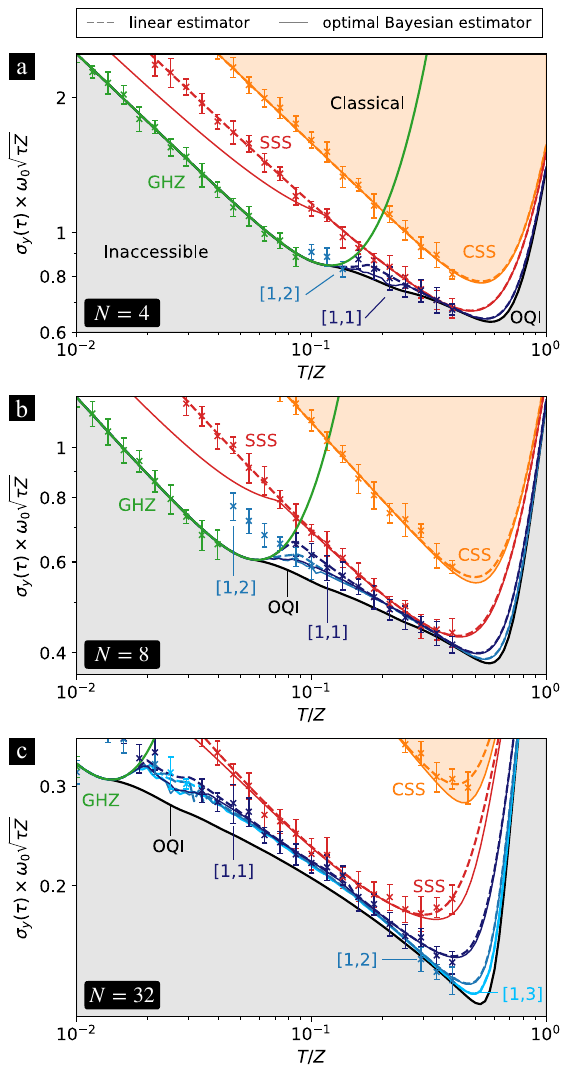}
         
            \caption{Numerical simulations of the full feedback loop in an atomic clock with a linear estimation strategy for ensemble sizes (a) $N=4$, (b) $N=8$ and (c) $N=32$ characterized by the Allan deviation. Lines depict theoretical predictions with the linear (dashed) and optimal Bayesian estimator (solid). Symbols represent the mean clock stability, while error bars indicate fluctuations over independent clock runs, arising from the stochastic nature of the process. Fringe hops limit stability in the plateau regime and at long interrogation times, as discussed in the main text. Further details on the numerical simulations are provided in App.~\ref{app:clock_simulation}.}
                \label{fig:estimator_comparison}
        \end{figure}

        In Fig.~\ref{fig:estimator_comparison}, we compare theoretical predictions of clock stability for optimized variational $[1,m]$ protocols employing both estimation strategies. Surprisingly, the linear estimator effectively achieves the same stability as the optimal Bayesian estimator. In particular, the optimal Bayesian estimator does not extend the dynamic range at long interrogation times and correspondingly does not enhance the minimal Allan deviation, while offering only a marginal enhancement in the plateau regime of the OQI, where GHZ protocols become ineffective. However, this gain is negligible, especially when considering the stability issues in this regime discussed in the previous sections. Consequently, in theory, the optimal Bayesian estimator does not provide a relevant improvement over the linear estimator, which is consistent with findings in Ref.~\cite{Marciniak2022} for exclusively anti-symmetric signals (cf. Supplementary Discussion S9 in Ref.~\cite{Marciniak2022}). 

        While theoretical predictions offer valuable insights, their validation in realistic scenarios is essential for a comprehensive analysis, as discussed before. Fig.~\ref{fig:estimator_comparison} additionally presents numerical simulations of the full feedback loop. The standard protocols perform as predicted by theory, exhibiting the same limitation imposed by fringe hops at long interrogation times, as observed with the optimal Bayesian estimator. For larger ensemble sizes $N\gtrsim 20$, where the coherence time limit and fringe hops constrain clock stability at the same level, the reduced dynamic range of the linear estimator for sinusoidal signals becomes relevant. As a results, the optimal Bayesian estimator achieves higher stabilities for CSS and SSS, as discussed in Sec.~\ref{sec:standard_protocols}. Again, variational protocols are constrained by fringe hops in two distinct regimes. (i) At the OQI plateau, the susceptibility to fringe hops is significantly enhanced for the linear estimator compared to the optimal Bayesian estimator, as indicated by larger deviations between theoretical predictions and numerical simulations, as well as broader regions where deeper quantum circuits are required to achieve theoretical expectations. However, given the strong variation in optimal Ramsey schemes and the emergence of fringe hops, operating a clock in these regimes may be experimentally unfavorable anyway, as discussed in previous sections. Thus, the stronger limitation imposed by fringe hops in this regime is of minor practical relevance. (ii) For $N\lesssim 20$, clock stability is limited by fringe hops at the same level for both estimation strategies, leading to comparable maximal interrogation times $T_\mathrm{sim}$ (cf. Fig.~\ref{fig:estimator_comparison}(a) and (b)). In contrast to the optimal Bayesian estimator, for larger ensembles with $N\gtrsim 20$, fringe hops remain the dominant limitation when using the linear estimation strategy. As a consequence, the minimal Allan deviation $\sigma_\mathrm{min}$ is not achieved for $[1,m]$ protocols, as illustrated in Fig.~\ref{fig:estimator_comparison}(c). Therefore, in clocks with a few tens of atoms, typically realized in tweezer-arrays, the linear estimation strategy causes fringe hops to impose a stricter constraint on clock stability than the coherence time of the local oscillator.
        
        In summary, the optimal Bayesian estimator guarantees to saturate the BCRB, thereby maximizing the use of the measurement data. Whether the linear estimation strategy can achieve comparable performance depends strongly on the particular interrogation scheme and must be evaluated for each specific scenario. For variational Ramsey protocols, as considered in this work, the optimal Bayesian estimator proves to be less susceptible to fringe hops. While this difference may be negligible in the regime of the OQI plateau, where these protocols are potentially unfavorable for experimental implementation, the critical ensemble size at which fringe hops and the coherence time limit constrain the clock stability at the same level is larger when using the linear estimation strategy.

        Moreover, it is important to note that the estimation strategy primarily affects the classical post-processing of the measurement outcome. Consequently, the complexity of the Ramsey sequence remains unchanged for both estimation strategies. The quantum circuit itself is only indirectly influenced, as the choice of estimator affects the optimal variational parameters. Typically, the optimal Bayesian estimator leads to smaller total twisting strengths $\mu = \sum_j|\mu_j|$, particularly for variational classes $[1,m]$ with $m>1$. Hence, the linear estimation strategy effectively requires larger twisting strengths to compensate for the non-linearity of the optimal Bayesian estimator. As a result, quantum circuits employing the optimal Bayesian estimator achieve shorter gate durations, which may be of practical interest.

\section{Dead time}\label{sec:deadtime}
    In the previous sections, we extensively discussed the trade-off between quantum projection noise (QPN), which decreases with the interrogation time $T$ (cf. Eq.~\eqref{eq:ADEV_local}), and the coherence time limit (CTL) of the local oscillator, which constraints clock stability at long interrogation times. Here, we extend the discussion to account for the effect of dead time $T_D$ in atomic clock operation. Dead time typically arises from processes such as probe preparation, measurement, and the application of feedback. During this period, frequency fluctuations of the local oscillator remain unmonitored by the Ramsey interrogation and therefore cannot be measured or corrected. The cumulative effect of this lack of information degrades the long-term stability of the atomic clock, a phenomenon first described by G. J. Dick~\cite{Dick1987,Dick1990}, therefore commonly referred to as the Dick effect. The contribution of the Dick effect to the clock stability is directly inferred from the spectral noise density $S_y(f)$ of the local oscillator and is given by~\cite{Dick1990,Santarelli1998}
    \begin{align}
        \sigma_{y,\text{Dick}}^2(\tau) = \frac{1}{\tau}\frac{T_C^2}{T^2}\sum_{k=1}^\infty S_y\left(\frac{k}{T_C}\right) \frac{\sin^2(\pi k T/T_C)}{\pi^2 k^2},\label{eq:DickEffect}
    \end{align}
    where $T_C=T_D+T$ is the clock cycle duration. The impact of the Dick effect diminishes with longer interrogation times $T$, as it depends on the ratio $T/T_C$ which decreases when the relative contribution of dead time is reduced. Taking dead time into account, the overall clock stability is determined by the interplay between QPN, CTL, and Dick noise.  Specifically, it is characterized by the total Allan deviation
    \begin{align}
        \sigma_{y,\text{tot}}(\tau) = \sqrt{\sigma_{y,\mathrm{QPN}}^2(\tau)+\sigma_{y,\mathrm{CTL}}^2(\tau) + \sigma_{y,\text{Dick}}^2(\tau)},\label{eq:totalADEV}
    \end{align}
    where QPN and the CTL are combined in the Bayesian framework as $\sigma_y^2(\tau) = \sigma_{y,\mathrm{QPN}}^2(\tau)+\sigma_{y,\mathrm{CTL}}^2(\tau)$. While the Bayesian approach generally does not  permit a strict separation of QPN and CTL contributions, except in specific cases such as the OQI or for CSS and SSS with a linear estimator (cf. Sec.~\ref{sec:standard_protocols}), it is nevertheless advantageous to treat them formally as independent components to discuss their general scaling quantitatively. The trade-off characterized by $\sigma_{y,\text{tot}}(\tau)$ has been thoroughly studied for CSS and SSS with a linear estimator in Ref.~\cite{Schulte2020}, where QPN was characterized  using local phase estimation theory, while the CTL was modeled via a stochastic differential equation describing the stabilized frequency of the local oscillator. In contrast, in this work we adapt the discussion to the Bayesian framework, which provides an intuitive and comprehensive approach to treating these effects. Additionally, for comparison, we include the ultimate lower bound on clock stability represented by the OQI. After analyzing the general scaling of $\sigma_{y,\text{tot}}(\tau)$ for standard protocols, we consider various experimental platforms and discuss the effect of dead times characteristic of each setup. Furthermore, we explore the potential benefits of variational quantum circuits in these regimes.

    \subsection{Dead time in Bayesian frequency metrology}\label{sec:deadtime_adjustments}
        In addition to the contribution described by Eq.\eqref{eq:DickEffect}, dead time affects Bayesian frequency metrology in two distinct ways. First, and most notably, it modifies the scaling of the Allan deviation, associated with QPN and the CTL, as a function of the interrogation time $T$. Instead of the ideal $\sim 1/\sqrt{T}$ scaling, dead time reduces it to $\sim \sqrt{T_C/T^2}$, as apparent in Eq.~\eqref{eq:ADEV_BMSE}. Second, dead time broadens the prior distribution of the phase due to unmonitored frequency fluctuations during $T_D$. Among these two effects, the modified scaling with $T$ has a substantially larger impact, whereas the broadening of the prior distribution introduces only a relatively minor correction. Nevertheless, incorporating the implicit broadening is crucial for accurate modeling and for identifying optimal Ramsey protocols and estimation strategies, as the prior width strongly influences the optimal interrogation sequence, as explored in previous sections.

        Although the prior width could, in principle, be adjusted iteratively to include dead time as for $T_D=0$, this approach is computationally demanding. Moreover, our goal is to establish a direct connection between scenarios with ($T_D > 0$) and without ($T_D=0$) dead time. Since the additional frequency fluctuations during dead time are unmonitored by the Ramsey interrogation, the broadening of the prior distribution during dead time and during the Ramsey sequence $T$ are independent processes. Treating the broadening of the phase distribution during dead time as a phase diffusion process, the modified prior distribution $\mathcal{P}(\phi) = \left(\mathcal{P}_D* \mathcal{P}_T\right)(\phi)$ is obtained by a convolution of the initial prior distribution $\mathcal{P}_T(\phi)$, resulting from the Ramsey interrogation time $T$ with corresponding width $\delta\phi_T$ (cf. Sec.~\ref{sec:Prior}), and the distribution $\mathcal{P}_D(\phi)$ associated with dead time. In this context, $\mathcal{P}_D(\phi)$ effectively acts as a Green's function~\cite{Gardiner2010}. Although local oscillator noise in general is correlated, the additional noise introduced during dead time within the full feedback loop is well approximated as white noise in the asymptotic limit of many clock cycles. Consequently, $\mathcal{P}_D(\phi)$ is modeled as a Gaussian distribution with zero mean and a width $\delta\phi_D$. As a result, the modified prior distribution $\mathcal{P}(\phi)$ remains Gaussian with zero mean and variance
        \begin{align}
            (\delta\phi)^2 = (\delta\phi_D)^2 + (\delta\phi_T)^2.\label{eq:prior_width_deadtime}
        \end{align}

        To fully incorporate the impact of dead time into the Bayesian framework, we now relate the broadening of the phase distribution, characterized by $\delta\phi_D$, to the dead time $T_D$, akin to the dead time-free case in Sec.~\ref{sec:Prior}. Rather than deriving a comprehensive model for arbitrary scenarios, we establish a relation $\delta\phi_D(T_D)$ that primarily aims to accurately predict behavior in the vicinity of the minimal Allan deviation $\sigma_\mathrm{min}$ at interrogation time $T_\mathrm{min}$. To this end, the broadening of the prior width during dead time can be effectively modeled by translating the additional frequency fluctuations into hypothetical phase shifts, as if they had occurred during a Ramsey interrogation of duration $T_D$. In this context, the associated prior width is given by (cf. App.~\ref{app:prior_deadtime})
        \begin{align}
                    (\delta\phi_D)^2 \simeq 2\left(\frac{T_D}{Z}\right)^{2+\alpha},\label{eq:varianceDeadtime}
        \end{align}
        reflecting a power-law dependence, analogous to Eq.~\eqref{eq:PowerLaw}. Here, the parameter $\alpha$ again characterizes the nature of the frequency noise, with values $\alpha=-1, 0, 1$ corresponding to white, flicker, and random walk frequency noise, respectively.
        
        As a consequence, adjusting the prior width according to Eq.~\eqref{eq:prior_width_deadtime} extends the Bayesian framework to incorporate dead time within the clock cycle, accounting for both the Ramsey interrogation time $T$ and the dead time $T_D$. Therefore, aside from adapting the prior width to reflect dead time $T_D$, the findings from the previous sections remain directly applicable. Therefore, the primary remaining task is to analyze the impact of the Dick effect $\sigma_{y,\mathrm{Dick}}(\tau)$ on overall clock stability.

    \subsection{General results}
        In general, the total clock stability reflects a trade-off between quantum projection noise (QPN), the coherence time limit (CTL) and the Dick effect, as described by Eq.~\eqref{eq:totalADEV}. While the CTL emerges at long interrogation times, limiting the clock stability as $T$ approaches the laser coherence time $Z$, both QPN and Dick noise decrease monotonically with the interrogation time $T$. Unlike QPN, which reduces with larger ensembles, the CTL and Dick noise are independent of $N$. As a result, whether the minimal Allan deviation $\sigma_\mathrm{min}$, achieved at optimal interrogation time $T_\mathrm{min}$, arises from a trade-off between QPN and CTL or between the Dick effect and CTL depends on the particular dead time, ensemble size and Ramsey protocol~\cite{Schulte2020}.

\begin{figure*}[tbp]
            \centering
                \includegraphics[width=1\textwidth]{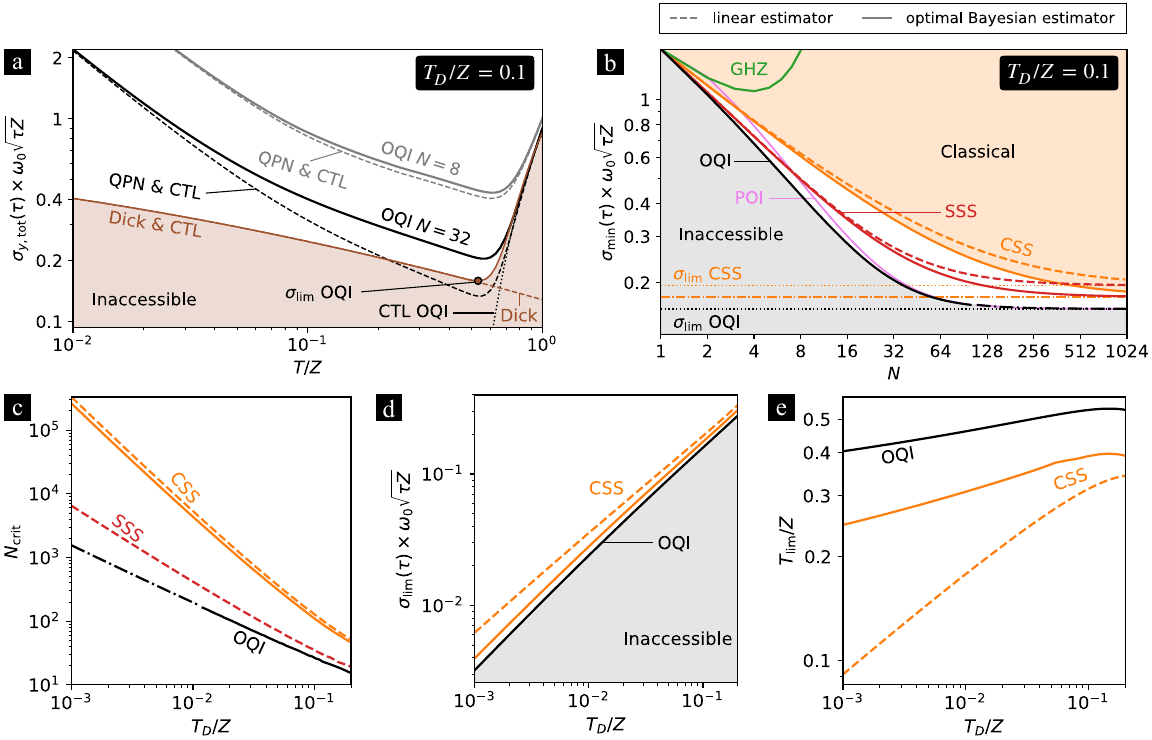}
         
            \caption{(a) Generic scaling of the dimensionless total Allan deviation $\sigma_{y,\mathrm{tot}}(\tau)\times \omega_0\sqrt{\tau Z}$ with the interrogation time $T$ for dead time $T_D/Z=0.1$. The total stability (solid) of the OQI for $N=8$ (gray) and $N=32$ (black) is shown in comparison to the trade-off between QPN and CTL (dashed). The $N$-independent lower limit $\sigma_\mathrm{lim}$ (symbol) is imposed by a trade-off (solid brown) between Dick noise (dashed brown) and CTL (dotted black). Consequently, the brown shaded area is inaccessible. (b) Scaling of the total dimensionless minimal Allan deviation $\sigma_\mathrm{min}\times \omega_0\sqrt{\tau Z}$ with the ensemble size $N$ for $T_D/Z=0.1$. GHZ protocols (green) achieve no gain compared to CSS. For CSS (orange) and SSS (red), both the linear (dashed) and optimal Bayesian estimator (solid) are depicted. The gray shaded area represents the inaccessible stability region set by the OQI (black), while the orange shaded area indicates achievable stabilities using uncorrelated atoms. Dotted lines correspond to the lower limit $\sigma_\mathrm{lim}$ for the OQI and CSS with linear estimator, while the dashed dotted line denotes the lower limit for CSS using the optimal Bayesian estimator. SSS exhibit the same lower limit as discussed in the main text and seen from the convergence. The POI (violet) saturates the OQI for $N\gtrsim 50$. For the OQI and POI, numerical optimization is performed for $N\leq 100$, while the asymptotic behavior, represented by the $\pi$HL (black dashed-dotted), is shown for $N>100$. (c) Critical ensemble size $N_\mathrm{crit}$ as a function of the dead time $T_D/Z$ for the OQI (black), CSS (orange) and SSS (red). Again, dashed lines correspond to the linear estimator, while solid lines represent the optimal Bayesian estimator. The evaluation of the SSS with optimal Bayesian estimator requires the computation of the conditional probabilities (cf. Sec.~\ref{sec:Estimators}) and thus is unfeasible for large $N$. For $N\gtrsim 100$, the asymptotic OQI, imposed by the $\pi$HL (black dashed-dotted), is shown. (d) Scaling of the total dimensionless lower limit $\sigma_\mathrm{lim}\times \omega_0\sqrt{\tau Z}$ with dead time for the OQI (black) and CSS (orange). For the CSS, linear (dashed) and optimal Bayesian estimator (solid) are displayed. (e) Corresponding interrogation times $T_\mathrm{lim}$, effectively characterizing the dynamic range. SSS result in the same $\sigma_\mathrm{lim}$ and $T_\mathrm{lim}$ as CSS, cf. discussion in the main text.}
                \label{fig:deadtime_general}
        \end{figure*}

        For short dead times or small ensembles, QPN typically dominates Dick noise, leading to behavior that closely resembles the dead time-free case (cf. $N=8$ in Fig.~\ref{fig:deadtime_general}(a)). Here, the clock stability is primarily determined by a trade-off between QPN and CTL and, therefore, depends on the ensemble size as well as the choice of Ramsey sequence. However, as dead time increases or QPN decreases, at some point, QPN is reduced to the level of Dick noise. Since Dick noise typically decreases more slowly with the interrogation time than QPN, first, it becomes dominant at long interrogation times, limiting the minimal Allan deviation $\sigma_\mathrm{min}$ (cf. $N=32$ in Fig.~\ref{fig:deadtime_general}(a)). Reducing QPN further, by either increasing the ensemble size or adapting the Ramsey interrogation, improves $\sigma_\mathrm{min}$ only marginally. In the regime where dead time effects strictly dominate over QPN, as is the case for large ensembles or long dead times, no further improvements in clock stability are possible, as Dick noise is independent of the particular Ramsey sequence and ensemble size. Therefore, we can define a lower limit $\sigma_\mathrm{lim}$ on the clock stability, at corresponding interrogation time $T_\mathrm{lim}$, which is characterized by a trade-off between Dick noise and CTL. Since the CTL is protocol-dependent, $\sigma_\mathrm{lim}$ in general differs for distinct Ramsey protocols and is primarily determined by their respective dynamic range.
        
        Fig.~\ref{fig:deadtime_general}(b) illustrates the scaling of $\sigma_\mathrm{min}$ with the ensemble size $N$ for the standard Ramsey protocols. For small ensembles, where QPN dominates, clock stability improves as $N$ increases, as in the ideal scenario ($T_D=0$). However, as the ensemble size grows, Dick noise becomes relevant, reducing the $N$-scaling and causing the clock stability to converge to $\sigma_\mathrm{lim}$. Unfortunately, explicit expressions for $\sigma_\mathrm{lim}$ can only be derived for protocols where QPN and CTL are separable, such as the OQI or CSS and SSS with a linear estimator. Otherwise, the convergence towards $\sigma_\mathrm{lim}$ with $N$ has to be evaluated numerically. As argued before, CSS and SSS with a linear estimator exhibit the same CTL and, consequently, identical lower limits. A similar behavior is observed for both protocols using the optimal Bayesian estimator, which however, achieves an improved $\sigma_\mathrm{lim}$ due to the larger dynamic range (cf. Sec.~\ref{sec:standard_protocols}). GHZ protocols, already highly susceptible to local oscillator noise in dead time-free scenarios, are further constrained by dead time, making them suitable only for small ensembles and short dead times. In the asymptotic limit, the performance of the POI again saturates the OQI.

        To characterize the transition between the regimes dominated by either QPN or Dick noise, we define a critical ensemble size $N_\mathrm{crit}$, at which the Allan deviation $\sigma_y(\tau)$ (cf. Eq.~\eqref{eq:ADEV_BMSE}), arising from QPN and the CTL, saturates $\sigma_\mathrm{lim}$ at $T_\mathrm{lim}$. Beyond $N_\mathrm{crit}$, Dick noise dominates over QPN and thus spoils the $N$-scaling of $\sigma_\mathrm{min}$, which ultimately converges towards $\sigma_\mathrm{lim}$, without substantial improvements as $N$ increases. Since $N_\mathrm{crit}$ depends explicitly on QPN, it differs for distinct Ramsey protocols and estimation strategies, as generically illustrated in Fig.~\ref{fig:deadtime_general}(c). For instance, $N_\mathrm{crit}$ for SSS is substantially smaller than for CSS, since they exhibit the same CTL, but SSS have a substantially smaller QPN. Consequently, the required ensemble size to approach $\sigma_\mathrm{lim}$ is substantially smaller for SSS compared to CSS, with a reduction of up to two orders of magnitude for short $T_D$, while still maintaining a significant difference even at long dead times. In contrast, the difference between OQI and SSS is relatively small, amounting to less than one order of magnitude for short dead times and becoming effectively negligible as $T_D$ increases.

        For a particular dead time $T_D$, the lower limit $\sigma_\mathrm{lim}$ is determined solely by the CTL of the Ramsey protocol and estimation strategy, essentially reflecting the dynamic range. Consequently, CSS and SSS achieve the same lower limit for a specific estimator. Moreover, as shown in Fig.~\ref{fig:deadtime_general}(d), the enhancement of $\sigma_\mathrm{lim}$ for the OQI compared to CSS or SSS is relatively small. Interestingly, increasing the dynamic range of CSS and SSS by substituting the linear by the optimal Bayesian estimation strategy yields a greater gain than the advantage provided by the OQI over CSS or SSS with the optimal Bayesian estimator.
        
        As $T_D$ increases, the potential enhancement of $\sigma_\mathrm{lim}$ diminishes further. This can be understood as follows: In general, dead time shifts $T_\mathrm{min}$, where the minimal Allan deviation $\sigma_\mathrm{min}$ is achieved, to longer interrogation times. This is shown in Fig.~\ref{fig:deadtime_general}(e) and primarily results from the impact of $\sigma_{y,\mathrm{Dick}}$ (cf. Fig.~\ref{fig:deadtime_general}(a)). However, in this regime, the difference in the CTL for distinct Ramsey schemes decreases with increasing $T$ (cf. Fig.~\ref{fig:standard_protocols}(a)), thereby reducing the advantage associated with a larger dynamic range. While the OQI allows unbiased phase estimation over $[-\pi,+\pi]$, the optimal Bayesian strategy for the CSS and SSS resembles the \textit{arcsin} estimator and thus covers the range $[-\pi/2,+\pi/2]$ (cf. Sec.~\ref{sec:standard_protocols}). As a result, the OQI and CSS or SSS with optimal Bayesian estimator exhibit a similar behavior, where the corresponding gain only marginally reduces with $T_D$. In contrast, the deviation between the linear and optimal Bayesian estimators for CSS and SSS reduces substantially with $T_D$, since the corresponding $N_\mathrm{crit}$ becomes smaller, leading to a reduced gain in dynamic range for the optimal Bayesian estimator, as discussed in Sec.~\ref{sec:standard_protocols}.

        To summarize, for small ensembles $N$ or short dead times $T_D$, clock stability  is primarily limited by QPN, closely resembling the dead time-free case. However, as $N$ or $T_D$ increases, the Dick effect becomes the dominant noise and ultimately limits the clock stability. Beyond the critical ensemble size $N_\mathrm{crit}$, which decreases with $T_D$, the minimal Allan deviation $\sigma_\mathrm{min}$ converges to the lower limit $\sigma_\mathrm{lim}$. In this regime, further improvements in clock stability by increasing the ensemble size or adapting the Ramsey sequence are marginal. As a consequence, clocks with large ensembles $N\gg 1$ limited by Dick noise sufficiently well approximate the lower limit $\sigma_\mathrm{lim}$ by employing CSS or SSS.

    \subsection{Setup specific dead times}\label{sec:deadtime_setups}
        Building on the general discussion of dead time effects on  clock stability in standard protocols, this section focuses on examining specific examples relevant to particular experimental setups, such as ion-traps, tweezer-arrays and lattice clocks, ranging from a few to thousands of atoms.
        
        In general, atomic clock operation involves three key time scales: the laser coherence time $Z$, the dead time $T_D$ and the interrogation time $T$. In a given experimental setup, $Z$ and $T_D$ are primarily independent but fixed, defining a specific ratio $T_D/Z$. In contrast, $T$ remains an adjustable parameter, which is implicitly constrained by $Z$. As a consequence, findings on clock stability cannot be trivially rescaled with respect to various laser coherence times $Z$, as in the dead time-free scenario,  or dead times $T_D$, since a modification of $Z$ or $T_D$ results in a change of the ratio $T_D/Z$, which in turn has a substantial impact on the clock stability, as discussed in the previous section. 
        
        In experimental settings, the dead time $T_D$ and interrogation time $T$ commonly are expressed in terms of the dimensionless duty cycle 
        \begin{align}
            \eta = \frac{T}{T_C} = \frac{T}{T + T_D},
        \end{align}
        which quantifies the relative contribution of the interrogation time $T$ to the total duration of the clock cycle $T_C=T_D+T$. Hence, a larger duty cycle $\eta$ corresponds to a reduced relative impact of dead time. However, it is crucial to note that nevertheless a specific ratio of $T_D$ and $Z$ is always assumed implicitly.

        While the laser coherence time $Z$ is independent of the atomic reference, dead time strongly depends on the particular experimental platform. To address this, we investigate the clock stability for typical dead times across the three predominant regimes: ion-traps, tweezer-arrays, and lattice clocks. Each of these platforms exhibits distinct time scale dynamics and operational characteristics that significantly influence clock performance. Ion-traps provide the highest degree of control, including rapid cooling and no need for reloading due to the deep trap depths, resulting in relatively short dead times~\cite{Ludlow2015}. Although recent advancements in Coulomb crystals have facilitated multi-ion clocks~\cite{pelzer2023,Hausser2024}, ion-traps remain inherently limited in scalability, typically operating with only a few ions. In contrast, optical lattice clocks employ large ensembles of hundreds to thousands of atoms, enabling high precision at the cost of experimental challenges such as atom number fluctuations and interatomic collisions~\cite{Ludlow2015}. Furthermore, these systems exhibit substantially longer dead times due to processes such as loading the lattice and cooling the atoms~\cite{Takamoto2015,Katori2003,Masoudi2015,Katori2011}. Additionally, dead time has a particularly pronounced impact on clock stability in lattice clocks, as QPN is typically suppressed well below the Dick effect, as discussed in the previous section. Tweezer-arrays bridge between these two contrary approaches, offering a balance between the high control in ion-traps and the scalability inherent in lattice clocks~\cite{madjarov2019,norcia2019,young2020,shaw2024}. By incorporating ensembles of several tens of atoms, they offer a promising compromise between precision, scalability, and operational efficiency. 

        In the following, we investigate the clock stability for representative dead times $T_D$ and laser coherence times $Z$ across these three distinct regimes. As discussed before,  this is equivalently expressed by fixing the ratio $T_D/Z$. Starting with ion-traps, which feature relatively short dead times, we explicitly examine the interplay between  laser coherence time $Z$, dead time $T_D$, and interrogation time $T$, or equivalently the duty cycle $\eta$, using state-of-the-art parameter values to develop an intuitive understanding of the relationship between these time scales. Subsequently, we progressively increase the dead time for setups representing tweezer-arrays and lattice clocks, illustrating how dead time increasingly constrains clock performance and how the optimal Ramsey protocols change accordingly.

        \paragraph*{Ion-traps---} In ion-traps, dead times of about $T_D=100~\mathrm{ms}$ are routinely implemented in experiments. Moreover, state-of-the-art clock lasers achieve laser coherence times $Z$ of several seconds. However, in practice, this impressive level of coherence is often not entirely maintained as the laser propagates between the cavity, which not necessarily is located close to the trap or even in the same laboratory, and the ions. While optical path-length stabilization should, in principle, preserve coherence all the way to the ions, experimental imperfections, such as phase noise within the vacuum chamber, typically lead to a degradation of this quality. Therefore, we assume a laser coherence time of $Z=2~\mathrm{s}$ at the location of the ions. Hence, in this exemplary scenario we obtain a ratio $T_D/Z=0.05$. Furthermore, Fig.~\ref{fig:simulations} demonstrates that fringe hops limit the clock stability at interrogation times around $T_\mathrm{sim}\sim 0.4-0.5\times Z$. Consequently, the maximal achievable duty cycle, given by $\eta_\mathrm{max} = T_\mathrm{sim} / (T_\mathrm{sim} + T_D)$, is on the order of $90\%$.

\begin{figure*}[tbp]
            \centering
                \includegraphics[width=1\textwidth]{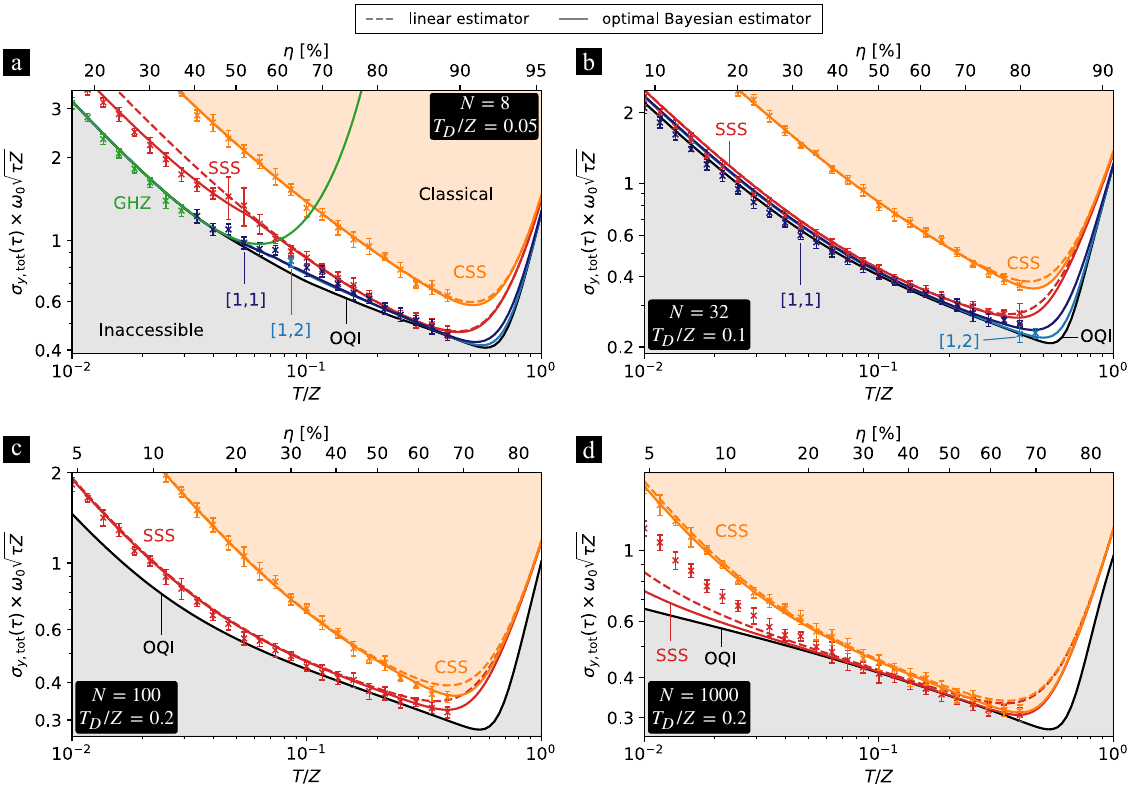}
         
            \caption{Theoretical predictions and numercial simulations of various Ramsey protocols for (a) $N=8$ and $T_D/Z=0.05$, (b) $N=32$ and $T_D/Z=0.1$, (c) $N=100$ and $T_D/Z=0.2$, (d) $N=1000$ and $T_D/Z=0.2$. Theory curves (lines) are displayed for the linear (dashed) and optimal Bayesian estimator (solid). Symbols represent numerical simulations in the full feedback loop of an atomic clock employing the optimal Bayesian estimation strategy. In both cases, the total Allan deviation is rescaled with respect to the atomic transition frequency $\omega_0$, total averaging time $\tau$ and laser coherence time $Z$. The lower $x$-axis represents the interrogation time $T$ relative to $Z$, while the upper $x$-axis denotes the dimensionless duty cycle $\eta$. The gray shaded area represents the inaccessible stability region set by the OQI limit (black), while the orange shaded area indicates achievable stabilities using uncorrelated atoms. For $N=8$ (a) and $N=32$ (b) the performance of variational quantum circuits (blue) is shown in addition to the standard protocols, namely GHZ states (green), CSS (orange) and SSS (red).}
                \label{fig:deadtime}
        \end{figure*}

        As discussed in the previous section, for small ensembles and short dead times, clock stability typically resembles the dead time-free scenario ($T_D=0$), as illustrated in Fig.~\ref{fig:deadtime}(a) for $N=8$ and $T_D/Z=0.05$.  In this case, QPN remains the primary limitation, while the Dick effect has only a marginal impact, leading to a behavior similar to that shown in Fig.~\ref{fig:simulations}(b). However, in comparison, the plateau of the OQI in the presence of dead time is substantially less pronounced and thus, fringe hops impose a less stringent limitation in this regime. As in the $T_D=0$ case, SSS approximate the OQI in the transition region between the plateau and $\sigma_\mathrm{min}$. At long interrogation times, fringe hops remain the primary constraint, limiting the interrogation time to $T_\mathrm{sim} < T_\mathrm{min}$. Furthermore, variational protocols effectively provide no significant enhancement in clock stability around $T_\mathrm{sim}$. As a result, for optical atomic clocks based on ion-traps, GHZ states and SSS approach the  OQI over a broad range of interrogation times, while the deviation from the OQI or variational classes within the plateau are reduced compared to the dead time-free scenario.

        \paragraph*{Tweezer-arrays---} For tweezer arrays, we consider a representative case with $N=32$ in Fig.~\ref{fig:deadtime}(b), assuming an increased dead time of $T_D/Z=0.1$. Within the framework of the previous example, this corresponds to an absolute dead time of $T_D=200\mathrm{ms}$ and a maximal achievable duty cycle of approximately $\eta_\mathrm{max} = 80\%$. As already evident in Fig.~\ref{fig:deadtime_general}(a), dead time in this regime imposes a significant limitation on clock stability. While GHZ states essentially are ineffective, SSS already perform close to the OQI for short and intermediate interrogation times. Variational protocols offer only marginal improvements in stability, with a noticeable enhancement observed only for $[1,m]$ protocols in the vicinity of $T_\mathrm{min}$. However, this gain is significantly smaller than in the dead time-free case, and unlike the $T_D=0$ scenario, fringe hops constrain clock stability at interrogation times $T_\mathrm{sim} \lesssim T_\mathrm{min}$. Additionally considering a safety margin for fringe hops, as discussed in Sec.~\ref{sec:Simulations}, the improvement becomes effectively negligible. Consequently, SSS emerge as a robust Ramsey sequence, achieving clock stabilities close to the OQI in this regime.

        Interestingly, for short interrogation times, deviations between theoretical predictions and numerical simulations appear. These discrepancies stem from the assumed prior width in the presence of dead time, which is intended to provide a reliable model primarily for interrogation times in the vicinity of $\sigma_\mathrm{min}$.

        \paragraph*{Crossover regime---} Typically, the boundaries with respect to the ensemble size $N$ between different platforms are not sharply defined.  To explore the transition between tweezer-arrays and lattice clocks, we examine the case of $N=100$ in Fig.~\ref{fig:deadtime_general}(c), with an increased dead time $T_D/Z=0.2$. In the example above, this corresponds to $T_D=400~\mathrm{ms}$ and an associated maximal duty cycle of approximately $\eta_\mathrm{max} = 65\%$. Such an increase in dead times is characteristic of lattice clocks, as discussed before, but can also result from various processes such as the overhead of operating multiple tweezer arrays simultaneously, the potential need for reloading due to shallower trap depths or extended cooling times. Moreover, inhomogeneous interactions may be relevant, as addressed in Ref.~\cite{Kaubruegger2021}. In this regime, variational classes are no longer favorable, as discussed in previous sections. A key characteristic of this regime is that dead time becomes the dominant limitation. However, while CSS have not yet fully converged to the lower bound $\sigma_\mathrm{lim}$, SSS already approximate it. As a result, SSS perform close to the OQI across all interrogation times, except at $T_\mathrm{min}$, where their limited dynamic range becomes apparent. Additionally, the choice of estimation strategy for standard protocols gains importance, as the optimal Bayesian estimator yields significantly higher clock stability at long interrogation times compared to the linear estimator.

        \paragraph*{Lattice clocks---} Finally, we investigate the regime of lattice clocks with large ensembles $N\gg 1$, where QPN is reduced well below the Dick noise. Fig.~\ref{fig:deadtime}(d) illustrates the case of $N=1000$ with $T_D/Z=0.2$. In this regime, both CSS and SSS closely approximate the lower limit $\sigma_\mathrm{lim}$. As a result, at long interrogation times, both protocols achieve comparable clock stability, whereas SSS provide a significant advantage at short interrogation times. Furthermore, the optimal Bayesian estimator results in a substantially higher stability in the vicinity of $T_\mathrm{min}$ compared to the linear estimation strategy. Notably, deviations from theoretical predictions and numerical simulations appear for the SSS at short interrogation times due to the choice of prior width (cf. Sec.~\ref{sec:deadtime_adjustments}). Additionally, since SSS introduce correlations between atoms, unlike CSS, numerical approximations are required to simulate the full feedback loop for $N\gg 1$, which can further contribute to discrepancies.\\

        In summary, for small ensembles representing ion-traps, the behavior closely resembles the dead time-free case. Here, standard protocols as GHZ states or SSS already achieve clock stabilities comparable to the OQI for a wide range of interrogation times. As the ensemble size $N$ or dead time $T_D$ increases, Dick noise becomes the dominant limitation, effectively reducing the potential enhancement offered by variational quantum circuits compared to SSS. In particular, dead time results in SSS performing close to the OQI for a variety of scenarios. In the regime of large ensembles $N\gg 1$, characteristic for lattice clocks, CSS likewise converge to the lower limit $\sigma_\mathrm{lim}$ at long interrogation times and thus, are sufficient to approximate the OQI. As a consequence, dead time significantly constraints clock stability, where the degree of limitation increases with the ensemble size.

\section{Outlook}
    At this point, rather than reiterating the detailed insights and results from the previous sections, we refer the reader to the introduction for a comprehensive overview. Here, we briefly discuss promising avenues for further improvements in optical atomic clocks.

    Overall, laser noise constrains clock stability in three distinct ways: via the laser coherence time limit (CTL), the emergence of fringe hops and dead time effects. In addition to ongoing technological improvements in laser stability~\cite{Matei2017,Robinson2019,Kedar2023}, several interrogation schemes have been proposed that go beyond the conventional single-ensemble approach, where the atomic reference is interrogated with the same protocol in every clock cycle, by employing adaptive schemes~\cite{Mullan2014,Han2024,Zhou2024} and multi-ensemble strategies. For instance, dynamical decoupling sequences~\cite{Drscher2020} and synchronous differential clock comparisons~\cite{Nardelli21,Hume2016,Zheng2022,Clements2020} have been demonstrated to extend interrogation times well beyond the laser coherence time. Other approaches involve active feedback and feedforward on the laser~\cite{Pezze2020,Kim2022}, or cascaded clock operation that allows for increasingly long interrogation times~\cite{Kessler2014,Kim2022,Borregaard2013,Bowden2020,Cao2024}. Furthermore, dead time free clock operation can be achieved by asynchronously interrogating two atomic ensembles~\cite{Biedermann2013,Takamoto2011,Schioppo2016}. As proposed by Rosenband and Leibrandt in Ref.~\cite{Rosenband2013}, partitioning atoms into multiple ensembles with distinct interrogation times can exponentially improve clock stability relative to the atom number. Moreover, synchronous out-of-phase interrogations expand the invertible phase range and enhance the dynamic range~\cite{Li2022,Zheng2024}. Although these approaches extend beyond the scope of the present work, many of their underlying principles can be integrated with the Ramsey protocols discussed here, potentially mitigating the limitations imposed by laser noise and enabling longer interrogation times.

    Beyond the effects of laser noise, other decoherence processes can have a significant impact on the clock performance. For instance, Ref.~\cite{Macieszczak2014} considers additional collective dephasing that is not associated with laser noise. Since collective dephasing is phenomenologically similar to the treatment of laser noise within the Bayesian framework, it affects stability in much the same way. The impact of uncorrelated single-atom dephasing in the Bayesian framework has been explored in Ref.~\cite{Kaubruegger2021}, where it was observed that for moderate dephasing strengths the overall behavior remains qualitatively unchanged, although stability is naturally degraded.  However, the benefit provided by variational quantum circuits, or more complex Ramsey schemes in general, over SSS diminishes substantially as the dephasing strength increases, leading to a behavior akin to that observed for dead time. A comparable pattern was reported in Ref.~\cite{Thurtell2024} for correlated single-atom dephasing. In scenarios where single-atom dephasing imposes the dominant noise source, the regime explored in the seminal work by Huelga \textit{et al.} in Ref.~\cite{Huelga1997} is entered.  In the asymptotic limit of large ensembles, SSS are known to be optimal~\cite{UlamOrgikh2001}, whereas for small ensembles, the optimal interrogation scheme can be efficiently determined numerically by exploiting permutational invariance~\cite{Fröwis2014}. Likewise, SSS remain asymptotically optimal for setups limited by spontaneous decay. In contrast, for small ensembles, highly entangled GHZ-like states approach the ultimate bounds, as demonstrated in Ref.~\cite{Kielinski2024}. Beyond decoherence during the Ramsey dark time, noise affecting the twisting operations has also been considered in Refs.~\cite{Thurtell2024,Scharnagl2023}. As expected, deeper quantum circuits, which generally require stronger total twisting strengths, exhibit higher susceptibility to noise. Overall, although the optimal Ramsey sequence ultimately depends on the specific system parameters, these observations reinforce the conclusions of this work: SSS provide a robust scheme with low operational complexity that ensures optimal or near-optimal stability across a wide range of scenarios, whereas deeper quantum circuits offer a significant advantage only in specific parameter regimes.

\section*{Acknowledgments}
We thank P. Schmidt, T. Mehlstäubler, N. Huntemann, U. Sterr, C. Lisdat, R. Kaubrügger, D. Vasilyev and P. Zoller for valuable discussions on the topics of this progress article. We acknowledge funding by the Deutsche Forschungsgemeinschaft (DFG, German Research Foundation) through Project-ID 274200144 – SFB 1227 (projects A06 and A07) and Project-ID 390837967 - EXC 2123, and by the Quantum Valley Lower Saxony Q1 project (QVLS- Q1) through the Volkswagen foundation and the ministry for science and culture of Lower Saxony.


%


\newpage
\ \\
\newpage
\onecolumngrid
\section*{Appendix}
\setcounter{section}{0}  
\renewcommand{\thesection}{A\arabic{section}}
\setcounter{figure}{0}  
\renewcommand{\thefigure}{A\arabic{figure}}
\setcounter{equation}{0}  
\renewcommand{\theequation}{A\arabic{equation}}

\section{Derivations of Bayesian bounds}\label{app:bounds}
    In the following, we derive the Bayesian bounds summarized in the main text.\\

    \subsection{Bayesian Cram\'{e}r-Rao Bound (BCRB)}\label{app:BCRB}
    To derive the Bayesian Cram\'{e}r-Rao bound (BCRB), following Ref.~\cite{Gill1995}, we assume standard regularity conditions
    \begin{align}
        \sum_x \partial_\phi P(x|\phi)= \partial_\phi \sum_x P(x|\phi)=0,\label{eq:regularity}
    \end{align}
    where the last equality results from normalization of the conditional probabilities $P(x|\phi)$, and vanishing of the prior at the boundaries
    \begin{align}
        \lim_{\phi\to \pm\infty} \mathcal{P}(\phi)=0.\label{eq:boundary}
    \end{align}
    By defining the function
    \begin{align}
        f(\phi,x) = \sqrt{\mathcal{P}(\phi)P(x|\phi)}\left[\phi-\phi_\mathrm{est}(x)\right],
    \end{align}
    the BMSE can be expressed as a squared norm
    \begin{align}
        (\Delta\phi)^2 = \int \dd\phi\, \sum_x f^2(\phi,x).
    \end{align}
    Furthermore, we define
    \begin{align}
        g(\phi, x) = \frac{1}{\sqrt{\mathcal{P}(\phi)P(x|\phi)}} \frac{\dd \mathcal{P}(\phi)P(x|\phi)}{ \dd\phi\,}
    \end{align}
    such that
    \begin{align}
        \int \dd\phi\, \sum_x g^2(\phi, x) &= \int \dd\phi\, \sum_x \frac{1}{\mathcal{P}(\phi)P(x|\phi)} \left(\mathcal{P}(\phi)\frac{\dd P(x|\phi)}{ \dd\phi\,} + P(x|\phi)\frac{\dd \mathcal{P}(\phi)}{ \dd\phi\,} \right)^2\\
        &= \int \dd\phi\, \mathcal{P}(\phi)\sum_x \frac{1}{P(x|\phi)} \left(\frac{\dd P(x|\phi)}{ \dd\phi\,} \right)^2 + \int \dd\phi\,  \frac{1}{\mathcal{P}(\phi)} \left(\frac{\dd \mathcal{P}(\phi)}{ \dd\phi\,} \right)^2 \sum_x P(x|\phi)\\
        &\quad+ 2\int \dd\phi\,\frac{\dd \mathcal{P}(\phi)}{ \dd\phi\,} \sum_x \frac{\dd P(x|\phi)}{ \dd\phi\,}\\
        &= \overline{\mathcal{F}}+ \mathcal{I}.
    \end{align}
    In the last step, we introduced the average Fisher information $\overline{\mathcal{F}}$ and prior knowledge $\mathcal{I}$ defined in Eq.~\eqref{eq:FI_average} and Eq.~\eqref{eq:prior_knowledge}, respectively. Furthermore, the last term vanishes as a consequence of the regularity condition Eq.~\eqref{eq:regularity}. Using partial integration, Eq.~\eqref{eq:boundary} and normalization of the probability distributions, we evaluate
    \begin{align}
        \int \dd\phi\, \sum_x f(\phi, x)g(\phi, x) &= \int \dd\phi\, \sum_x \left[\phi-\phi_\mathrm{est}(x)\right]\frac{\dd \mathcal{P}(\phi)P(x|\phi)}{ \dd\phi\,}\\
        &= \left[\sum_x\left[\phi-\phi_\mathrm{est}(x)\right]\mathcal{P}(\phi)P(x|\phi)\right]^{+\infty}_{-\infty} - \int \dd\phi\, \mathcal{P}(\phi) \sum_x P(x|\phi)\\
        &=-1.
    \end{align}
    Finally, application of the Cauchy-Schwarz inequality yields
    \begin{align}
        \left(\int \dd\phi\, \sum_x f(\phi, x)g(\phi, x)\right)^2 &\leq \left(\int \dd\phi\, \sum_x f^2(\phi, x)\right)\left(\int \dd\phi\, \sum_x g^2(\phi, x)\right)
    \end{align}
    which, with the definitions from above, is equivalent to $1\leq (\Delta\phi)^2\left[\overline{\mathcal{F}}+ \mathcal{I}\right] $ and ultimately results in the van Trees inequality Eq.~\eqref{eq:BCRB}.

    \subsection{Bayesian Quantum Cram\'{e}r-Rao Bound (BQCRB)}\label{app:BQCRB}
    To derive the Bayesian quantum Cram\'{e}r-Rao bound (BQCRB), we follow Ref.~\cite{Macieszczak2014}. We start by rewriting the BMSE as
    \begin{align}
        (\Delta\phi)^2 &= \int  \dd\phi\, \mathcal{P}(\phi) \sum_x \tr\left(\Lambda_{\phi,T}[\rho_\mathrm{in}]\Pi_x \right) \left[\phi - \phi_\mathrm{est}(x)\right]^2\\
        &= (\delta\phi)^2 +\tr\left(\int \dd\phi\, \mathcal{P}(\phi)\Lambda_{\phi,T}[\rho_\mathrm{in}]\sum_x \Pi_x \phi_\mathrm{est}^2(x)\right) - 2\tr\left(\int \dd\phi\, \mathcal{P}(\phi)\phi\Lambda_{\phi,T}[\rho_\mathrm{in}]\sum_x \Pi_x \phi_\mathrm{est}(x)\right)\\
        &= (\delta\phi)^2 + \tr(\overline{\rho}L_2)-2\tr(\overline{\rho}'L_1)\label{eq:pre_BQCRB}
    \end{align}
    where $(\delta\phi)^2$ represents the variance of the prior distribution, $\overline{\rho} = \int \dd\phi\, \mathcal{P}(\phi)\Lambda_{\phi,T}[\rho_\mathrm{in}]$ denotes the average state and $\overline{\rho}' = \int \dd\phi\, \mathcal{P}(\phi)\Lambda_{\phi,T}[\rho_\mathrm{in}] \phi$. Furthermore, we have combined the measurement $\{\Pi_x\}$  and estimator $\phi_\mathrm{est}$ by defining the operators $L_1 = \sum_x \Pi_x \phi_\mathrm{est}(x)$ and $L_2 = \sum_x \Pi_x \phi_\mathrm{est}^2(x)$.

    In a first step, following Refs.~\cite{Macieszczak2014,Helstrom1969,Personick1971}, we will demonstrate that, without loss of optimality, the measurement can be restricted to the class of projection-valued measures (PVM), i.e. projective von Neumann measurements $\Pi_x = \ket{x}\bra{x}$, with orthonormal eigenstates $\ket{x}$, $\langle x|x'\rangle=\delta_{x,x'}$, of the observable $X$ with eigenvalue $x$. We denote the projective strategy by $L_{1,2}^\mathrm{PVM}$, where $L_{1}^\mathrm{PVM} = L_1 = \sum_x \phi_\mathrm{est}(x) \ket{x}\bra{x}$ effectively corresponds to the eigendecomposition. Based on Eq.~\eqref{eq:pre_BQCRB}, we have to show that $\tr(\overline{\rho}L_2^\mathrm{PVM})\leq \tr(\overline{\rho}L_2)$ to prove that we do not lose any optimality by restricting to the projective strategy. Using that $L_1$ is hermitian and $\Pi_x\geq 0 $, we can consider the inequality
    \begin{align}
        \sum_x (\phi_\mathrm{est}(x) - L_1)\Pi_x(\phi_\mathrm{est}(x) - L_1) &\geq 0\\
        \sum_x \phi_\mathrm{est}^2(x)\Pi_x -\sum_x \phi_\mathrm{est}(x) \Pi_x L_1 - L_1\sum_x  \phi_\mathrm{est}(x) \Pi_x  +  L_1 \sum_x\Pi_x L_1&\geq 0\label{eq:PVM_proof1_dummy}\\
        L_2 - L_1^2 &\geq 0\\
        L_2 &\geq L_1^2,\label{eq:PVM_proof1}
    \end{align}
    where we have identified $L_1$ and $L_2$ in Eq.~\eqref{eq:PVM_proof1_dummy} and used the completeness relation $\sum_x\Pi_x = \id$. However, equality in Eq.\eqref{eq:PVM_proof1} is achieved specifically for the projective strategy, since $L_{2}^\mathrm{PVM} = \sum_x \phi_\mathrm{est}^2(x) \ket{x}\bra{x} = (L_{1}^\mathrm{PVM})^2$. Therefore, $\tr(\overline{\rho}L_2^\mathrm{PVM})\leq \tr(\overline{\rho}L_2)$ and it is always optimal to choose the measurement to be projective.

    In a second step, we derive the BQCRB Eq.~\eqref{eq:BQCRB}. Choosing the projective strategy discussed above and accordingly labeling $L=L_1$ and thus $L_2=L^2$, the BMSE reads
    \begin{align}
                (\Delta\phi)^2 = (\delta\phi)^2 + \tr(\overline{\rho}L^2)-2\tr(\overline{\rho}'L).\label{eq:L_opt}
    \end{align}
    Hence, the task of finding the optimal measurement and estimation maps to the optimization of the operator $L$, containing both. Variation of $L$ according to $L \mapsto L + \epsilon \delta L$ with infinitesimal parameter $\epsilon$ and hermitian $\delta L$ yields
    \begin{align}
        (\Delta\phi)^2 &= (\delta\phi)^2 + \tr(\overline{\rho}[L^2 + \epsilon L \delta L + \epsilon \delta L L + \epsilon^2\delta L^2])-2\tr(\overline{\rho}'[L + \epsilon \delta L]).
    \end{align}
    Differentiation with respect to $\epsilon$ and evaluation at $\epsilon=0$ results in
    \begin{align}
        0 =  \tr([ \overline{\rho}L  +   L\overline{\rho} -2\overline{\rho}' ]\delta L).\label{eq:BQCRB_dummy1}
    \end{align}
    Since Eq.~\eqref{eq:BQCRB_dummy1} must hold for any $\delta L$, it implies
    \begin{align}
        \overline{\rho}'&= \frac{1}{2}\left(L\overline{\rho} + \overline{\rho} L \right),\label{eq:Bayesian_SLDdef}
    \end{align}
    reproducing Eq.~\eqref{eq:L_equation}. Substituting this expression for $\overline{\rho}'$ in Eq. \eqref{eq:L_opt}, we find
    \begin{align}
        (\Delta\phi)^2 &= (\delta\phi)^2 + \tr(\overline{\rho}L^2)-2\tr(\overline{\rho}'L)\\
        &= (\delta\phi)^2 + \tr(\overline{\rho}L^2)-\tr(\overline{\rho}L^2) - \tr(L\overline{\rho}L)\\
        &= (\delta\phi)^2 - \tr(\overline{\rho}L^2),\label{eq:BQCRB_app}
    \end{align}
    which corresponds to the BQCRB Eq.~\eqref{eq:BQCRB}.

    Finally, we will express the BQCRB in terms of the quantum Fisher information (QFI) of the average state $\overline{\rho}$. Assuming a unitary phase evolution according to Eq.~\eqref{eq:unitaryEvolution}, corresponding to the von Neumann equation
    \begin{align}
        \partial_\phi \Lambda_{\phi,T}[\rho_\mathrm{in}] = -i[S_z, \Lambda_{\phi,T}[\rho_\mathrm{in}]],\label{eq:vonNeumann}
    \end{align}
    and a Gaussian prior distribution as defined in Eq.~\eqref{eq:prior}, we can rewrite $\overline{\rho}'$ as
    \begin{align}
        \overline{\rho}' &= \int  \dd\phi\, \mathcal{P}(\phi) \phi \Lambda_{\phi,T}[\rho_\mathrm{in}]\\
        &= -(\delta\phi)^2 \int  \dd\phi\, (\partial_\phi \mathcal{P}(\phi))  \Lambda_{\phi,T}[\rho_\mathrm{in}]\\
        &= -(\delta\phi)^2 \left[\mathcal{P}(\phi) \Lambda_{\phi,T}[\rho_\mathrm{in}]\right]_{-\infty}^{+\infty}  + (\delta\phi)^2\int  \dd\phi\, \mathcal{P}(\phi)\partial_\phi \Lambda_{\phi,T}[\rho_\mathrm{in}] \label{eq:rhoprime_Gaussian_dummy}\\
        &= -i(\delta\phi)^2\left[S_z,\int  \dd\phi\, \mathcal{P}(\phi)\Lambda_{\phi,T}[\rho_\mathrm{in}]\right]\\
        &= -i(\delta\phi)^2\left[S_z,\overline{\rho}\right]\label{eq:rhoprime_Gaussian}
    \end{align}
    where we exploited the property $\partial_\phi \mathcal{P}(\phi)= - (\delta\phi)^{-2}\phi \mathcal{P}(\phi)$ of a Gaussian prior distribution. Furthermore, we used partial integration in the second step and Eq.~\eqref{eq:boundary} as well as Eq.~\eqref{eq:vonNeumann} in Eq.~\eqref{eq:rhoprime_Gaussian_dummy}. With Eq.~\eqref{eq:Bayesian_SLDdef} and Eq.~\eqref{eq:rhoprime_Gaussian}, we obtain
    \begin{align}
         \frac{1}{2}\left(L\overline{\rho} + \overline{\rho} L \right) &=-i(\delta\phi)^2 \left[S_z,\overline{\rho}\right].\label{eq:SLD_L}
    \end{align}
    Substituting $L:= (\delta\phi)^2 L_\mathrm{local}$, the BQCRB Eq.~\eqref{eq:BQCRB_app} and the implicit equation Eq.~\eqref{eq:SLD_L} become
    \begin{align}
        (\Delta\phi)^2 &= (\delta\phi)^2 \left[1- (\delta\phi)^2 \tr(\overline{\rho}L_\mathrm{local}^2)\right]\\
        \frac{1}{2}\left(L_\mathrm{local}\overline{\rho} + \overline{\rho} L_\mathrm{local} \right) &=-i \left[S_z,\overline{\rho}\right].
    \end{align}
    Comparison to the QFI approach in local phase estimation shows that $L_\mathrm{local}$ defines the symmetric logarithmic derivative (SLD)~\cite{Helstrom1969,Holevo2011,Paris2009} and thus, $\tr(\overline{\rho}L_\mathrm{local}^2)=\mathcal{F}_Q[\overline{\rho}]$ corresponds to the quantum Fisher information of the average state $\overline{\rho}$, resulting in Eq.~\eqref{eq:BQCRB_QFI}
    \begin{align}
            (\Delta\phi_\mathrm{BQCRB})^2 = (\delta\phi)^2\left[1 - (\delta\phi)^2\mathcal{F}_Q[\overline{\rho}]\right].
    \end{align}

    \subsection{Optimal Quantum Interferometer (OQI)}\label{app:OQI}
    The optimal quantum interferometer (OQI) represents the ultimate lower bound of the BMSE. However, no general expressions for arbitrary ensemble sizes exist, but rather complex optimization procedures are required. 
    
    \paragraph*{Iterative algorithm---} In the following, we outline an algorithm introduced in Refs.~\cite{Macieszczak2013,Macieszczak2014}, which iteratively optimizes the initial probe state $\rho_\mathrm{in}$ and measurement $\{\Pi_x\}$. Although numerical optimization becomes challenging with increasing ensemble size, this algorithm enables efficient computation at least for small $N$. For a given input probe state $\rho_\mathrm{in}$, the optimal projective measurement and estimation strategy $L$ can be determined according to the previous discussion on the BQCRB. Conversely, for a given $L$, the optimal $\rho_\mathrm{in}$ can be evaluated as follows. Rewriting Eq.~\eqref{eq:pre_BQCRB}, the BMSE can be expressed as
    \begin{align}
        (\Delta\phi)^2 &= (\delta\phi)^2 +\tr\left(\int \dd\phi\, \mathcal{P}(\phi)\Lambda_{\phi,T}[\rho_\mathrm{in}](L^2 -2\phi L)\right).
    \end{align}
    Defining the adjoint quantum channel $\Lambda_{\phi,T}^\dag$ via $\tr(\Lambda_{\phi,T}[\rho]A) = \tr(\rho\Lambda_{\phi,T}^\dag[A])$ for an arbitrary operator $A$, the BMSE becomes
    \begin{align}
        (\Delta\phi)^2 &= (\delta\phi)^2 +\tr\left(\rho_\mathrm{in}\int \dd\phi\, \mathcal{P}(\phi)\Lambda_{\phi,T}^\dag[L^2 -2\phi L]\right).\label{eq:OQI_operator}
    \end{align}
    Consequently, the optimal input probe state $\rho_\mathrm{in}=\ket{\psi_\mathrm{in}}\bra{\psi_\mathrm{in}}$ corresponds to the eigenvector $\ket{\psi_\mathrm{in}}$ of the operator $\int \dd\phi\, \mathcal{P}(\phi)\Lambda_{\phi,T}^\dag[L^2 -2\phi L]$ associated with its most negative eigenvalue. In the iterative algorithm, starting from an arbitrary state, repeatedly the optimal measurement and the corresponding optimal probe state are determined, until the BMSE converges to the OQI.

    \paragraph*{Coherence time limit (CTL)---} In the following, we derive Eq.~\eqref{eq:CTL_OQI}. Considering a $2\pi$-periodic quantum channel with respect to the phase $\phi$ as imposed by Eq.~\eqref{eq:unitaryEvolution}, the OQI allows for unambiguous phase estimation within the range $[-\pi,+\pi]$. Exceeding this invertible regime, an estimation error is accumulated which increases with the distance from the primary Ramsey fringe. In particular, an estimation error of $\epsilon_k = (2\pi k)^2$ is accumulated if the phase slips in the region $[-(2k+1)\pi , -(2k-1)\pi]$ or $[+(2k-1)\pi , +(2k+1)\pi]$ for $k\in\mathbb{N}$. The estimation error associated with these events can be modeled by 
    \begin{align}
        (\Delta\phi_{\mathrm{CTL}}^{\mathrm{OQI}})^2 =& \sum_{k=1}^\infty \epsilon_k P_k\label{eq:CTL_General}
    \end{align}
    which effectively represents the average of the estimation error $\epsilon_k$ weighted with its corresponding probability
    \begin{align}
        P_k = \int_{-(2k+1)\pi}^{-(2k-1)\pi} \dd\phi\,  \mathcal{P}(\phi) + \int_{(2k-1)\pi}^{(2k+1)\pi} \dd\phi\,  \mathcal{P}(\phi).
    \end{align}
    Consequently, Eq.~\eqref{eq:CTL_General} constitutes an asymptotic limit for broad prior distributions. In the context of atomic clocks, this regime corresponds to long interrogation times, where the coherence time of the local oscillator will become relevant and ultimately limits the clock stability. Therefore, we will denote Eq.~\eqref{eq:CTL_OQI} as the coherence time limit (CTL) of the OQI. Assuming a Gaussian prior distribution as defined in Eq.~\eqref{eq:prior}, the probabilities $P_k$ can be evaluated explicitly to read
    \begin{align}
        P_k &= 2 \int_{(2k-1)\pi}^{(2k+1)\pi} \dd\phi\,  \mathcal{P}(\phi)= 2 \int_{0}^{(2k+1)\pi} \dd\phi\,  \mathcal{P}(\phi) - 2 \int_0^{(2k-1)\pi} \dd\phi\,  \mathcal{P}(\phi)= \mathrm{erf}\left(\frac{(2k+1)\pi}{\sqrt{2}\delta\phi}\right)- \mathrm{erf}\left(\frac{(2k-1)\pi}{\sqrt{2}\delta\phi}\right)
    \end{align}
    where we substituted $t = \frac{\phi}{\sqrt{2}\delta\phi}$ and used the error function $\mathrm{erf}(z) = \int_0^z\dd t  e^{-t^2}$. In the relevant regime of prior widths considered in this work, where typically only the adjacent fringes around $\phi=0$ contribute, the prior distribution $\mathcal{P}(\phi)$ is effectively limited to the region $[-3\pi,+3\pi]$. As a result, the CTL simplifies significantly compared to the general form in Eq.~\eqref{eq:CTL_General}, which accounts for contributions from all Ramsey fringes. In this restricted regime, the CTL reduces to
    \begin{align}
        (\Delta\phi_{\mathrm{CTL}}^{\mathrm{OQI}})^2 &= 4\pi^2 \left[\int_{-\infty}^{-\pi} \dd\phi\,  \mathcal{P}(\phi) + \int^{\infty}_{\pi} \dd\phi\,  \mathcal{P}(\phi)\right]= 4\pi^2 \left[1-\int_{-\pi}^\pi \dd\phi\,  \mathcal{P}(\phi)\right]= 4\pi^2 \left[1-\mathrm{erf}\left(\frac{\pi}{\sqrt{2}\delta\phi}\right)\right].
    \end{align}

    \paragraph*{Asymptotic scaling ($\pi$HL)---} With increasing ensemble size, the numerical algorithm presented above becomes computationally challenging. However, in the asymptotic limit ($N \gg 1$), an explicit analytical expression for the OQI can be derived. Assuming unitary phase evolution as described by Eq.~\eqref{eq:unitaryEvolution} and restricting to the invertible range $[-\pi,+\pi]$, it has been shown for arbitrary prior distributions~\cite{Gorecki2020,Jarzyna2015,Buzek1999,Berry2000} that the ultimate lower bound is given by $\pi$-corrected Heisenberg limit ($\pi$HL), as defined in Eq.~\eqref{eq:piHL}. An intuitive derivation for Gaussian prior distributions is given in Ref.~\cite{Jarzyna2015} and is reproduced here. Based on Eq.~\eqref{eq:BQCRB_QFI}, the optimization of the BQCRB over all input probe states $\rho_\mathrm{in}$ is equivalent to evaluating the QFI for the averaged state $\overline{\rho}$. This averaging can be formally associated with a collective dephasing process, where the dephasing rate is identified with the variance of the prior distribution~\cite{Jarzyna2015,Macieszczak2014}. Combining this perspective with the asymptotic result for collective dephasing in Ref.~\cite{Knysh2014}, the asymptotic OQI can be expressed as
    \begin{align}
        (\Delta\phi_\mathrm{OQI})^2 \overset{N\gg 1}{\simeq}  (\delta\phi)^2\left[1-\frac{1}{1 + \frac{\pi^2}{N^2(\delta\phi)^2}}\right] \overset{N\gg 1}{\simeq}(\delta\phi)^2\left[1-\left(1-\frac{\pi^2}{N^2(\delta\phi)^2}\right)\right] = \frac{\pi^2}{N^2}
    \end{align}
    where we used the expansion $\frac{1}{1+x}\overset{x\ll 1}{\simeq} 1 - x$. This result is valid for Gaussian prior distributions with widths $\delta\phi \ll N$, which encompasses all relevant widths in the asymptotic regime $N\gg 1$. It is  therefore reasonable to expect that this result generalizes to arbitrary prior distributions, as the fundamental characteristics of the estimation problem in this regime remain largely unaffected by the specific shape of the prior~\cite{Jarzyna2015,Gorecki2020}.

    \paragraph*{Phase operator based interferometer (POI)---} Finally, we aim to identify the protocol that saturates the asymptotic limit of the OQI. As discussed above, simultaneously determining the optimal measurement, input state, and estimation strategy is a highly non-trivial problem. However, assuming a flat prior distribution and a periodic cost function in the interval $[-\pi,+\pi]$, the concept of covariant measurements~\cite{Holevo2011,Buzek1999} provides an explicit solution for the optimal measurement operator, the so-called phase operator~\cite{Summy1990,Berry2000,Buzek1999,Pegg1988,Sanders1995,Derka1998,Luis1996,Kaubruegger2021,Demkowicz_Dobrza_ski_2015}. The phase operator $\Phi$ is defined as
    \begin{align}
        \Phi &= \sum_{s=-N/2}^{N/2}\phi_s\ket{s}\bra{s}\nonumber\\
        \phi_s &= \frac{2\pi s}{N+1}\label{eq:phaseOperator}\\
        \ket{s} &= \frac{1}{\sqrt{N+1}}\sum_{M=-N/2}^{N/2}e^{-i\phi_s M}\ket{M}\nonumber
    \end{align}
    where $\phi_s$ are the eigenvalues with corresponding eigenstates $\ket{s}$, constructed from the eigenstates $\ket{M}$ of $S_z$ with eigenvalue $M$ and total spin $N/2$. An interferometer based on $\Phi$ is referred to as phase operator based interferometer (POI). Furthermore, under these assumptions, the optimal input states in the asymptotic regime ($N\gg 1$), known as \textit{sine} states \cite{Summy1990,Berry2000,Buzek1999,Pegg1988,Sanders1995,Derka1998,Luis1996,Kaubruegger2021,Demkowicz_Dobrza_ski_2015} and saturating the $\pi$HL, can also be explicitly determined
    \begin{align}
        \ket{\psi_\Phi} &= \sqrt{\frac{2}{N+1}}\sum_{M=-N/2}^{N/2} \sin\left(\frac{\pi(M+1/2)}{N+1}\right)\ket{M}.
    \end{align}
    However, since the assumptions of a periodic cost function and flat prior distribution are contrary to the framework introduced in Sec.~\ref{sec:BayesianPhaseEstimation}, namely a global BMSE with phases $-\infty < \phi < +\infty$ and Gaussian prior distributions, these states are not necessarily optimal in the approach pursued in this work. Therefore, the optimal initial states and measurements must be explicitly evaluated. Nevertheless, it is instructive to investigate the performance of the POI and compare to the standard protocols as well as variational classes discussed in the main text. Notably, this scenario is contrary to the BQCRB, since the measurement is fixed by $\Phi$, while we aim to optimize over all initial states. For a fixed prior width, the optimal state $\ket{\psi_\Phi}$ can be identified by adapting the iterative algorithm presented above and building on methods from Ref.~\cite{Kaubruegger2021}: Starting with an arbitrary initial state $\ket{\psi_\mathrm{in}^{(0)}}$, such as $\ket{\psi_\mathrm{in}^{(0)}} = \ket{s=0}$, the optimal Bayesian estimator (cf. Sec.~\ref{sec:Estimators}) $\phi_{est}^{(0)}(s)$ is computed. Based on $\phi_{est}^{(0)}(s)$, the subsequent input probe state $\ket{\psi_\mathrm{in}^{(1)}}$ in the iterative algorithm is evaluated by selecting the eigenstate corresponding to the most negative eigenvalue of the operator $\int \dd\phi\, \mathcal{P}(\phi)\Lambda_{\phi,T}^\dag[L^2 -2\phi L]$ defined in Eq.~\eqref{eq:OQI_operator}. This ensures that the state $\ket{\psi_\mathrm{in}^{(1)}}$ is optimal for a given measurement and estimator. This process is repeated until convergence to the optimal state $\ket{\psi_\Phi}$, tailored to the framework considered in this work, is achieved. Numerical evaluation of this iterative algorithm shows that the POI saturates the OQI in the limit of large ensembles within the framework of this work, as discussed in the main text and depicted in Fig.~\ref{fig:standard_protocols}(b).

\section{Estimators}\label{app:estimatros}
    In the following, we derive explicit expressions for the linear and optimal Bayesian estimators, as presented in Sec.~\ref{sec:Estimators}.

    \subsection{Linear estimator}\label{app:linear_estimator}
    With the linear estimator defined in Eq.~\eqref{eq:linear_est} as $\phi_\mathrm{est}^\mathrm{linear}(x) = a\cdot x$, the BMSE is expressed as
    \begin{align}
        (\Delta\phi)^2 &= (\delta\phi)^2 - 2a\int  \dd\phi\, \mathcal{P}(\phi) \phi\sum_x xP(x|\phi)  + a^2\int  \dd\phi\, \mathcal{P}(\phi) \sum_x x^2 P(x|\phi)\\
        &= (\delta\phi)^2 - 2a\int  \dd\phi\, \mathcal{P}(\phi) \phi\braket{X(\phi)}  + a^2\int  \dd\phi\, \mathcal{P}(\phi)\braket{X^2(\phi)}.\label{eq:linear_deriv1}
    \end{align}
    Here, the moments of the observable $X$ are defined by $\braket{X^n(\phi)}=\sum_x x^n P(x|\phi)$. The optimal scaling factor $a$ is determined by minimizing the BMSE. Differentiating Eq.~\eqref{eq:linear_deriv1} and solving for $a$ yields
    \begin{align}
        a&= \frac{\int  \dd\phi\, \mathcal{P}(\phi) \phi\braket{X(\phi)}}{\int  \dd\phi\, \mathcal{P}(\phi)\braket{X^2(\phi)}}.
    \end{align}
    Hence, from Eq.~\eqref{eq:linear_deriv1}, the corresponding BMSE is given by Eq.~\eqref{eq:BMSE_linear}, i.e.
    \begin{align}
        (\Delta\phi)^2 &= (\delta\phi)^2 - \frac{\left[\int  \dd\phi\, \mathcal{P}(\phi) \phi\braket{X(\phi)}\right]^2}{\int  \dd\phi\, \mathcal{P}(\phi)\braket{X^2(\phi)}}.
    \end{align}
    Due to the linearity of the estimator, the scaling factor and BMSE only depend on the first and second moments of the observable $X$, rather than the full statistical model $P(x|\phi)$. This dependence significantly simplifies practical computations, while retaining reliable performance in several situations. Nevertheless, the linear estimation strategy is not optimal in general.

    \subsection{Optimal Bayesian estimator}\label{app:mmse_estimator}
    To start with, we expand Eq.~\eqref{eq:BMSE2}
    \begin{align}
        (\Delta\phi)^2 &= \sum_x P(x) \left[\int  \dd\phi\,  P(\phi|x)\phi^2 -2\phi_\mathrm{est}(x)\int  \dd\phi\, P(\phi|x)\phi + \phi_\mathrm{est}^2(x)\int  \dd\phi\,  P(\phi|x)\right].
    \end{align}
    As before, the first term results in the prior variance $(\delta\phi)^2$, while the last integral simplifies to unity due to the normalization of the posterior distribution. To minimize the BMSE, the optimal Bayesian estimator has to minimize the term in the brackets for each measurement outcome $x$, since $P(x)\geq 0$ and $\phi_\mathrm{est}(x)$ is independent for different  $x$. Differentiation and solving for the estimator yields the optimal Bayesian estimator given in Eq.~\eqref{eq:MMSE_est}, $\phi_\mathrm{est}^\mathrm{opt}(x) = \int  \dd\phi\,  P(\phi|x)\phi$~\cite{Demkowicz_Dobrza_ski_2015}. Thus, the optimal Bayesian estimator corresponds to the mean posterior phase. With this result, the BMSE becomes
    \begin{align}
        (\Delta\phi)^2 &=(\delta\phi)^2 - \sum_x P(x)\left(\phi_\mathrm{est}^\mathrm{opt}(x)\right)^2.
    \end{align}
    Equivalently, the BMSE can be expressed in terms of the statistical model $P(x|\phi)$ and prior distribution $\mathcal{P}(\phi)$ according to Bayes theorem Eq.~\eqref{eq:BayesTheorem}, resulting in Eq.~\eqref{eq:BMSE_MMSE}. Unlike the linear estimator, the optimal Bayesian estimator as well as the corresponding BMSE depend explicitly on the statistical model, rather than just the first and second moments of the observable. While this dependence ensures optimality, it also increases computational complexity.

\section{Allan deviation}\label{app:ADEV}
    In the following, we characterize the frequency fluctuations $\omega(t)=\omega_0 - \omega_\mathrm{LO}(t)$, arising from deviations of the atomic transition frequency $\omega_0$ and local oscillator (LO) frequency $\omega_\mathrm{LO}(t)$. To facilitate comparisons between LO at different $\omega_0$, it is convenient to introduce the dimensionless fractional frequency deviation
    \begin{align}
        y(t) = \frac{\omega(t)}{\omega_0} = \frac{\omega_0 - \omega_\mathrm{LO}(t)}{\omega_0}.\label{eq:fractionalFrequency}
    \end{align}
    In general, the LO produces a continuous noisy frequency trace $y(t)$. However, in many applications, including the operation of an atomic clock, only a sequence of discrete frequency measurements averaged over individual clock cycles of duration $T_C$ is recorded. We assume that each clock cycle consists of dead time $T_D$, during which the atomic reference is not interrogated, followed by an interrogation time $T$. Consequently, the frequency trace is divided into equal intervals of duration $T_C=T_D+T$. The fractional frequency value recorded at the end of cycle $k$ is obtained by
    \begin{align}
        y_k = \frac{1}{T_C}\int_{(k-1)T_C}^{kT_C}\dd t\,y(t) = \frac{1}{T_C}\left[\int_{(k-1)T_C}^{(k-1)T_C+T_D}\dd t\,y(t) + \int_{(k-1)T_C+T_D}^{kT_C}\dd t\,y(t)\right].
    \end{align}
    A common method to characterizing statistical processes involves calculating the mean value $\overline{y}$ and the variance $s^2$, defined by
    \begin{align}
        \overline{y} &= \frac{1}{n}\sum_{k=1}^n y_k\\
        s^2 &= \frac{1}{n-1}\sum_{k=1}^n (y_k-\overline{y})^2,
    \end{align}
    where $n$ denotes the total number of fractional frequency values $y_k$. However, the standard variance is only a meaningful measure for uncorrelated noise. If the noise is correlated, the deviation from its mean value is no longer stationary~\cite{Handbook} and thus, the standard variance might be non-convergent. Consequently, it is not recommended to characterize frequency standards or atomic clocks using the standard variance.

    To determine the clock stability, we consider the fractional frequency fluctuations averaged over a total measurement time $\tau=mT_C$, corresponding to $m$ individual clock cycles,
    \begin{align}
        \overline{y}_j = \frac{1}{\tau}\int_{(j-1)\tau}^{j\tau}\dd t\,y(t)= \frac{1}{\tau}\sum_{k=(j-1)m+1}^{jm} \int_{(k-1)T_C}^{kT_C}\dd t\,y(t) = \frac{1}{m}\sum_{k=(j-1)m+1}^{jm} y_k.
    \end{align}
    The most widely used time-domain metric for evaluating the stability of frequency standards and atomic clocks~\cite{Barnes1971,Handbook,Rutman1991,Riehle2003} is represented by the Allan deviation (ADEV)~\cite{Allan1966,Allan1987}. It is defined as the square root of the Allan variance~\cite{Allan1966,Riehle2003}
    \begin{align}
        \sigma_y^2(\tau) &= \frac{1}{2}\langle{(\overline{y}_{j+1}-\overline{y}_{j})^2}\rangle,\label{eq:formalADEV}
    \end{align}
    where $\langle\cdot\rangle$ denotes statistical averaging. While the standard deviation measures the spread of uncorrelated fluctuations around a mean value, the Allan deviation focuses on differences between consecutive measurements, making it sensitive to time-correlated noise. This distinction allows the Allan deviation to reveal trends and patterns in the temporal behavior of frequency fluctuations, which the standard deviation would average out or fail to capture. Unlike the standard variance, the Allan variance converges for most types of noise encountered in frequency standards. As it quantifies the fractional frequency fluctuations, a lower value indicates reduced instability, or equivalently, improved stability. The Allan deviation $\sigma_y(\tau)$ depends on the averaging time $\tau$, providing insights into noise characteristics on different timescales. Short averaging times $\tau \sim T_C$ reveal short-term stability, while large $\tau \gg T_C$ describe long-term stability. Therefore, analyzing the full $\tau$-dependence of $\sigma_y(\tau)$ is essential to assess the performance of different oscillators.

    In practice, for finite data sets, the statistical averaging is realized as~\cite{Handbook}
    \begin{align}
        \sigma_y^2(\tau) &= \frac{1}{2(M-1)}\sum_{j=1}^{M-1}(\overline{y}_{j+1}-\overline{y}_{j})^2
    \end{align}
    where $M = \frac{m}{n}$ represents the number of consecutive frequency intervals with length $\tau=mT_C$. The quantity usually addressed is the square root of the Allan variance, namely the Allan deviation (ADEV).

    To summarize, the Allan deviation is an essential metric for assessing the stability of atomic clocks and frequency standards. It provides a measure of fractional frequency fluctuations over various timescales, distinguishing between short-term and long-term stability. Unlike the standard variance, which is only meaningful for uncorrelated noise, the Allan deviation is well-suited to analyze correlated noise, capturing trends and temporal patterns in the fluctuations. This makes it an indispensable tool for comparing and optimizing the performance of different oscillators.

\section{Sensitivity of standard protocols}\label{app:standard_protocols}
    In the following, we derive the sensitivities of the CSS, SSS and GHZ protocols introduced in the main text.

    \subsection{Coherent spin states (CSS)}\label{app:CSS}
    For a measurement of the collective spin operator $S_y$ and unitary phase evolution through a rotation around the $z$-axis, according to Eq.~\eqref{eq:unitaryEvolution}, the first and second moments of the observable are given by
    \begin{align}
        \braket{X(\phi)} &= \braket{S_y(\phi)} = \braket{S_y}\cos(\phi) + \braket{S_x}\sin(\phi)\\
        \braket{X^2(\phi)} &= \braket{S_y^2(\phi)} = \braket{S_y^2}\cos^2(\phi) + \braket{S_y S_x + S_x S_y}\sin(\phi)\cos(\phi)+\braket{S_x^2}\sin^2(\phi),
    \end{align}
    where the expectation values $\braket{\cdot}$ are evaluated with respect to the initial state $\ket{\psi_\mathrm{in}}$, independent of the phase $\phi$. Assuming a Gaussian prior distribution, as defined in Eq.~\eqref{eq:prior}, the integrals in Eq.~\eqref{eq:BMSE_linear} become 
    \begin{align}
         \int  \dd\phi\,  \mathcal{P}(\phi) \phi\braket{X(\phi)} &= \int  \dd\phi\,  \mathcal{P}(\phi) \phi \left[\braket{S_y}\cos(\phi) + \braket{S_x}\sin(\phi) \right] = \braket{S_x}(\delta\phi)^2e^{-(\delta\phi)^2/2}\\
         \int  \dd\phi\,  \mathcal{P}(\phi)\braket{X^2(\phi)} &=\int  \dd\phi\,  \mathcal{P}(\phi) \left[\braket{S_y^2}\cos^2(\phi) + \braket{S_y S_x + S_x S_y}\sin(\phi)\cos(\phi)+\braket{S_x^2}\sin^2(\phi)\right]\\
         &=e^{-(\delta\phi)^2}\left[\braket{S_y^2}\cosh\left((\delta\phi)^2\right) + \braket{S_x^2} \sinh\left((\delta\phi)^2\right)\right],
     \end{align}
     where terms with odd integrands vanish directly. Thus, the optimal linear scaling factor, corresponding BMSE and effective measurement uncertainty are given by
     \begin{align}
         a &= \frac{\braket{S_x}(\delta\phi)^2e^{(\delta\phi)^2/2}}{\braket{S_y^2}\cosh\left((\delta\phi)^2\right) + \braket{S_x^2} \sinh\left((\delta\phi)^2\right)}\nonumber\\
         (\Delta\phi)^2 &= (\delta\phi)^2\left[1-(\delta\phi)^2\frac{\braket{S_x}^2}{\braket{S_y^2}\cosh((\delta\phi)^2)+ \braket{S_x^2}\sinh((\delta\phi)^2)} \right]\label{eq:BMSE_Sy}\\
         (\Delta\phi_M)^2 &=  \frac{\braket{S_y^2}}{\braket{S_x}^2}\cosh((\delta\phi)^2) + \frac{\braket{S_x^2}}{\braket{S_x}^2}\sinh((\delta\phi)^2)-(\delta\phi)^2.\nonumber
     \end{align}
    
    For the conventional Ramsey protocol, a coherent spin state (CSS) polarized in $x$-direction is prepared by a $\pi/2$-pulse applied to the collective ground state
    \begin{align}
        \ket{\mathrm{CSS}}=\mathcal{R}_y\left(-\tfrac{\pi}{2}\right) \kett{\downarrow}^{\otimes N} = \ket{+}^{\otimes N} = \left[\frac{1}{\sqrt{2}}(\kett{\downarrow} + \kett{\uparrow})\right]^{\otimes N},\label{eq:CSS_state}
    \end{align}
    which represents $N$ uncorrelated atoms, each in an equal superposition of the ground and excited state. CSS and their properties are discussed in detail in Refs.~\cite{Arecchi1972,Radcliffe1971,Hioe1974}. With expectation values
    \begin{align}
        &\braket{S_x} = \frac{N}{2}, \qquad  \braket{S_y} = \braket{S_z} = 0, \qquad\braket{S_x^2} = \frac{N^2}{4}, \qquad\braket{S_y^2}  = \frac{N}{4}= \braket{S_z^2}, \qquad\braket{S_x S_y} = 0 = \braket{S_x S_z},
    \end{align}
    we derive
     \begin{align}
        a_\mathrm{CSS}&= \frac{2e^{(\delta\phi)^2/2}}{\cosh((\delta\phi)^2)+ N\sinh((\delta\phi)^2)}\\
        (\Delta\phi_\mathrm{CSS})^2 &= (\delta\phi)^2\left[1-(\delta\phi)^2\frac{N}{\cosh((\delta\phi)^2)+ N\sinh((\delta\phi)^2)} \right]\\
    (\Delta\phi_M^\mathrm{CSS})^2 &=\frac{\cosh((\delta\phi)^2)}{N} + \sinh((\delta\phi)^2)-(\delta\phi)^2.
    \end{align}
    Rewriting the first term, we recover the result from Ref.~\cite{Leroux2017}
    \begin{align}
        (\Delta\phi_M^\mathrm{CSS})^2 = \frac{e^{(\delta\phi)^2} }{N} + \left(1-\frac{1}{N}\right)\sinh((\delta\phi)^2)-(\delta\phi)^2.
    \end{align}

    \subsection{Squeezed spin states (SSS)}\label{app:SSS}
    The application of an one-axis-twisting (OAT) interaction $\mathcal{T}_z(\mu) = \exp\left(-i\tfrac{\mu}{2}S_z^2\right)$ with small twisting strength $\mu$ to the CSS, defined in Eq.~\eqref{eq:CSS_state}, generates a squeezed spin state (SSS). To align the minimal spin variance along the $y$-axis, an additional rotation $\mathcal{R}_x(\theta)$ around the $x$-axis by an angle $\theta$ is applied. Thus, the initial state reads
    \begin{align}
        \ket{\mathrm{SSS}}= \mathcal{R}_x(\theta)\mathcal{T}_z(\mu)\ket{\mathrm{CSS}}.\label{eq:SSS_state}
    \end{align}
    These states are introduced and discussed in detail in Ref.~\cite{Kitagawa1993}. In comparison to CSS, the SSS differs primarily in its polarization and spin variances, while other properties remain unchanged. Hence, the optimal linear scaling factor, BMSE and effective measurement uncertainty are given by Eq.~\eqref{eq:BMSE_Sy} with expectation values
    \begin{align}
        \braket{S_x} =& \frac{N}{2} \cos^{N-1}\left(\tfrac{\mu}{2}\right)\\
        \braket{S_x^2} =& \frac{N}{4}\left\{N\left[1-\cos^{2N-2}\left(\tfrac{\mu}{2}\right)\right] -\frac{1}{2}(N-1)A\right\} + \braket{S_x}^2\\
        \braket{S_y^2} =& \frac{N}{4}\left\{ 1 + \frac{1}{4}\left(N-1\right) \left[ A - \sqrt{A^2+B^2}\right]\right\},
    \end{align}
    where $A =1 - \cos^{N-2}(\mu)$ and $B = 4\sin\left(\tfrac{\mu}{2}\right)\cos^{N-2}\left(\tfrac{\mu}{2}\right)$.

    \subsection{GHZ states}\label{app:GHZ}
    The GHZ state~\cite{GHZ} is defined by
    \begin{align}
        \ket{\mathrm{GHZ}} &= \frac{1}{\sqrt{2}} \left[\kett{\downarrow}^{\otimes N} + \kett{\uparrow}^{\otimes N}\right],
    \end{align}
    which represents an equal superposition of the collective ground and excited states and thus, is maximally entangled. After the free evolution, the state reads
    \begin{align}
        \ket{\psi_\phi} &= \mathcal{R}_z\left(-\tfrac{\pi}{2N}\right)\mathcal{R}_z(\phi)\ket{\mathrm{GHZ}} = \frac{1}{\sqrt{2}}\left[e^{i\frac{N}{2}\phi-i\frac{\pi}{4}}\kett{\downarrow}^{\otimes N} + e^{-i\frac{N}{2}\phi+i\frac{\pi}{4}}\kett{\uparrow}^{\otimes N}\right],\label{eq:GHZ_phi}
    \end{align}
    where the additional rotation $\mathcal{R}_z\left(-\tfrac{\pi}{2N}\right)$ is applied to shift the optimal working point to $\phi_0=0$, since the prior is centered around $\phi=0$. Equivalently, the prior distribution could be shifted by $\pi/2N$. The expectation value of the parity $\Pi = (-1)^N\sigma_x^{\otimes N}$ is given by
    \begin{align}
        \braket{\Pi(\phi)} =(-1)^N \sin(N\phi).
    \end{align}
    Since $\sigma_x^2 = \id$, the second moment directly yields $\braket{\Pi^2(\phi)}= 1$. Hence, the integrals in Eq.~\eqref{eq:BMSE_linear} become
    \begin{align}
            \int  \dd\phi\,  \mathcal{P}(\phi) \phi\braket{X(\phi)} &= (-1)^N \int  \dd\phi\,  \mathcal{P}(\phi) \phi\sin(N\phi)=(-1)^N N (\delta\phi)^2e^{-N^2 (\delta\phi)^2/2}\\
            \int  \dd\phi\,  \mathcal{P}(\phi)\braket{X^2(\phi)}  &= 1.
    \end{align}
    Consequently, the corresponding optimal linear scaling factor, BMSE and effective measurement uncertainty are given by
     \begin{align}
         a_\mathrm{GHZ} &= (-1)^N N (\delta\phi)^2e^{-N^2(\delta\phi)^2/2}\nonumber\\
         (\Delta\phi_\mathrm{GHZ})^2 &= (\delta\phi)^2\left[ 1 - N^2 (\delta\phi)^2 e^{-N^2(\delta\phi)^2}\right]\label{eq:BMSE_GHZ_Parity}\\
         (\Delta\phi_M^\mathrm{GHZ})^2 &=  \frac{e^{N^2(\delta\phi)^2}}{N^2} - (\delta\phi)^2.\nonumber
     \end{align}
     Due to the binary nature of the parity measurement, the linear estimator is already optimal and thus, saturates the BCRB and coincides with the optimal Bayesian estimator. 
     
     However, the parity measurement can also be mimicked by a projective spin measurement and application of the corresponding optimal Bayesian estimator: For $N$ even, a Ramsey pulse is applied after the free evolution time, implemented by a rotation of $\pi/2$ around the $x$-axis, resulting in the final state
    \begin{align}
            \ket{\psi_f} &= \mathcal{R}_x\left(\tfrac{\pi}{2}\right) \ket{\psi_\phi} = \frac{1}{\sqrt{2}}\frac{1}{\sqrt{2}^N}\left[e^{i\frac{N}{2}\phi-i\frac{\pi}{4}}\left(\kett{\downarrow}-i\kett{\uparrow}\right)^{\otimes N} + e^{-i\frac{N}{2}\phi+i\frac{\pi}{4}}\left(\kett{\uparrow}-i\kett{\downarrow}\right)^{\otimes N}\right].
    \end{align}
     For $N$ odd, calculations are analogous with final rotation around the $y$-axis. Finally, a projective measurement of $S_z$ is performed. Note that the final Ramsey pulse can equivalently be absorbed in the observable, leading to an effective measurement of $S_y$, as for the CSS and SSS protocol. The conditional probabilities are evaluated to read
    \begin{align}
        P\left(x=+\tfrac{N}{2}-N_-|\phi\right) =\frac{1}{2^N}\binom{N}{N_-} \left[1 + (-1)^{\frac{N}{2}+N_-}\sin\left(N\phi\right)\right]
    \end{align}
    where $N_-$ denotes the number of atoms in the ground state. Interestingly, the conditional probabilities for $N_-$ and $N-N_-$ are equal (since $N$ is even), resulting in a vanishing signal $\braket{X(\phi)}\equiv 0$. Nevertheless, with
    \begin{align}
        P\left(x=+\tfrac{N}{2}-N_-\right) &= \int \dd\phi\,  P\left(x|\phi\right)\mathcal{P}(\phi) =  \frac{1}{2^N}\binom{N}{N_-},
    \end{align}
    the optimal Bayesian estimator is given by
    \begin{align}
        \phi_\mathrm{est}\left(x=+\tfrac{N}{2}-N_-\right) &= \frac{1}{P\left(x\right)}\int \dd\phi\,  P\left(x|\phi\right)\mathcal{P}(\phi) \phi= (-1)^{\frac{N}{2}+N_-}N (\delta\phi)^2e^{-N(\delta\phi)^2/2}.
    \end{align}
    and an efficient estimation is possible. Consequently, the optimal estimation strategy distinguishes between even and odd numbers of atoms in the ground state and thus, effectively mimics a parity measurement.

    As a final step, we determine the BQCRB for the GHZ state. Using Eq.~\eqref{eq:GHZ_phi}, we find
    \begin{align}
        \Lambda_{\phi,T}[\rho_\mathrm{in}] = \frac{1}{2}\left[\kett{\downarrow}\braa{\downarrow}^{\otimes N} + e^{iN\phi}\kett{\downarrow}\braa{\uparrow}^{\otimes N} + e^{-iN\phi}\kett{\uparrow}\braa{\downarrow}^{\otimes N} +\kett{\uparrow}\braa{\uparrow}^{\otimes N} \right],
    \end{align}
    which leads to the average state
    \begin{align}
        \overline{\rho} &= \frac{1}{2}\left[\kett{\downarrow}\braa{\downarrow}^{\otimes N} + e^{-N^2(\delta\phi)^2/2}\left(\kett{\downarrow}\braa{\uparrow}^{\otimes N} + \kett{\uparrow}\braa{\downarrow}^{\otimes N} \right)+\kett{\uparrow}\braa{\uparrow}^{\otimes N} \right].\label{eq:rho_av_GHZ}
    \end{align}
    Interestingly, Eq.~\eqref{eq:rho_av_GHZ} is no longer pure due to the averaging and effectively corresponds to a real $2$x$2$-matrix. Hence, using Eq.~\eqref{eq:BQCRB_QFI}, the BQCRB directly follows from
    \begin{align}
        \mathcal{F}_Q[\overline{\rho}] = N^2 e^{-N^2(\delta\phi)^2}
    \end{align}
    and results in the same value as in Eq.~\eqref{eq:BMSE_GHZ_Parity}.

\section{Numerical Methods}
    This section presents the numerical methods employed in this work, including the optimization of the variational quantum circuits, the simulation of the full feedback loop in an atomic clock, the iterative determination of the prior width and the incorporation of dead time noise into the prior phase distribution.

    \subsection{Optimization}\label{app:optimization}
\begin{figure*}[tbp]
            \centering
                \includegraphics[width=\textwidth]{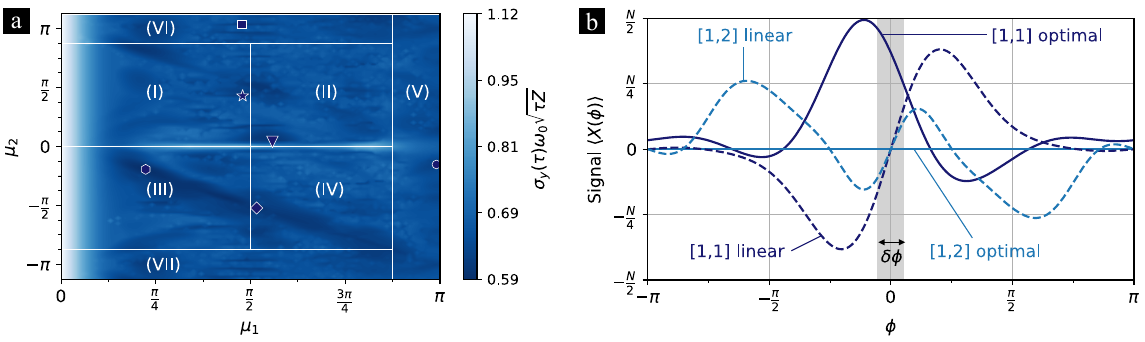}
         
            \caption{(a) Optimization landscape in the $\mu_1$-$\mu_2$-plane for the $[1,1]$ class with $N = 8$ at an interrogation time of $T/Z = 0.1$. The Allan deviation $\sigma_y(\tau)$ is rescaled by the averaging time $\tau$, coherence time $Z$ and transition frequency $\omega_0$. Darker areas correspond to better stability. The optimization areas (I-VII) are separated by white lines, while local minima within these regions are illustrates by symbols. In theory, the lowest instability is achieved by the protocol indicated by the hexagon in area (III), while the other local minima result in a comparable clock stability. (b) Comparison of signals $\braket{X(\phi)}$ of the optimal $[1,1]$ and $[1,2]$ protocols for the linear (dashed) and optimal Bayesian estimator (solid), with $N = 8$ at an interrogation time of $T/Z = 0.1$. The gray shaded region represents the spread of the prior distribution, with its width $\delta\phi$ corresponding to the specific interrogation time.}
                \label{fig:landscape_and_signals}
        \end{figure*}

        The $2+3(n+m)$ variational parameters of the quantum circuits introduced in Sec.~\ref{sec:VariationalClass} are optimized using the \textit{scipy} library in Python~\cite{SciPy}. Due to the vast number of local minima, a global optimization method is required. Specifically, we employ a \textit{differential evolution} approach. As discussed in Sec.~\ref{sec:Simulations}, for fixed interrogation time $T$ and class $[n,m]$, multiple local minima are identified in different regions to mitigate limitations caused by fringe hops. Following Refs.~\cite{Schulte2020,Scharnagl2023}, Fig.~\ref{fig:landscape_and_signals}(a) illustrates a generic optimization landscape in the $\mu_1$-$\mu_2$-plane for the $[1,1]$-class with $N = 8$ at an interrogation time of $T/Z = 0.1$. The $\mu_1$-$\mu_2$-plane is divided into four quadrants (I-IV), with regimes of strong twistings (V-VII) considered separately. Notably, region (VII) is effectively equivalent to (VI) due to the periodic nature of OAT interactions. For variational protocols $[1,m]$ with deeper circuits ($m=2,3$), the twisting strength landscape expands and is divided into eight quadrants considering the twisting strengths $\mu_j$. Local minima identified in these landscapes, represented by distinct symbols, correspond to specific protocols simulated within the full feedback loop of an atomic clock (see Sec.~\ref{sec:limiation_FH}).

    \subsection{Ramsey signals}\label{app:signals}
        The standard protocols, namely CSS, SSS and GHZ protocols, exhibit sinusoidal signals. While CSS and SSS have a dynamic range of $[-\pi/2, +\pi/2]$, allowing for unbiased phase estimation within this interval, the phase is imprinted $N$ times faster for the GHZ state, leading to a correspondingly $N$ times smaller dynamic range. In contrast, the application of variational quantum circuits with multiple layers for state preparation and measurement can generate arbitrary signals. Within the variational classes introduced in Sec.~\ref{sec:VariationalClass}, there are no restrictions on the geometry, and therefore the signal shape is not constrained. While $[n,0]$-protocols yield collective spin measurements that result in sinusoidal signals, increasing $m$ allows for arbitrary signal shapes. Typically, when $m \neq 0$, highly non-sinusoidal signals are generated, where the dynamic range adapts to the interrogation time and the associated frequency fluctuations of the local oscillator.

        Generic Ramsey signals for the $[1,1]$ and $[1,2]$ protocols are illustrated in Fig.~\ref{fig:landscape_and_signals}(b) for $N=8$ at interrogation time $T/Z=0.1$. The signals, associated with the optimal variational parameters, are compared for both the linear and optimal Bayesian estimation strategies. Specifically, the signal for the $[1,1]$-class with the optimal Bayesian estimator corresponds to the hexagon in Fig.~\ref{fig:landscape_and_signals}(a).

        In principle, the estimation strategy does not affect the signal directly, as it is determined solely by the initial state, phase imprint, and measurement. However, the choice of the estimator influences the optimization of the variational parameters, which in turn affects the signal. Consequently, the linear estimation strategy typically results in anti-symmetric signals, at least within the range of the prior distribution. In contrast, the optimal Bayesian estimator can become highly non-linear. As a result, these signals often exhibit strongly non-sinusoidal shapes, lacking symmetry and any apparent relation to the phase. While this may initially seem counterintuitive, this approach achieves lower phase estimation uncertainty when combined with the corresponding estimator, as discussed in Sec.~\ref{sec:VariationalClass}. An example is presented in Sec.~\ref{sec:standard_protocols} for the GHZ protocol with projective spin measurement, where the estimator effectively mimics a parity measurement, while the signal itself vanishes. A similar behavior is observed for the $[1,2]$ protocol with optimal Bayesian estimator in Fig.~\ref{fig:landscape_and_signals}(b).

    \subsection{Clock simulation}\label{app:clock_simulation}
    This section provides a brief overview of the methods used to simulate an atomic clock. The Monte Carlo simulation of the full feedback loop builds on Refs.~\cite{Leroux2017,Schulte2020}. Throughout this work, we assume clock operations with identical interrogation sequences in each clock cycle. Specifically, the interrogation time $T$, dead time $T_D$ and Ramsey protocol are fixed in a single clock run. Consequently, the trace of average frequency deviations $\{\omega_k\}$, where $k$ denotes the index of the clock cycle, can be generated in advance for a given spectral noise density $S_y(f)$ or local oscillator Allan deviation $\sigma_{y,\mathrm{LO}}(\tau)$. In principle, the frequency traces can be obtained by Fourier transformation of the corresponding noise in the Fourier frequency domain. However, this approach becomes computationally expensive for long traces with length $n\gg 1$. While white and random walk frequency noise can be generated using standard methods~\cite{Handbook}, flicker frequency noise is efficiently generated by a sum of multiple damped random walks~\cite{Leroux2017}. Given the frequency trace $\{\omega_k\}$, the clock operation is simulated by implementing the full feedback loop, closely following the basic clock operation outlined in Sec.~\ref{sec:Motivation}. In each clock cycle $k$, the accumulated phase $\phi_k=\omega_k T$ is used to evaluate the conditional probabilities $P(x|\phi_k)$. A random measurement outcome $x_k$ is then sampled based on this statistics. Finally, the frequency is corrected by the servo according to $\omega_{\mathrm{corr},k}$ inferred from the phase estimate $\phi_\mathrm{est}(x_k)$. The servo is implemented by a general linear predictor, presented in Ref.~\cite{Leroux2017}, taking into account 50 previous frequency estimates. This procedure is repeated in each clock cycle for $n=10^7$ cycles in a single clock run. Afterwards, the clock stability is determined for the stabilized frequency trace. 

    Although the Allan deviation $\sigma_y(\tau)$ depends on the total averaging time $\tau$, clock stability is typically characterized by a single value, assuming that the Allan deviation scales as $\sim 1/\sqrt{\tau}$ for $\tau\gg 1~\mathrm{s}$ (cf. Eq.~\eqref{eq:ADEV_BMSE}). Conventionally, this is the Allan deviation at unit averaging time $\sigma_y(\tau=1~\mathrm{s})$. Equivalently, the Allan deviation can be rescaled with the total averaging time, i.e. $\sigma_y(\tau)\sqrt{\tau}$. However, the delayed feedback in a clock operation leads to a large deviation of theoretical prediction and simulation or experiment at $\tau\sim 1~\mathrm{s}$, while the long term stability $\sigma_y(\tau)\propto 1/\sqrt{\tau}$ is achieved for $\tau \gg 1~\mathrm{s}$. Therefore, the stability is evaluated at $\tau\gg 1$ and extrapolated to its hypothetical value at $\tau = 1~\mathrm{s}$ based on the scaling $\sigma_y(\tau)\propto 1/\sqrt{\tau}$. 

    Since clock operation involves stochastic processes, such as frequency trace generation and measurement sampling, results vary across different clock runs.  To ensure robust stability estimates, each setup, defined by fixed ensemble size $N$, interrogation time $T$, dead time $T_D$, and Ramsey sequence, is simulated over 10 independent clock runs. Symbols in the corresponding figures therefore represent mean values, while error bars indicate standard deviations. To include a protocol in the results, we impose the stringent criterion that no fringe hops occur across $10^8$ total clock cycles, since even a single fringe hop results in a complete loss of clock stability.

    Fringe hops represent a substantial limitation at long interrogation times,  where they can dominate even over the coherence time limit for small ensembles. Additionally, fringe hops arise for the variational protocols at the plateau of the OQI (cf. Sec.~\ref{sec:limiation_FH}). To address this, we simulate several protocols from the optimization (see above) for a given interrogation time $T$ and variational class $[1,m]$. As a result, the theoretically optimal protocol may be limited by fringe hops in the full feedback loop and thus, a different protocol performs best in simulations, leading to substantial deviations between theoretical predictions and numerical simulations. To give an example, the variational protocol associated with the hexagon in area (III) in Fig.~\ref{fig:landscape_and_signals} achieves the lowest Allan deviation in theory, while it is limited by fringe hops in the numerically simulated full feedback loop. Instead, the protocol corresponding to the circle in area (V) performs best in numerical simulations, resulting in a significant deviation to theoretical prediction. In extreme cases, all simulated protocols may suffer from fringe hops, and no data points are shown at these interrogation times. However, the susceptibility to fringe hops at the OQI plateau decreases with increasing circuit complexity $m$ (cf. Sec.~\ref{sec:limiation_FH}). Thus, simulation results for protocols with the lowest depth $m\leq 3$ that are not limited by fringe hops are shown.

    \subsection{Iterative prior width}\label{app:iterated_prior}
    Eq.~\eqref{eq:PowerLaw} provides a good approximation of the prior phase width $\delta\phi$ in the regime of large ensembles $N$ and long interrogation times $T$, as demonstrated in Refs.~\cite{Leroux2017,Kaubruegger2021,Schulte2020}. However, as discussed in the main text, the prior width $\delta\phi$ and estimation error $\Delta\phi$ mutually influence each other in the full feedback loop of an atomic clock. Moreover, any model of the prior width can only capture the true residual noise to a certain degree. Consequently, an on-device optimization, as utilized in Ref.~\cite{Marciniak2022}, would most accurately reflect the experimental conditions and thus, yield the best results. However, this approach has several disadvantages. First, it precludes theoretical predictions and \textit{ab initio} studies of clock stability, making it impossible to exclude protocols prone to fringe hops, for instance. Second, it is exceptionally demanding in terms of experimental time. While the variational parameters need only be optimized for individual clock runs, evaluating the Allan deviation as a cost function requires a sufficiently long averaging time $\tau$ for each optimization step to achieve the long-term scaling according to $1/\sqrt{\tau}$. Unlike Bayesian phase estimation, which can focus on single interrogation cycles, on-device optimization for clock stability must account for time-varying frequency deviations $\omega$ across cycles. As a result, on-device optimization using the Allan deviation as a cost function is impractical. 
    
    To overcome these challenges, we focus on modeling the prior knowledge according to Eq.~\eqref{eq:prior} and iteratively adjust the prior width $\delta\phi$ to account for the closed feedback loop dynamics. The general strategy involves simulating the full feedback loop multiple times and using the results from previous simulations to estimate the prior width for the subsequent iteration stage. This procedure is repeated until convergence is achieved. In each iteration stage, the frequency deviation $\omega$ at the end of the Ramsey dark time is recorded and interpolated as a function of the interrogation time at a fixed ensemble size. Directly applying this iterative method to the variational protocols would lead to the same issues discussed above. Therefore, it is advantageous to use fixed and reliable protocols, such as CSS and SSS, to ensure consistency. Additionally, comparing results across protocols would be cumbersome, as each protocol yields a distinct prior width and corresponding OQI. Instead, we approximate the prior width $\delta\phi$ for a fixed ensemble size through the following iterative stages:
    \begin{itemize}
        \item Stage 0 (Initialization): Start with a heuristic prior width, where $\delta\phi$ is interpolated linearly on a log-log scale between $(\delta\phi)^2 = (T/Z)^{4/3}N^{-1/4}$ for $T/Z=0.01$ and the value given by Eq.~\eqref{eq:PowerLaw} for $T/Z=1$. Using this prior width, simulate the CSS protocol with the optimal Bayesian estimator and record the resulting frequency deviations $\{\omega_k\}$.
        \item Stage 1 (Refinement): Use the recorded $\{\omega_k\}$ from the previous simulation to determine the corresponding prior phase distribution. Fit this distribution to a Gaussian as described in Eq.\eqref{eq:prior} to obtain an updated prior width $\delta\phi$. Plot $\delta\phi$ as a function of interrogation time and fit it with a fifth-order polynomial. Exclude prior widths for interrogation times where fringe hops limit stability and additionally add the value from Eq.\eqref{eq:PowerLaw} at $T/Z=1$. Simulate the SSS protocol with the updated prior width.
        \item Stages $2,3,\ldots$ (Iteration): Repeat the refinement process. 
    \end{itemize}
    Convergence is typically achieved after stage 3, even for small ensembles, as the prior width from stage 4 introduces only negligible adjustments. This convergence is generically illustrated in Fig.~\ref{fig:prior_variance}(a). Hence, the prior width from stage 3 is adopted to model a realistic atomic clock scenario used in Sec.~\ref{sec:Simulations} and Sec.~\ref{sec:deadtime_setups}. While this iterative approach provides a reasonable approximation of the closed feedback loop dynamics, it remains a simplification. Consequently, deviations between theoretical predictions and numerical simulations may still arise, as discussed above and in the main text.

    \subsection{Prior width with dead time}\label{app:prior_deadtime}
\begin{figure*}[tbp]
            \centering
                \includegraphics[width=1\textwidth]{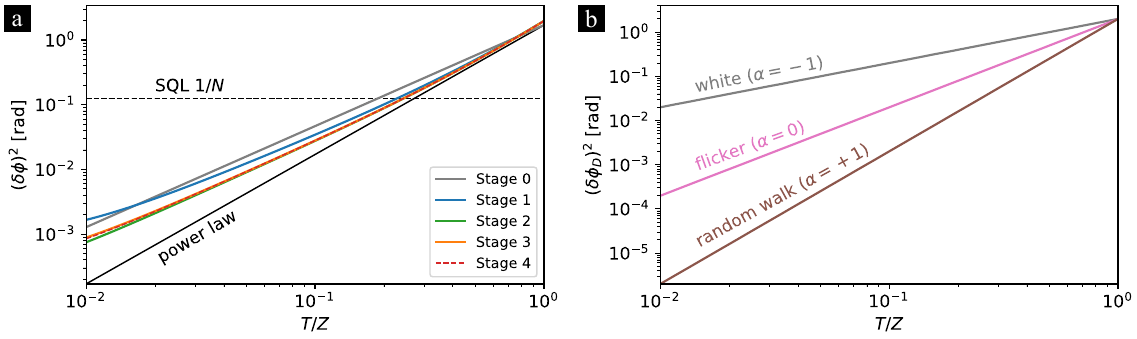}
         
            \caption{(a) Convergence of the prior variance $(\delta\phi)^2$ in the iterative approach for $N=8$. The distinct iteration stages are illustrated by colored lines. Additionally, the power law scaling Eq.~\eqref{eq:PowerLaw} and the SQL $1/N$ are shown for comparison. (b) Additional noise due to dead time characterized by the associated prior variance $(\delta\phi_D)^2$ given in Eq.~\eqref{eq:varianceDeadtime} for white (gray), flicker (pink) and random walk frequency noise (brown). Mean values are averaged over 10 independent runs.}
                \label{fig:prior_variance}
        \end{figure*}

    As discussed in the main text, the additional noise introduced during dead time can be approximated as white noise in the asymptotic limit of many clock cycles. The corresponding prior width $\delta\phi_D$ is determined by simulating the uncorrected frequency trace of the local oscillator and quantifying the noise accumulated during the cycle duration $T_D$. Specifically, the new frequency deviations $\omega_{\mathrm{new},k} = \omega_{k}-\omega_{k-1}$ are recorded for each cycle $k$, representing the differences between consecutive cycles. Using the recorded $\omega_{\mathrm{new},k}$, the phase distribution associated with a hypothetical phase shift during $T_D$ is evaluated, and the  corresponding prior width $\delta\phi_D$ is extracted. Simulations confirm the power-law scaling predicted by Eq.~\eqref{eq:varianceDeadtime}, as illustrated in Fig.~\ref{fig:prior_variance}(b).

\end{document}